\documentclass[iop]{emulateapj}
\newcommand{\kms}{km\ s$^{-1}$}
\newcommand{\kmsmpc}{km\ s$^{-1}$\ Mpc$^{-1}$}
\newcommand{\ergscm}{erg\ s$^{-1}$ cm$^{-2}$} 
 
\newcommand{\OII}{[\ion{O}{2}]}

\shorttitle{Analysis of X-ray AGN in High-z LSS}
\shortauthors{Rumbaugh et al.}

\begin{document}

\title{The Evolution and Environments of X-ray Emitting Active
  Galactic Nuclei in High-Redshift Large-Scale Structures}

\author{
  N. Rumbaugh\altaffilmark{1}, D. D. Kocevski\altaffilmark{2}, R. R. Gal\altaffilmark{3}, B. C. Lemaux\altaffilmark{1}, L. M. Lubin\altaffilmark{1}, C. D. Fassnacht\altaffilmark{1}, E. J. McGrath\altaffilmark{2}, G. K. Squires\altaffilmark{4}
}

\email{Electronic address: narumbaugh@ucdavis.edu}

\altaffiltext{1}{Department of Physics, University of California,
  Davis, 1 Shields Avenue, Davis CA 95616, USA}
\altaffiltext{2}{University of California
    Observatories/Lick Observatory, University of California, Santa
    Cruz, CA 95064, USA}
\altaffiltext{3}{Institute for Astronomy, University of Hawai'i,
     2680 Woodlawn Drive, HI 96822, USA}
\altaffiltext{4}{Spitzer Space Center, California Institute of Technology, M/S
    220-6, 1200 E. California Blvd., Pasadena, CA 91125, USA}

\begin{abstract}

We use deep Chandra imaging and an extensive optical spectroscopy
campaign on the Keck 10-m telescopes to study the properties of X-ray
point sources in two isolated X-ray selected clusters, two
superclusters, and one `supergroup' at redshifts of $z\sim0.7-0.9$. We
first study X-ray point sources using the statistical measure of
cumulative source counts, finding that the measured overdensities are
consistent with previous results, but we recommend caution
in overestimating the precision of the technique. Optical spectroscopy
of objects matched to X-ray point sources confirms a total of 27 AGNs
within the five structures, and we find that their host galaxies tend
to be located away from dense cluster cores. More than $36\%$ of the
host galaxies are located in the `green valley' on a color magnitude
diagram, which suggests they are a transitional population. Based on
analysis of \OII\ and H$\delta$ line strengths, the average spectral properties of the AGN host
galaxies in all structures indicate either on-going star formation or
a starburst within $\sim 1$ Gyr, and that the host galaxies are
younger than the average galaxy in the parent population. These
results indicate a clear connection between starburst and nuclear
activity.  We use
composite spectra of the spectroscopically confirmed members in each
structure (cluster, supergroup, or supercluster) to separate them
based on a measure of the overall evolutionary state of their
constituent galaxies. We define structures as having more evolved
populations if their average galaxy has lower EW(\OII) and
EW(H$\delta$). The AGNs in the more evolved structures have lower
rest-frame 0.5--8 keV X-ray luminosities (all below $10^{43.3}$ erg
s$^{-1}$) and longer times since a starburst than those in the unevolved
structures, suggesting that the peak of both star formation and AGN
activity has occurred at earlier times. With the wide range of
evolutionary states and timeframes in the structures, we use our
results to analyze the evolution of X-ray AGNs and evaluate potential
triggering mechanisms.

\end{abstract}


\section{Introduction}

In recent years, research has illuminated a link between active
galactic nuclei (AGNs) and the evolution of their host
galaxies. Studies have found that a galaxy's central black hole mass
is correlated with the mass of the central bulge \citep[e.g.,
][]{FM00,geb00,trem02,shields03,MH03,HR04}. In the local universe,
high-luminosity AGNs are preferentially found in early-type galaxies
with young mean stellar ages \citep{kauf03,heck04}. As the redshift
increases, the cosmic rates of both star formation and AGN activity
increase \citep{BoyTer98,bluck11}, and, at high redshift, intensely star-forming submillimeter galaxies have been found to have an AGN fraction of $\sim$20-30$\%$ \citep{laird10,georg11}.

The connection between AGNs and host evolution is also observed in
optical colors. Studies have shown that galaxies are organized based
on rest-frame color, separating into a `red sequence' and a `blue
cloud' on a color-magnitude diagram \citep[CMD; e.g.,][]{strat01,baldry04,weiner05}. The sparsely populated
`green valley' between these regions is thought to be a transitional
area, where blue cloud galaxies are rapidly migrating onto the red
sequence after the cessation of star formation \citep{faber07}. A
proposed mechanism for this quick transition is AGN feedback, where
the AGN activity, driven by major mergers or tidal interactions,
causes a truncation of star formation and leads to the color evolution
onto the red sequence
\citep{hopkins05,spring05,hopkins07,somer08}. This theory is supported
by a number of studies which find an association between AGN activity
and the green valley
\citep{nandra07,georg08,silver08,koc09b,hickox09,scha09}. However, some studies using mass-selected samples have found a more uniform color distribution of AGN hosts \citep{silver09b,xue10}. In addition, \citet{card10} have found that $\sim75\%\ $of green valley AGN hosts are dust-reddened members of the blue cloud, although these results are apparently in conflict with the more recent studies of \citet{ros11}. Studies examining AGNs for recent merger activity have also found mixed
results \citep[e.g.,][]{sanchez05,pierce07,georg08,koc09a,koc11a}.

\citet{men01} offer an alternative to major merger driven AGNs, where AGNs are instead triggered by minor mergers, mainly in
red galaxies. Star formation in this scenario would only be briefly re-ignited at the cores of many
of these systems and the galaxies would return to the red sequence
after its cessation. These AGNs would then represent a red sequence
population evolving in luminosity space rather than blue cloud
galaxies migrating onto the red sequence. In support of this theory,
several studies have found AGN activity associated with blue
early-type galaxies or red galaxies with blue cores
\citep{lee06,martel07,koc09b}. Additionally, \citet{CO07} propose that recycled stellar material in elliptical galaxies could fuel central starbursts and black hole accretion. Without precise observations of morphologies, the effects of these two models would appear similar. 

Other plausible models exist as well, some important only in certain regimes. For example, at high redshift ($z \gtrsim 2$), \citet{bour11} propose that violent disk instabilities could lead to powerful nuclear activity, while \citet{HH06} present a model where lower luminosity AGNs could be triggered by mild disk instabilities or gas funneled through bars. Note that the properties of individual AGNs are widely varying, and it could be the case that any number of modes contribute to AGN triggering, with different contributions in different mass and/or redshift regimes. 

To better understand how AGN are triggered and the role they play in
galaxy evolution, it is useful to study large-scale structures (LSSs) at high
redshift. These environments contain a large number of galaxies in the
process of transitioning from actively star-forming to passive, and
there is evidence that nuclear activity increases at higher redshifts \citep{BoyTer98,east07,kart07,bluck11}.

To this end, we have examined the X-ray selected AGN population within
five LSSs at redshifts of
$z\approx0.7$-0.9. These structures are in varying states of evolution
and include complex superclusters, an interacting supergroup, and
isolated clusters. Each has been studied extensively by the
Observations of Redshift Evolution in Large-Scale Environments
(ORELSE) survey, which is searching for LSSs in the
vicinities of 20 known clusters between $z=0.6$-1.3. The survey has
compiled extensive multi-wavelength datasets for each structure, which
include multi-band optical, radio, and X-ray imaging, as well as
thousands of spectroscopic redshifts \citep{lub09}.

In this paper, we present analyses of X-ray point sources and the
properties of the AGN population within the following five LSSs: the Cl1604 and Cl1324 superclusters at $z\approx0.9$ and
$z\approx0.76$, respectively, the Cl0023+0423 supergroup at $z\approx0.82$,
and two X-ray selected and relatively relaxed, isolated clusters,
RXJ1757.3+6631 at $z=0.69$ and RXJ1821.6+6827 at $z=0.84$. We study
this sample with a combination of Chandra X-ray data, optical imaging,
and near-IR and optical spectroscopy.  For our cosmological model, we
assume $\Omega_m = 0.3$, $\Omega_\Lambda = 0.7$, and $h_{70} =
H_0/70\ $\kmsmpc. 

We discuss the clusters and superclusters in our survey in Section
\ref{sec:samp}. Observations, data reduction, and techniques are
discussed in Section \ref{sec:red}. The global properties of our
sample are discussed in Section \ref{sec:globchar}. The statistical
measurements of cumulative source counts are covered in Section
\ref{sec:CSC}. Analysis of the AGNs is presented in Section
\ref{sec:AGN}.

\section{The ORELSE Structure Sample}
\label{sec:samp}

In this section we describe the five structures in our sample, which
are succintly summarized in Table \ref{strsumtab}. The redshift
boundaries used in the following analyses were determined by visually
examining each structure's redshift histogram. Delineating where
structures end is not straightforward, with some having associated
filaments or possible nearby sheets. The redshift boundaries are
chosen with the aim to include all galaxies which
could be part of each overall LSS.

\begin{deluxetable*}{lccccccccc}
\tablecaption{Properties of Observed ORELSE Structures}
\tablehead{
\colhead{\footnotesize{Structure}}
 & \colhead{\footnotesize{R.A.}\tablenotemark{a}}
 & \colhead{\footnotesize{Dec.}\tablenotemark{a}}
 & \colhead{\footnotesize{$\langle z\rangle$}}
 & \colhead{\footnotesize{$z$ Lower}}
 & \colhead{\footnotesize{$z$ Upper}}
 & \colhead{\footnotesize{Num. of}}
 & \colhead{\footnotesize{$\sigma$}}
 & \colhead{\footnotesize{Confirmed}} 
 & \colhead{\footnotesize{Confirmed}} \\
   \colhead{{ }}
 & \colhead{\footnotesize{(J2000)}}
 & \colhead{\footnotesize{(J2000)}}
 & \colhead{{ }}
 & \colhead{\footnotesize{Bound}}
 & \colhead{\footnotesize{Bound}}
 & \colhead{\footnotesize{Clusters/Groups}}
 & \colhead{\footnotesize{Range\tablenotemark{b}}}
 & \colhead{\footnotesize{Members\tablenotemark{c}}}
 & \colhead{\footnotesize{AGN\tablenotemark{c}}}
}
\startdata
Cl1604 & 16\ 04\ 15 & +43\ 16\ 24 & 0.90 & 0.84 & 0.96 & 10 & 300-800 & 531 & 10 \\
Cl0023 & 00\ 23\ 51 & +04\ 22\ 55 & 0.84 & 0.82 & 0.87 & \ 4 & 200-500  & 244 & \ 7 \\
Cl1324 & 13\ 24\ 45 & +30\ 34\ 18 & 0.76 & 0.65 & 0.79 & 10 & 200-900 & 393 & \ 6 \\
RXJ1821 & 18\ 21\ 32.4 & +68\ 27\ 56 & 0.82 & 0.80 & 0.84 & \ 1 & $910\pm80$  & \ 90 & \ 3 \\
RXJ1757 & 17\ 57\ 19.4 & +66\ 31\ 29 & 0.69 & 0.68 & 0.71 & \ 1 & $650\pm120$  & \ 42 & \ 1 \\
\enddata
\label{strsumtab}
\tablenotetext{a}{ \footnotesize{Coordinates for Cl1604, Cl0023, and Cl1324 are the approximate central position of these large-scale structures, while those for RXJ1821 and RXJ1757 are given as the centroid of the peak of diffuse X-ray emission associated with the respective cluster.}}
\tablenotetext{b}{\footnotesize{In units of \kms. For Cl0023, Cl1604, and Cl1604, this measurement is the range of velocity dispersions of groups and clusters within the structure. For RXJ1821 and RXJ1757, it is the dispersion of the single cluster. All velocity dispersions are measured within 1 Mpc.}}
\tablenotetext{c}{\footnotesize{Spectroscopically confirmed objects ($Q=3$,4) within the redshift bounds of the structure; see Section \ref{sec:optobs} for quality flag details.}}
\end{deluxetable*}

\subsection{The Cl1604 Supercluster}

The Cl1604 supercluster at $z\approx 0.9$ is one of the largest structures studied at high redshift. It consists of at least 10 clusters and groups and spans $100\ h_{70}^{-1}\ $Mpc along the line of sight and $13\ h_{70}^{-1}\ $Mpc in the plane of the sky \citep{lubin00,GalLub04,gal08,lemaux09}. The massive member clusters Cl1604+4304 and Cl1604+4321 were first discovered in the optical survey of \citet{gunn86}. The proximity of the clusters suggested that they were components of a larger structure. Further wide field imaging has revealed 10 distinct red galaxy overdensities, suggesting the existence of a supercluster \citep{lubin00,GalLub04,gal08}. Spectroscopic observations have confirmed four of the overdensities to be clusters with velocity dispersions in excess of 500 \kms, while four others were confirmed to be poor clusters or groups with dispersions in the range 300-500 \kms\ \citep{post98,post01,gal05,gal08}. 

The two most massive clusters in Cl1604 have associated diffuse X-ray emission. Cl1604+4304 and Cl1604+4314, hereafter Clusters A and B, have measured bolometric X-ray luminosities of $15.76 \pm 1.48\ $ and $11.64\pm  1.49 \times 10^{43}\ h_{70}^{-2}\ $erg s$^{-1}\ $ and X-ray temperatures of $3.50^{+1.82}_{-1.08}\ $ and $1.64^{+0.65}_{-0.45}\ $keV, respectively \citep{koc09a}. While these values place Cluster A on the $\sigma$-$T$ curve for virialized clusters, Cluster B is well off from it, suggesting that Cluster A is relaxed while Cluster B is not. All other clusters have only an upper limit on their bolometric luminosity of $7.4\times 10^{43}\ h_{70}^{-2}\ $erg s$^{-1}\ $\citep{koc09a}. 

While many galaxies in Cl1604 have substantial \OII\ emission, near-infrared spectroscopy has shown that a significant portion of this emission is due to contributions from low-ionization nuclear emission-line regions (LINERs) and Seyferts \citep{lemaux10}. Also, \citet{koc11b} studied 24$\mu$m selected galaxies in and around three clusters and three groups in Cl1604 using the Multiband Imaging Photometer for {\it Spitzer} \citep[MIPS;][]{rieke04} and found evidence for recent starburst activity and an infalling population. Analysis of the morphologies using the Advanced Camera for Surveys \citep[ACS;][]{ford03} on the Hubble Space Telescope (HST) revealed that many of these 24$\mu$m bright galaxies were disturbed, indicating mergers and interactions were likely responsible for starburst activity.

We refer the reader to \citet{koc09a}, \citet{gal08}, and \citet{lemaux11}, for more details on the data processing, supercluster properties, and observations. 

\subsection{Cl0023+0423}

The Cl0023+0423 structure at $z\approx 0.84$, hereafter Cl0023, was also discovered as an overdensity in the optical survey of \citet{gunn86}. The structure was later observed by \citet{oke98} using the Low-Resolution Imaging Spectrograph \citep[LRIS;][]{oke95} on the Keck 10m telescope, where the overdensity was resolved into two structures. Further study has shown that the structure consists of four merging galaxy groups separated by approximately 3000 \kms\ in radial velocity and $\sim0.25$ Mpc on the plane of the sky \citep{lub98,lub09}. The constituent groups have measured velocity dispersions within $1.0\ h_{70}^{-1}\ $Mpc of 480$\pm 170$, 430$\pm 70$, 290$\pm 80$, and 210$\pm 30\ $ \kms\ \citep{lub09}. Simulations suggest that the groups will merge to form a cluster of mass $\sim 5\times 10^{14} M_{\odot}$ within $\sim$1 Gyr \citep{lub98}. 

\citet{lub09} found Cl0023 to have a large blue population, with $51\%\ $of galaxies bluer than their red galaxy color-color cut, down to an $i'$-band magnitude of 24.5. Spectroscopic analysis found that $\sim80\%\ $of galaxies had measurable \OII\ emission, which, because of the large blue population, is most likely due to ongoing star formation\footnote{Refer to Section \ref{sec:globchar} for a discussion of \OII\ emission.}. 

We refer the reader to \citet{lub09} and \citet{koc09c} for more details on the supergroup properties and observations. 

\subsection{The Cl1324 Supercluster}

The Cl1324 supercluster is a LSS at $z\approx 0.76$. The two most
massive clusters in the structure, Cl1324+3011 at $z=0.76$ and
Cl1324+3059 at $z=0.69$, were first discovered in the optical survey
of \citet{gunn86}. Because of the proximity of the clusters on the sky
and in redshift space, this structure was chosen for the ORELSE survey
to investigate the possible existence of structure in the
field. Wide-field imaging has revealed a total of ten clusters and
groups, detected through red galaxy overdensities, and, despite
extensive spectroscopy, only four have been spectroscopically
confirmed as constituent clusters or groups (see Gal et al. 2012, in
preparation).

\begin{deluxetable}{lcc}
\tablecaption{Spectroscopic Observation Characteristics}
\tablehead{
\colhead{\footnotesize{Sructure}}
 & \colhead{\footnotesize{Central}}
 & \colhead{\footnotesize{ Approx. Spectral}}\\
 \colhead{\footnotesize{ }}
& \colhead{\footnotesize{$\lambda$ (\AA) }}
& \colhead{\footnotesize{Coverage (\AA)}}}
\startdata
Cl0023 & 7500-7850 & 6200-9150\\
Cl1604 & 7700 & 6385-9015\\
Cl1324 & 7200 & 5900-8500\\
RXJ1821 & 7500-7800 & 6200-9100\\
RXJ1757 & 7000-7100 & 5700-8400\\
\enddata
\label{speccovtab}
\end{deluxetable}

Cl1324+3011 was previously studied in \citet{lubin02} and \citet{lubin04}, where a velocity dispersion of $1016^{+126}_{-93}$ \kms\ and a temperature of $2.88_{-0.49}^{+0.71}\ $keV, using XMM-Newton, were measured for the cluster. According to these measurements, the cluster does not fall close to the $\sigma$-$T$ curve for virialized clusters, which would imply that it is not well relaxed. New Chandra results for Cl1324 are presented in N. Rumbaugh et al. (2012, in preparation), including new X-ray temperatures for Cl1324+3011 and Cl1324+3059. In addition, we present here updated velocity dispersions for these two clusters of $930\pm120$ and $870\pm120\ $\kms, respectively. The new measurements place Cl1324+3011 closer to the $\sigma$-T curve for virialized clusters, only offset by $\sim1\sigma$. While the older measurements suggested the cluster was not relaxed, the new measurements are more consistent with virialization. Similarly to Cl1324+3011, Cl1324+3059 is offset from the curve, but still by less than 1$\sigma$.

The photometric and spectroscopic observations of the Cl1324 supercluster will be covered in full in R. R. Gal et al. (2012, in preparation). In this paper, we present redshift histograms of the full structure and velocity dispersions for the four confirmed groups and clusters, as well as the Chandra observations. 

\begin{deluxetable*}{lcccccc}
\tablecaption{Numbers of Spectroscopic Targets, X-ray Sources, and Member Galaxies}
\tablehead{  
\colhead{\footnotesize{Structure}}
 & \colhead{\footnotesize{Spectroscopic}}
 & \colhead{\footnotesize{Spectroscopic}}
 & \colhead{\footnotesize{X-ray Sources,}}
 & \colhead{\footnotesize{X-ray Sources,}} & \colhead{\footnotesize{Attempted}}
 & \colhead{\footnotesize{Confirmed}}\\
\colhead{\footnotesize{ }}
 & \colhead{\footnotesize{Targets}}
 & \colhead{\footnotesize{Redshifts\tablenotemark{a}}}
 & \colhead{\footnotesize{$>$3$\sigma$ ($>$2$\sigma$)\tablenotemark{b}}}
 & \colhead{\footnotesize{Matched\tablenotemark{c}}}
 & \colhead{\footnotesize{Redshifts\tablenotemark{d}}}
 & \colhead{\footnotesize{Redshifts\tablenotemark{a}}}
}
\startdata
Cl1604	& 2465 	& 1785 	& 158(213)	& 112(128) 	& 43(48)	& 38(42)	\\
Cl0023	& 1136 	& \ 892 & \ 94(133) 	& \ \ 58(72)   	& 39(49)	& 26(32)	\\
Cl1324	& 1419 	& 1155 	& 174(217) 	& 126(133)	& 38(40)	& 28(30)	\\
RXJ1821	& \ 351 & \ 306 & 102(132) 	& \ \ 64(72)   	& 15(18)	& 10(13)	\\
RXJ1757	& \ 549 & \ 421 & \ 87(107) 	& \ \ 57(62)   	& 18(19)	& \ \ 9(9)	\\
\enddata
\label{srcsum}
\tablenotetext{a}{\footnotesize{Only includes redshifts with quality flag Q=-1,3 or 4; see Section \ref{sec:optobs} for flag details.}}
\tablenotetext{b}{\footnotesize{Includes sources with a significance $>$3$\sigma$ ($>$2$\sigma$) in at least one of the three bands: soft, hard, or full.}}
\tablenotetext{c}{\footnotesize{X-ray sources matched to optical counterparts.}}
\tablenotetext{d}{\footnotesize{Includes all X-ray sources that were targeted for spectroscopy, regardless of the quality of measured redshift.}}
\end{deluxetable*}

\subsection{RX J1821.6+6827}

The X-ray-selected cluster RXJ1821.6+6827, hereafter RXJ1821, at
$z=0.82$, was the highest redshift cluster discovered in the ROSAT
North Ecliptic Pole (NEP) survey, where it is also referred to as
NEP5281 \citep{gioia03,hen06}. Using XMM-Newton data, the cluster was
found to have slightly elongated diffuse X-ray emission with a
measured bolometric luminosity of $1.17^{+0.13}_{-0.18} \times
h_{70}^{-2}\ {\rm erg}\ s^{-1}$ and a temperature of
$4.7^{+1.2}_{-0.7}\ $keV \citep{gioia04}. The same study measured a
velocity dispersion of ${775}^{+182}_{-113}\ $\kms\ using 20 cluster
members. Later analysis by \citet{lub09} used 40 galaxies within 1 Mpc
to measure a velocity dispersion of $926\pm 77\ $\kms. Redshift
histograms of RXJ1821 are characteristic of a single, isolated
structure, although a small kinematically associated group has been
detected to the south \citep{lub09}. While the temperature and
dispersion measurements place the cluster near the $\sigma$-$T$
relation for virialized clusters, the elongated X-ray emission could
be indicative of still ongoing formation of the cluster.

\citet{lub09} measured a blue fraction of only $24\%\ $for RXJ1821, down to a magnitude limit of $i' = 24.5$. They found a population dominated by massive, old (formation epoch of $z_f \sim 2-3$) galaxies, along with fainter galaxies that were quenched more recently. They also found that $36\%\ $of galaxies had detectable \OII\ emission. Near-IR spectroscopy of a subset of these \OII\ emitting galaxies suggests that some of the emission is due to LINER or AGN activity \citep{lemaux10}. 

We refer the reader to \citet{lub09} for more details on the data processing, cluster properties, and observations.  

\subsection{RXJ1757.3+6631}

The $z=0.69$ cluster RXJ1757.3+6631, hereafter RXJ1757, was discovered as part of the ROSAT NEP survey, where it is also identified as NEP200 \citep{gioia03}. \citet{gioia03} found the structure to have an X-ray luminosity of 8.6$\times 10^{43}\ h^{-2}_{70}\ $erg s$^{-1}$ in the 0.5-2.0 keV band. The structure is dominated by a single, large cluster. In this paper, we present a velocity dispersion, redshift histograms, and analysis of X-ray point sources for RXJ1757, none of which have been previously published. 

\section{Observations and Reduction}
\label{sec:red}
\subsection{Optical and NIR Observations}
\label{sec:optobs}

Ground-based optical imaging data were obtained with the Large Format Camera \citep[LFC;][]{simcoe00} on the Palomar 5m telescope. Observations were taken using the Sloan Digital Sky Survey (SDSS) $r'$, $i'$, and $z'$ filters. The 5 $\sigma$ point source limiting magnitudes for the five fields ranged from 25.5-25.1, 25.0-24.5, and 23.6-23.3, in the $r\arcmin$, $i\arcmin$, and $z\arcmin$ bands, respectively.

Cl1604 was also imaged using ACS. The HST imaging for Cl1604 consists of 17 ACS pointings designed to image nine of the ten galaxy density peaks in the field. Observations were taken using the F606W and F814W bands. These bands roughly correspond to broadband V and I, respectively.

Our photometric catalog is complemented by an unprecedented amount of spectroscopic data. For this part of the study, we used the Deep Imaging Multi-Object Spectrograph \citep[DEIMOS;][]{faber03} on the Keck II 10m telescope. In addition, Cl1604 and RXJ1821 have some LRIS coverage \citep[see][]{oke98,GalLub04,gioia04}. DEIMOS has a wide field of view ($16\farcm9 \times 5\farcm0$), high efficiency, and is able to position over 120 targets per slit mask, which makes the instrument ideal for establishing an extensive spectroscopic catalog. We targeted objects down to an $i'$-band magnitude of 24.5. On DEIMOS, we used the 1200 line mm$^{-1}$ grating, blazed at 7500 \AA, and 1$\arcsec$ slits. These specifications create a pixel scale of 0.33 \AA\ pixel$^{-1}$ and a FWHM resolution of $\sim1.7$\AA, or 68 \kms. The central wavelength was varied from structure to structure and sometimes between different masks for the same field. Central wavelengths for the spectroscopic observations for the five fields and the approximate spectral coverages are displayed in Table \ref{speccovtab}. When more than one central wavelength was used per field, a range is given. Total exposure times for the observations are in the range of 1-4 hours per mask.

Spectroscopic targets were chosen based on color and magnitude. The number of spectroscopic targets in each field is shown in Table \ref{srcsum}. Redshifts were determined or measured for all targets and given a quality flag value, $Q$, where $Q = 1$ indicates that we could not determine a secure redshift, $Q = 2$ means a redshift was obtained using features that were only marginally detected, $Q = 3$ means one secure and one marginal feature were used to calculate the redshift, and $Q = 4$ means at least two secure features were used. Those sources determined to be stars were given a flag of $-1$. See \citet{gal08} for more details on quality flags and the spectral targeting method. For our analysis, redshifts with $Q = -1$,3,4 were deemed satisfactory, and the number of such sources in each field is shown in Table \ref{srcsum}. 

Spectroscopic data have been previously presented for the Cl1604 supercluster, Cl0023, and RXJ1821 as part of the ORELSE survey. We present new data for each of these structures, as well as for the Cl1324 supercluster and RXJ1757. We include ten DEIMOS masks for Cl1324, with exposure times ranging from 6635 s to 10800 s, and four DEIMOS masks for RXJ1757, with exposure times ranging from 7200 s to 14,730 s. We include a total of 18 spectroscopic masks for Cl1604, six more than what were included in \citet{gal08}\footnote{See \citet{lemaux11}) for a detailed description of all Cl1604 spectroscopic observations and results.}. We include nine total masks for Cl0023, four more than in \citet{koc09c}. We include three masks for RXJ1821, one more than in \citet{lub09}. Many of the new targets were optical counterparts to X-ray sources. From our measured redshifts with $Q=3$ or 4 from the new masks, we find 114 new galaxies in Cl1604 within $0.84 < z < 0.96$, 104 new galaxies in Cl0023 within $0.82 < z < 0.87$\footnote{Note that while we define the bounds of Cl0023 as $0.82 < z < 0.87$ here, \cite{koc09c} use $0.820 < z < 0.856$, excluding a sheet of galaxies at $z\approx0.86$-0.87. In the smaller range of $0.820 < z < 0.856$, we add 96 new galaxies. We use a wider redshift range in order to be consistently liberal in our structure boundaries.}, and five new galaxies in RXJ1821 within $0.80 < z < 0.84$.

\subsection{X-ray Observations}

All X-ray imaging of the clusters was conducted with the Advanced CCD Imaging Spectrometer (ACIS) of the Chandra X-ray Observatory, using the ACIS-I array (PI: L. M. Lubin)\footnote{Details of the Chandra observations, such as ID numbers, are described in N. Rumbaugh et al. (2012, in preparation).}. This array has a $16\farcm9 \times 16\farcm9$ field of view. Some of the five structures were imaged with one pointing and some with two, but every pointing had the same approximate total exposure time of 50 ks. Cl0023, RXJ1821, and RXJ1757 were each imaged with one pointing of the array. Cl1604 and Cl1324, with angular sizes in excess of 20$\arcmin$, were observed with two pointings each. For Cl1604, the two pointings are meant to cover as much of the structure as possible, and there is a small overlap ($\sim 30$ arcminutes$^2$). For Cl1324, the two pointings are centered near the two largest and originally discovered clusters, Cl1324+3011 and Cl1324+3059. There is an approximately 13$\arcmin$ gap between the north and south pointings. 

In this paper, we present new Chandra data for Cl1324, RXJ1821, and RXJ1757. 

\begin{deluxetable}{lccc}
\tablecaption{Count-rate to Flux Conversion Factors and \ion{H}{1} Column Densities}
\tablehead{
\colhead{\footnotesize{Structure}}
 & \colhead{\footnotesize{$N_{HI}$\tablenotemark{a}}}
 & \colhead{\footnotesize{Count Rate to}}
 & \colhead{\footnotesize{Count Rate to}} \\
   \colhead{{ }}
 & \colhead{\footnotesize{($10^{20}$cm$^{-2}$)}}
 & \colhead{\footnotesize{Flux Conversion,}}
 & \colhead{\footnotesize{Flux Conversion,}} \\
   \colhead{{ }}
 & \colhead{{ }}
 & \colhead{\footnotesize{Soft Band\tablenotemark{b}}}
 & \colhead{\footnotesize{Hard Band\tablenotemark{b}}}}
\startdata
Cl1604 & 1.2 & 6.12 & 22.2  \\
Cl0023 & 2.7 & 6.37 & 22.3 \\
Cl1324 & 1.1 & 6.16 & 22.2 \\
RXJ1821 & 5.6 & 8.23 & 22.3 \\
RXJ1757 & 4.0 & 6.60 & 22.2 \\
\enddata
\tablenotetext{a}{\footnotesize{Galactic neutral hydrogen column density, using the dataset of \citet{DicLock90}.}} 
\tablenotetext{b}{\footnotesize{X-ray net count rate to unabsorbed flux conversion factor, in units of 10$^{-12}$ erg cm$^{-2}$.}}
\label{nhc2ftab}
\end{deluxetable}

\subsection{X-ray Data Reduction and Photometry}
\label{sec:datared}

The reduction of the data was conducted using the {\it Chandra} Interactive Analysis of Observations 4.2 software \citep[CIAO;][]{frusc06}. Each observation was filtered by energy into three bands: 0.5-2 keV (soft), 2-8 keV (hard), and 0.5-8 keV (full). Data were checked for flares using {\it dmextract} and the {\it Chandra} Imaging and Plotting System (ChIPS) routine {\it lc\_clean}. Exposure maps were created using the routine {\it merge\_all}. For vignetting correction, exposure maps were normalized to their maximum value, then images were divided by this normalized exposure map. 

To locate point sources, the routine {\it wavdetect} was run on each observation, without vignetting correction, using wavelet scales of $2^{i/2}$ pixels, with $i$ ranging from 0 to 8. A threshold significance of ${10}^{-6}$ was used, which would imply fewer than one spurious detection per ACIS chip, which has dimensions of $1024 \times 1024$ pixels ($0\farcs492\ $per pixel). However, this assumes a uniform background, which is almost certainly not the case. To measure realistic detection significances, we instead used photometric results explained below. Point source detection was carried out on images in each of the soft, hard, and full bands separately. For Cl1604 and Cl1324, {\it wavdetect} was run on each of the two pointings separately. Output object positions from the three different bands were cross-correlated to create one final composite list for each field. 

We carried out follow up photometry on the point sources. Circular
apertures containing 95$\%$ of the flux were created for each point
source using the point-spread function (PSF) libraries in the {\it Chandra} calibration
database. For {\it Chandra}, the PSF depends on both energy and off-axis
angle. For the soft and hard bands, respectively, we used the PSF
libraries for energies of 1.497 and 4.510 keV. Photometry was carried
out on the vignetting-corrected images in the soft and hard
bands. Background counts for each source were calculated in annuli
with inner and outer radii of $1.2 \times R_{95}$ and $2.4 \times
R_{95}$, where $R_{95}$ is the radius of the circular aperture
containing 95$\%$ of the flux. Since only 95$\%$ of the flux is
enclosed, net counts calculated from the apertures were multiplied by
1/0.95 to recover all the counts. Full band counts were calculated by
summing those from the soft and hard bands. The results of the photometry were used to calculate detection significances for the sources in each of the three bands using \begin{equation}
\sigma = C/\left(1.0+\sqrt{0.75 + B}\right)
\label{eq:detsig}
\end{equation}
where $C$ is the net photon counts from the source and $B$ is the background counts \citep{geh86}. X-ray sources with significances $< 2\sigma$ were rejected as spurious. With $\sim 75\%\ $of accepted sources having detection significances $> 3\sigma$, and the remaining $\sim25\%\ $with detection significances between $2\sigma$ and $3\sigma$, we expect a spurious detection rate of $\lesssim 1.5\%$, based on a normal distribution.

Due to the low number of photons observed for many sources, we opted to normalize a power-law spectral model to the net count rate of individual point sources to determine fluxes. We assumed a photon index of $\gamma = 1.4$, which is the  approximate slope of the X-ray background in the hard band \citep{tozzi01,kush02}. Count rates were calculated by dividing net counts by the nominal exposure time at the aimpoint of the appropriate observation. The galactic neutral hydrogen column density was calculated at the aim point of each observation using the Colden tool from the {\it Chandra} proposal toolkit, which uses the dataset of \citet{DicLock90}. Hydrogen column densities and derived net count rate to unabsorbed flux conversion factors for each field are summarized in Table \ref{nhc2ftab}. Conversion factors were determined separately for the different pointings of Cl1604 and Cl1324, but did not differ to the three significant figures listed in the table. 

\subsection{Optical Matching}

\begin{figure*}
\begin{center}
\plotone{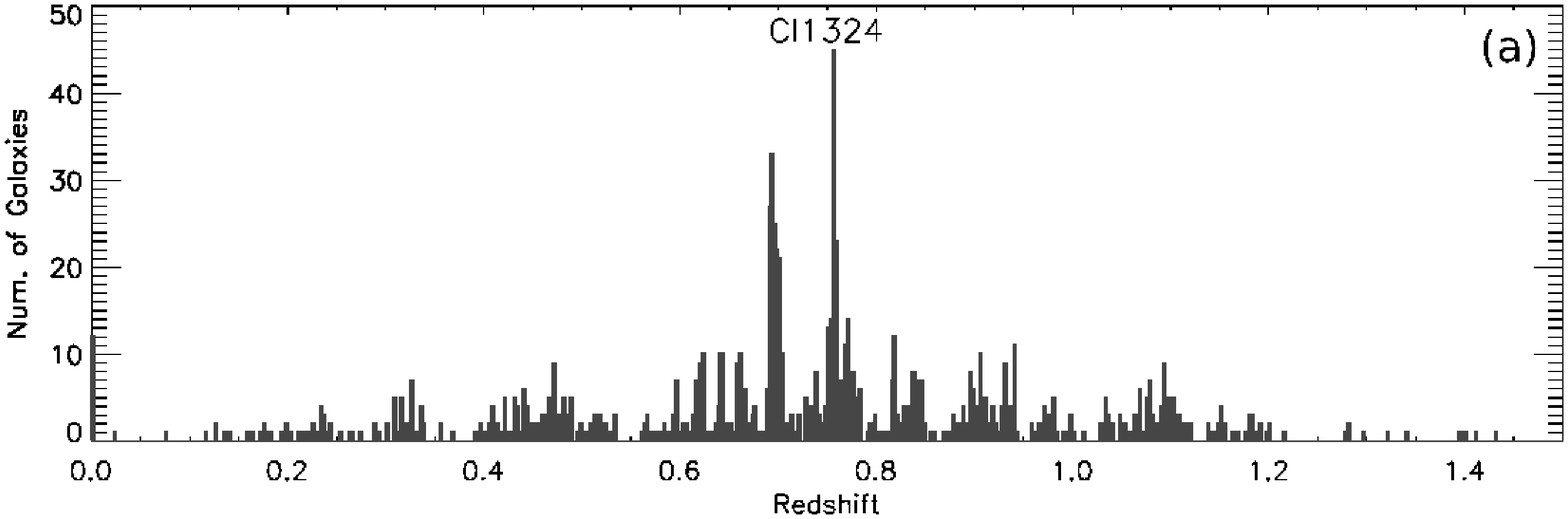}
\plotone{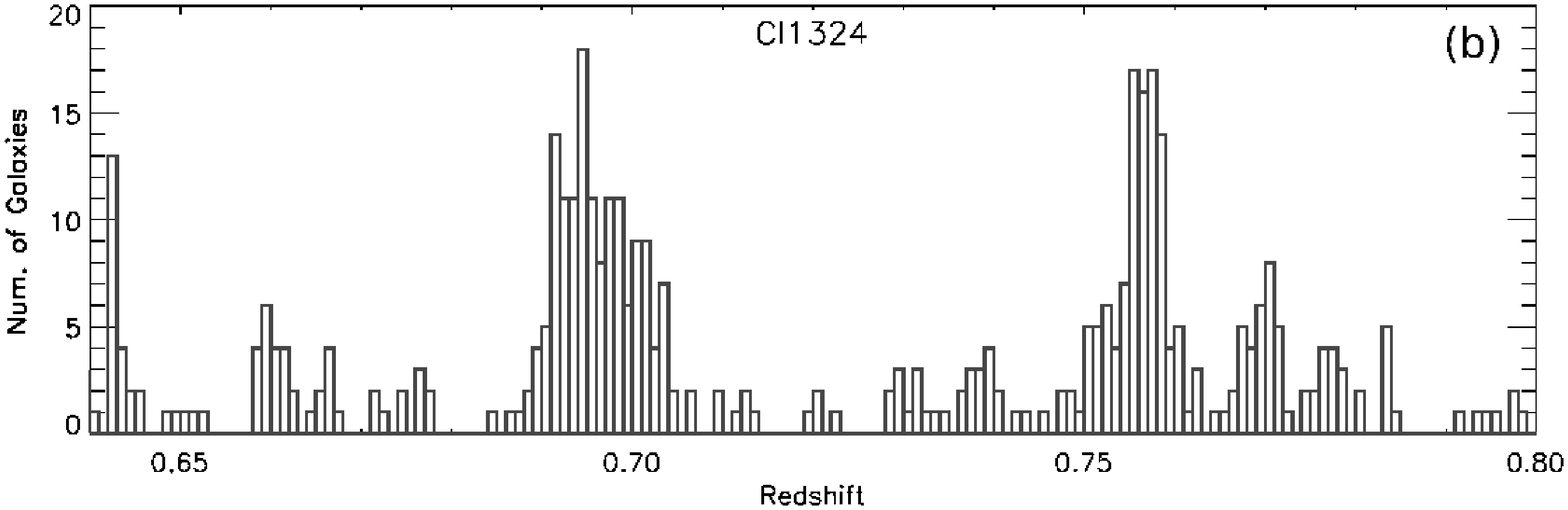}
\plotone{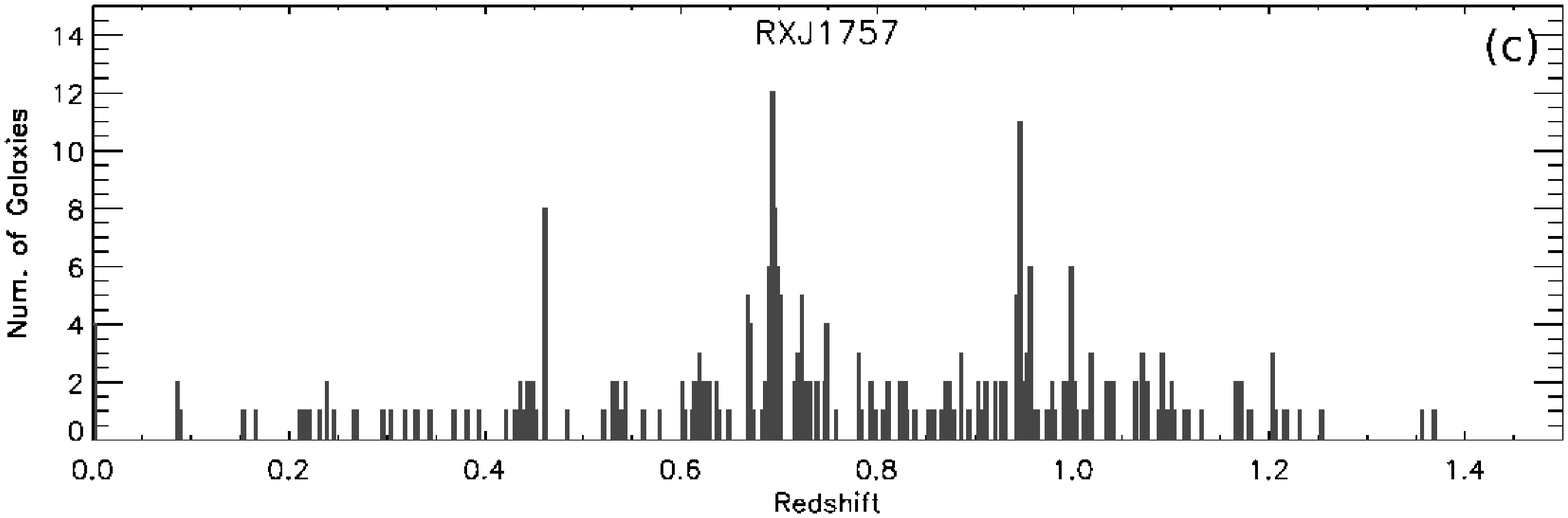}
\plotone{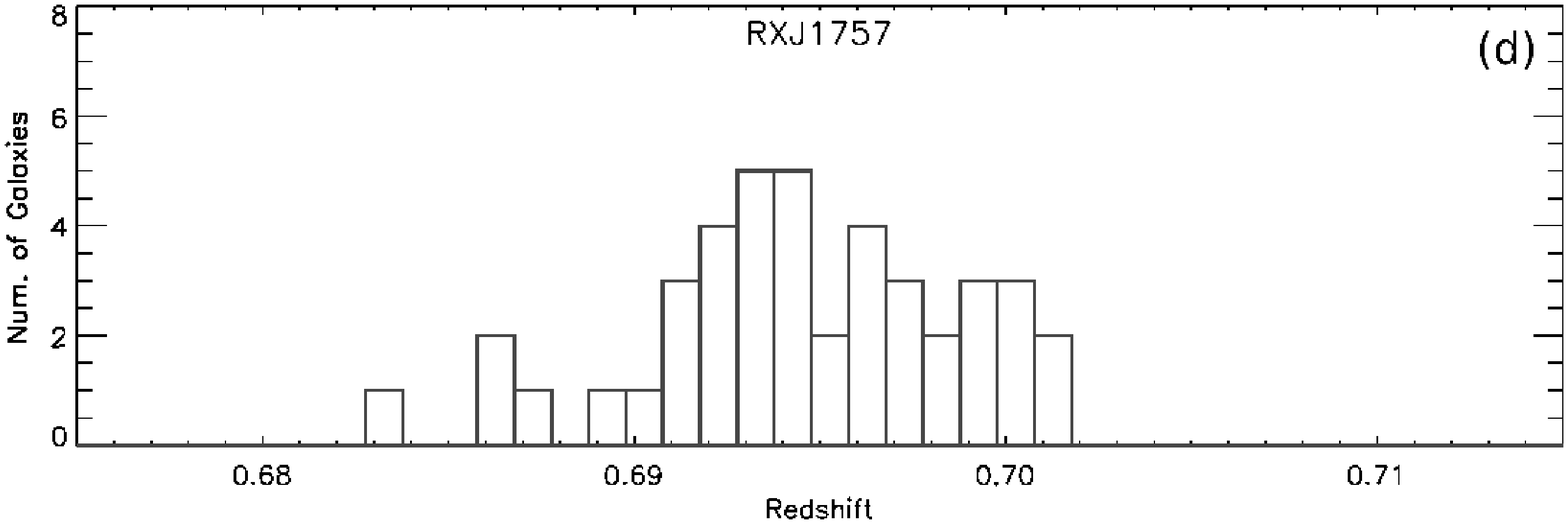}
\end{center}
\caption{\footnotesize{
Redshift distributions for the Cl1324 supercluster and RXJ1757. The first two panels include ({\it a}) all measured redshifts for Cl1324, with $Q=3$ or 4, up to $z=1.5$ and ({\it b}) zoomed in on the boundaries of the supercluster, $0.65 < z < 0.79$. The lower two panels include ({\it c}) all measured redshifts for RXJ1757, with $Q=3$ or 4, up to $z=1.5$  and ({\it d}) zoomed in on the boundaries of the cluster, $0.68 < z < 0.71$ . Although some $Q=3$,4 redshifts were measured above $z=1.5$, they have been omitted from these plots. 
}}
\label{1324+1757.rshists}
\end{figure*}

\begin{deluxetable}{llllll}
\tablecaption{Cl1324: Properties of Confirmed Groups and Clusters}
\tablehead{
\colhead{\footnotesize{Name}}
 & \colhead{\footnotesize{Alt.}}
 & \colhead{\footnotesize{R.A.}}
 & \colhead{\footnotesize{Dec.}}
 & \colhead{\footnotesize{$\sigma$}\tablenotemark{a}} \\
   \colhead{{ }}
 &  \colhead{{Name}}
 & \colhead{\footnotesize{(J2000)}}
 & \colhead{\footnotesize{(J2000)}}
 & \colhead{\footnotesize{}}
}
\startdata
Cl1324+3011 & A & 13 24 48.7 & +30 11 48 & $930\pm120$\\
Cl1324+3059 & I & 13 24 50.5 & +30 58 19 & $870\pm120$\\
Cl1324+3013 & B & 13 24 21.5 & +30 13 10 & $820\pm240$\\
Cl1324+3025 & C & 13 24 01.8 & +30 25 05 & $220\pm100$\tablenotemark{b}\\
\enddata
\tablenotetext{a}{\footnotesize{Velocity dispersion measured in km s$^{-1}$, within 1 Mpc}}
\tablenotetext{b}{\footnotesize{Dispersion calculated using only eight galaxies}}
\label{1324clusTab}
\end{deluxetable}

To search for AGNs within the individual clusters, we matched X-ray and
optical sources. In order to increase our completeness, the input to
the matching included all point sources detected by {\it wavdetect},
regardless of significance in any of the three bands. We used the
maximum likelihood ratio technique described in \cite{koc09a}, which
was developed by \citet{SS92} and also used by \citet{taylor05} and
\citet{gil07}. Our technique is similar to \citet{koc09a}, but with a
few key differences. The main statistic calculated in each case is the
likelihood ratio (LR), which estimates the probability that a given
optical source is the genuine match to a given point source relative
to the arrangement of the two sources arising by chance. The LR is
given by the equation
\begin{equation}
  LR_{i,j} = \frac{w_i \exp(-r_{i,j}^2/2\sigma_j^2)}{\sigma_j^2}
\end{equation}
Here, $r_{i,j}$ is the separation between objects $i$ and $j$, $\sigma_j$ is the positional error of object $j$\footnote{Positional errors of X-ray sources were calculated using the method of \citet{kim07}. Optical positional errors were small compared to those of X-ray objects, and were considered negligible.}, and $w_i = 1/n\left(<{m}_i\right)$ is the inverse of the number density of optical sources with magnitude brighter than ${m}_i$. The inclusion of the latter quantity is designed to weight against matching to fainter optical objects. However, in our analysis, we found that this particular weighting, used by \citet{koc09a}, of $1/n\left(<{m}_i\right)$ gave too much favor to bright objects, even when they were much farther from an X-ray object than a faint source. We adjusted the weighting to $w_i = {n\left(<{m}_i\right)}^{-1/2}$. We found that this change did not have a large overall effect but changed some borderline cases where a bright object with a large separation from the point had been chosen over a dimmer, much closer object. This includes one case, which prompted the adjustment, where spectroscopy showed that an M-type star was matched instead of a probable AGN, even though the M star was three times farther from the X-ray source.

For each field, except Cl1604, $n(<{m}_i)$ was measured using $i'$ magnitudes from our LFC catalogs. For Cl1604, ACS data were also available, but these observations did not cover the entire field. All objects were matched to the LFC catalogs. When possible, objects were also matched using the F814W magnitude from the ACS catalogs and matches to ACS objects took precedence over matches to LFC objects. 

From the LR, we use Monte Carlo simulations to derive the probability that a given match is genuine. We ran 10,000 trials for each X-ray object. In each trial, the object's position was randomized, and the LR was calculated based on nearby optical sources. The LR for a given X-ray object to optical counterpart pairing, ${LR}_{i,j}$, was compared against the distribution of LRs from the 10,000 Monte Carlo trials. We calculated the reliability as: \begin{equation}
R_{i,j} = 1 - \frac{N({LR}_j > {LR}_{i,j})}{10,000}
\end{equation} where $N({LR}_j > {LR}_{i,j})$ is the number of matches, to any optical source, across all 10,000 trials for that X-ray object with LR greater than ${LR}_{i,j}$. $R_{i,j}$ can be interpreted as the probability that optical source $i$ is the true match of X-ray source $j$, in the case of only one optical candidate. When there are multiple candidates, we used the method of \citet{rut00} to calculate the probability that optical source $i$ is the true match of X-ray source $j$ as
\begin{eqnarray} P_{i,j} = \frac{R_{i,j}\prod_{k \neq i}^N \left(1-R_{k,j}\right)}{S}\end{eqnarray}
The probability that no optical source is the true match is:
\begin{eqnarray}P_{none,j} = \frac{\prod_{k = 1}^N \left(1-R_{k,j}\right)}{S} \end{eqnarray}
where N is the total number of optical candidates and S is a normalization factor defined so that $\sum_{i=1}^N (P_{i,j}) + P_{none,j} = 1$. 

For an X-ray source with a single optical counterpart, a match was
considered genuine if $P_{none,j} < 0.15$. For X-ray sources with
multiple optical counterparts, a genuine match was chosen if $\sum_i
P_{i,j} > 0.85$ (which is equivalent to $P_{none,j} < 0.15$) and
$P_{i,j} > 4\sum_{k \neq i} P_{k,j}$ for any one object $i$. If the
first condition was true, but the second was not, all objects with
$P_{i,j} > 0.2$ were considered as matches. In subsequent sections, only one optical counterpart was considered for each X-ray source. In almost all cases, the highest probability match was used. However, in several cases, spectra of the primary and secondary matches indicated that the secondary match was an AGN, and thus more likely to be a genuine match. Note that our threshold is a
deviation from \citet{koc09a}. They used $P_{none,j} < 0.2$ instead of
0.15. We decided to use the more stringent threshold of 0.15, which
has been used by others \citep{mann02,taylor05}, to better limit the
number of false matches. The more restrictive cutoff omitted
$\lesssim10$ sources per field. We determined this threshold through
visual inspection of potential matches. This calibration entailed
determining at what approximate level of $P_{none,j}$ most matches
visually seemed spurious. However, optical candidates above our
threshold were also visually scrutinized ($\lesssim 3\%$ of the
total), and some were accepted after this inspection where we felt the
matching algorithm had failed. Note that setting a threshold for
genuine matches is not entirely objective, and a precedent has been
set for accomplishing this with visual inspection
\citep[e.g.,][]{mann97}. In Table \ref{srcsum}, we list, for each
field, the number of X-ray sources detected at $>$3$\sigma$
($>$2$\sigma$) in one of the three bands, as well as the number of
those sources matched to optical counterparts.

\section{Global Properties of the ORELSE Structures}
\label{sec:globchar}

The five structures in our sample span a range of evolutionary states. They include Cl0023, whose four constituent groups are still in the process of merging to form a single cluster; the two isolated X-ray selected clusters, RXJ1757 and RXJ1821, which appear to be in a more evolved and relaxed state; and the two superclusters, Cl1604 and Cl1324. We would like to compare the AGNs within our sample based on the evolutionary states of the structures to which they belong, which could shed light on how the AGNs in these systems are being triggered. In order to make such a comparison, we first present the global properties of the five ORELSE structures. 

\subsection{Redshift Distributions}

The Cl1604 supercluster, Cl0023, and RXJ1821 have all been studied previously as part of the ORELSE survey (see Section \ref{sec:samp} for individual references), although we have gathered new data on each, as described in Section \ref{sec:optobs}. While individual clusters in Cl1324 have been studied, the properties of the supercluster as a whole have not. In this section we do a preliminary exploration of this structure, which will be covered more thoroughly in an upcoming paper (R. R. Gal et al. 2012, in preparation). The cluster RXJ1757 was studied only as a part of the ROSAT NEP survey \citep{gioia03}, in little detail. Here, we present new redshift histograms of these last two structures derived from our ORELSE data. 
\newline
\newline
\newline
\subsubsection{Cl1324}

Figure \ref{1324+1757.rshists}(a) shows all confirmed redshifts in the spatial vicinity of Cl1324. We can see two peaks in the histogram, at $z\approx0.695$ and $z\approx0.755$. These peaks coincide with the two largest clusters in the structure, Cl1324+3011 and Cl1324+3059. From the distribution of red galaxies, we find ten overdensities in the supercluster, some of which can be observed in the redshift histogram. So far, we have confirmed four clusters and groups to be constituents, shown in Table \ref{1324clusTab}, with coordinates, redshifts, and measured velocity dispersions given. Additional multiobject spectroscopy to confirm the nature of the other red galaxy overdensities is planned.

\begin{figure*}
\begin{center}
\epsscale{1.0}
\plotone{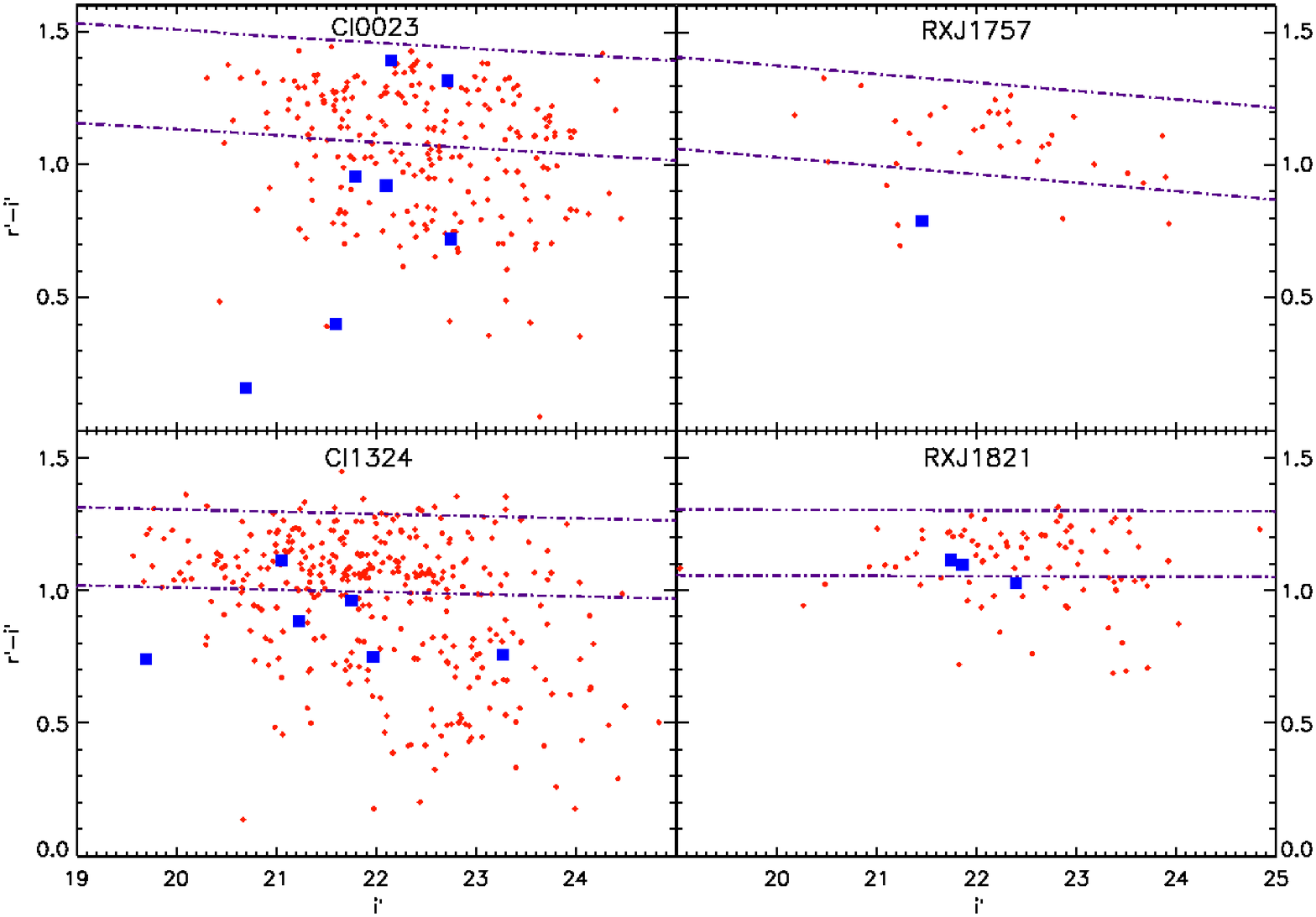}
\epsscale{0.65}
\plotone{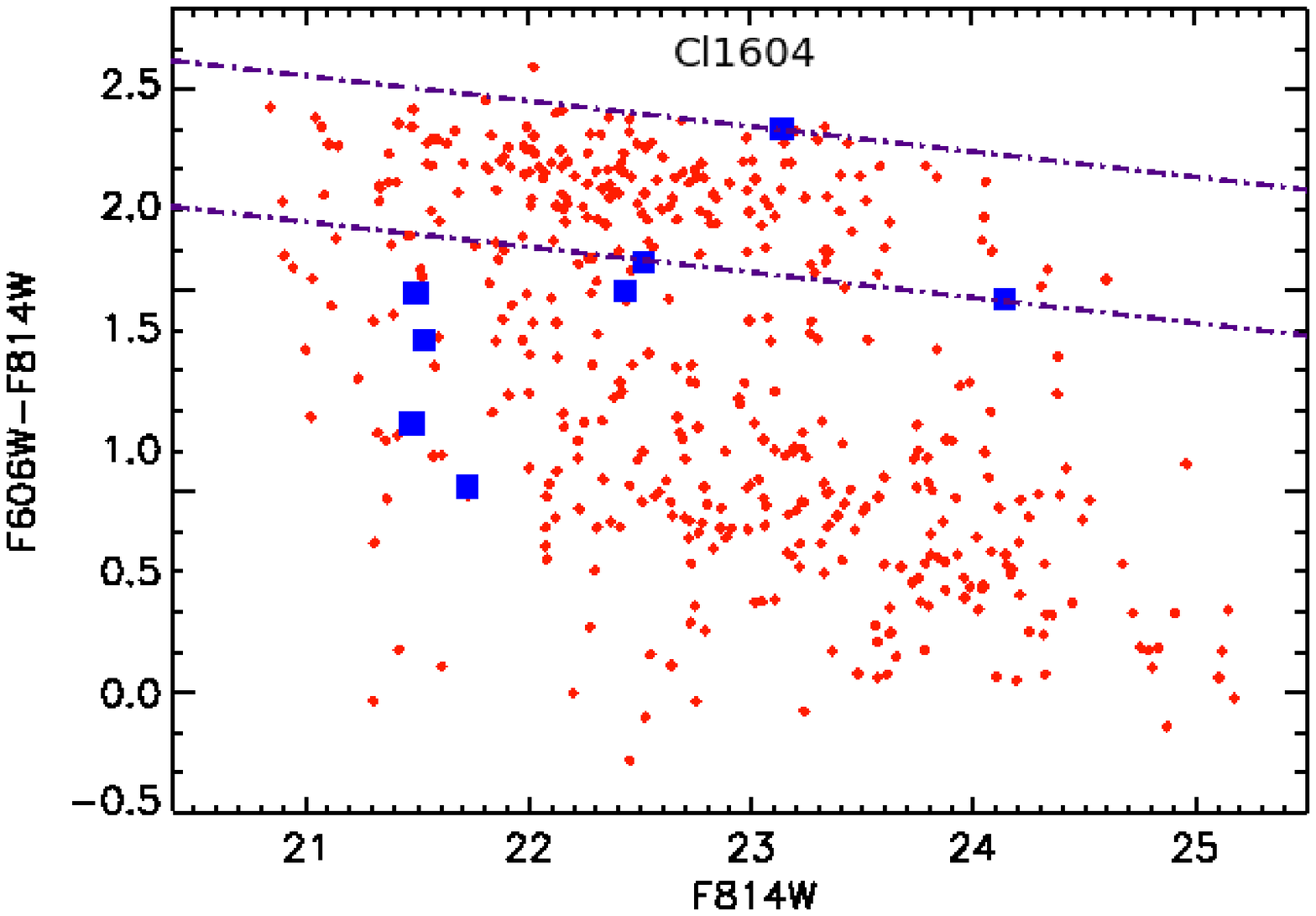}
\end{center}
\caption{
	\footnotesize{Color-magnitude diagrams for the five fields. LFC $r'-i'$ colors are displayed, except in the case of Cl1604, where ACS F606W-F814W colors are used. All confirmed supercluster/cluster members are shown and squares correspond to AGN hosts. Dashed lines indicate the boundaries of our red sequence fits. We can see a higher fraction of AGN are on the red sequence in the more evolved structures (Cl1324, RXJ1821, and RXJ1757) than in Cl1604 and Cl0023.}
}
\label{CMDS}
\end{figure*}

\subsubsection{RXJ1757}

The redshift histograms for RXJ1757 are displayed in Figures \ref{1324+1757.rshists}(c) and \ref{1324+1757.rshists}(d). At $z\approx0.69$ and $z\approx0.9$, we see two peaks in the distribution. The former is the overdensity associated with RXJ1757. When we examine the spatial distribution of the higher redshift peak, we find that the galaxies in its vicinity are distributed nearly uniformly across the field of view, implying a sheet of galaxies. 

Looking at the redshift distribution of confirmed galaxies within the bounds of RXJ1757 (Figure \ref{1324+1757.rshists}(d)), we can see the distribution is reasonably consistent with a Gaussian, confirmed by a Kolmogorov-Smirnov (K-S) test at a $99\%$ level, suggesting there is no significant substructure. However, we caution that we have a smaller sample of confirmed redshifts compared to the other fields. 

\subsection{The Red Sequence and the Blue Populations} 
\label{TRSATBP}

Figure \ref{CMDS} shows CMDs for all five fields. All spectroscopically confirmed supercluster/cluster members are shown. Squares indicate the confirmed X-ray AGNs within each structure, which are analyzed in Section \ref{sec:AGN}. The red sequence for each field is delineated by dotted lines. Red sequence fits for each field were calculated using a linear fitting and $\sigma$-clipping technique. First, a fit to a linear model, of the form \begin{equation} C = C_0 + m \times B \end{equation}
where $C$ is either $r' - i'$ or F606W-F814W and $B$ is either $i'$ or F814W, was carried out on member galaxies within a chosen magnitude and color range using a $\chi^2$ minimization \citep{gladders98,stott09}. The fit was initialized with a color range chosen ``by eye'' to conform to the apparent width of the red sequence of the structure. The magnitude bounds were defined as the range where the photometric errors were small ($\sigma_{i',F814W} \lesssim 0.05$). After the initial fit, colors were normalized to remove the slope. The color distribution was then fit to a single Gaussian using iterative $3\sigma$ clipping. At the conclusion of the algorithm, the boundaries of the red sequence were defined by a 3$\sigma\ $offset from the center, except for Cl1604 and Cl1324. For the two superclusters, the color dispersion was inflated due to the large redshift extent of these structures, and 2$\sigma$ offsets were used to achieve reasonable boundaries. For every field except Cl1604, the LFC $r'-i'$ color magnitude diagrams are shown. For Cl1604, ACS data were available and were used in place of LFC data because of their superior precision. Note, however, that two of the AGNs in Cl1604 are outside our ACS pointings, so that our analysis using ACS data only includes 8 AGNs in the Cl1604 supercluster.

\begin{deluxetable}{lcc}
\tablecaption{Global Blue Fractions}
\tablehead{
\colhead{\footnotesize{Structure}}
 & \colhead{\footnotesize{All Members\tablenotemark{a}}}
 & \colhead{\footnotesize{1$^{st}$ Two Masks\tablenotemark{b}}}
}
\startdata
Cl0023 & 0.47$\pm$0.06 & 0.51$\pm$0.13\\
Cl1604 & 0.57$\pm$0.05 & $...$ \\
Cl1324 & 0.42$\pm$0.04 & 0.42$\pm$0.12\\
RXJ1821 & 0.35$\pm$0.08 & 0.42$\pm$0.15\\
RXJ1757 & 0.17$\pm$0.08 & 0.09$\pm$0.10\\
\enddata
\tablenotetext{a}{\footnotesize{Includes all spectroscopically confirmed sources within the redshift bounds of the structure and with $i'$ or F814W $< 23.5$. Errors are Poissonian.}}
\tablenotetext{b}{\footnotesize{Only spectroscopically confirmed sources from the first two spectral masks taken for each field are included, excluding masks where X-ray matched targets were preferentially targeted. For Cl1324, the first two masks from each pointing are included. No measurement was made for Cl1604. See Section \ref{TRSATBP} for explanation.}}
\label{bftab}
\end{deluxetable}

\begin{figure*}[!t]
\epsscale{0.7}
\plotone{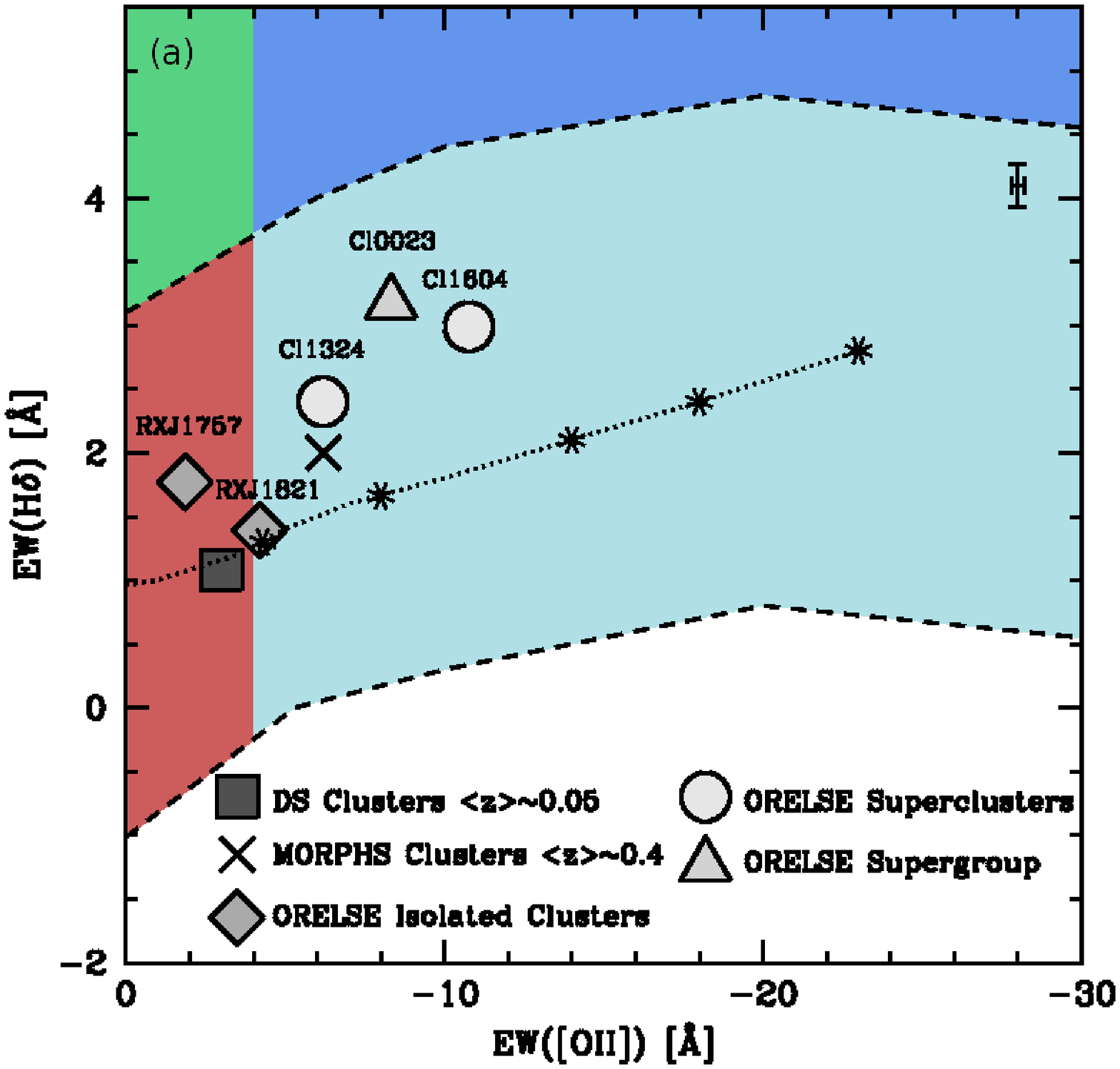}
\plotone{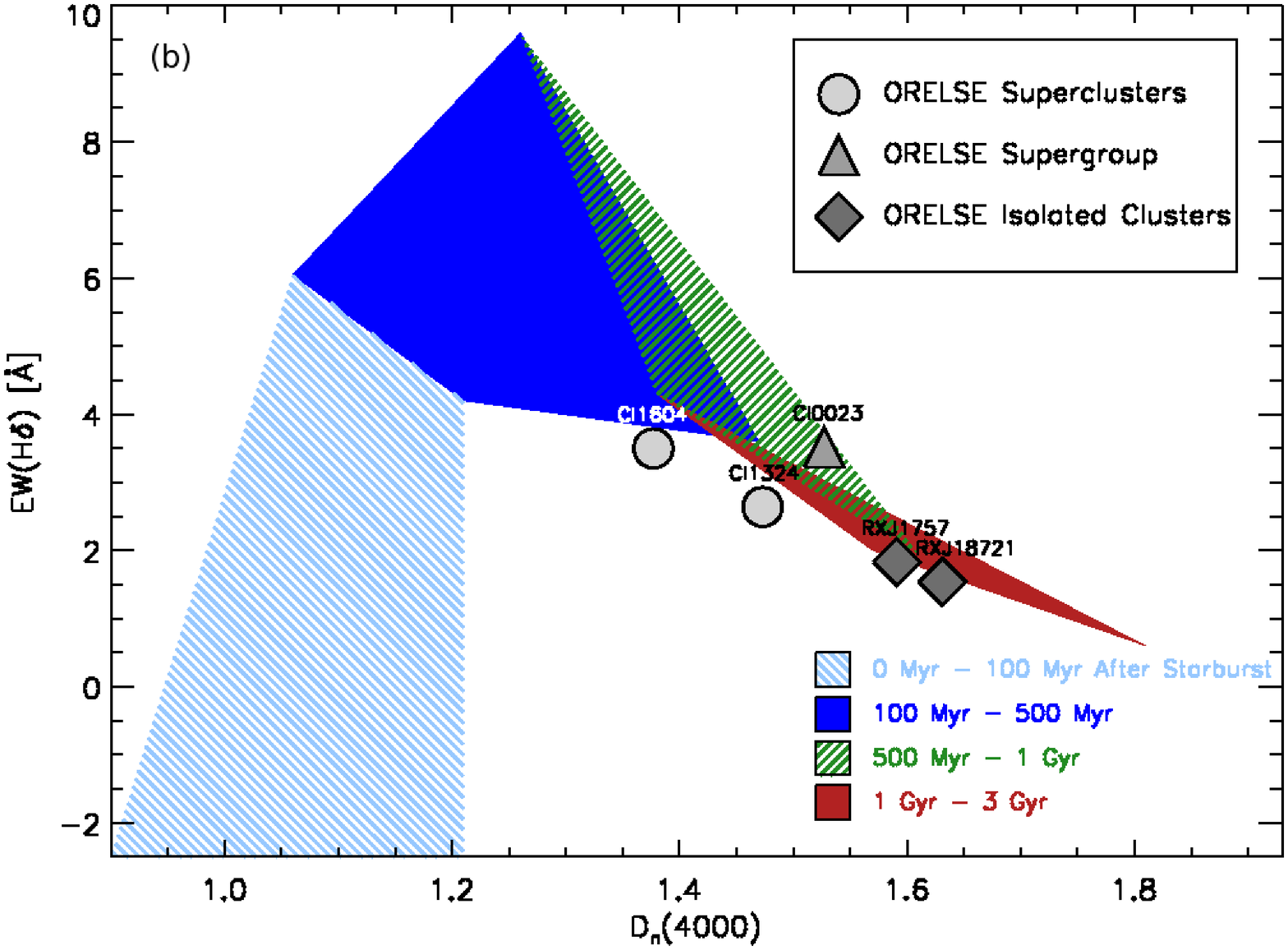}
\caption{\footnotesize{
    Spectral properties of the composite spectra of all spectroscopically confirmed galaxies in our five structures. In ({\it a}), EW(\OII) is used as a measure of star formation while EW(H$\delta$) is used to probe the post-starburst stellar population \citep{pogg97,pogg99}. The dotted line is a track for clusters made entirely of quiescent and continuously star-forming galaxies from 2dF \citep{dress04}, with a lower percentage of star forming galaxies to the left. The shaded regions represent galactic classifications, described in the text. The dashed lines enclose 95$\%$ of normal star-forming galaxies \citep{oemler09}. Spectral errors are shown in the upper right-hand corner. Also shown, for reference, are samples of cluster galaxies at $z\sim0.05$ \citep{DS88} and $z\sim0.4$ \citep{dress04}. In ({\it b}), we plot the average D$_{\rm n}$(4000) measurements of structure members versus the average EW(H$\delta$). For reference, we plot regions of EW(H$\delta$)--D$_{\rm n}$(4000) phase space covered by four post-starburst models of \citet{bc07}. Four different time ranges since starburst are shown. Because Bruzual \& Charlot models only incorporate stellar light, emission infill corrections were made for all EW(H$\delta$) measurements (see Section \ref{sec:xlums} for details). D$_{\rm n}$(4000) and EW(H$\delta$) errors are smaller than the points.
}} 
\label{OIIvsHd}
\end{figure*}

While all the structures show a substantial number of galaxies on the
red sequence, there are large differences in the blue
populations. Qualitatively, we observe a lower blue fraction in the
CMDs of RXJ1821 and RXJ1757 than in those of
Cl0023 and Cl1604. This is quantified in Table \ref{bftab}, where the
blue fraction\footnote{We define a blue galaxy as any with a color
  blueward of the lower boundary of the red sequence.} for all
confirmed cluster/supercluster members with $i'$ or F814W magnitudes
brighter than 23.5 are displayed in the first column. We can see that
Cl0023 and Cl1604 have the bluest galaxy populations, while RXJ1757
has the highest fraction of galaxies on the red sequence of all the
fields. Since this could be due to the low completeness of
spectroscopic coverage in this field, we attempted to make corrections
with two approaches: (1) using our efficiency of spectroscopically
confirming structure members to estimate the total number of member
galaxies, and (2) correcting using measurements of the background
galaxy density. All efforts to statistically estimate the true blue
fraction yielded large errors making the measurements highly
uncertain. Instead, we created a blue fraction measurement metric that
could be compared between fields \citep[see also ][]{lub09}. The
different fields have a varying number of spectral masks, with several
different spectroscopic priorities. However, the first several masks,
excluding those designed to specifically target X-ray matched sources,
have similar priorities for choosing
targets\footnote{That is, prioritizing red galaxies first followed by
  progressively bluer galaxies \citep[See
    e.g.][]{gal08,lub09}.}. Therefore, we chose to compare blue
fractions only among sources in the first two masks for each field,
since these sources should represent similar populations. We chose two
masks because this is the number of masks for RXJ1757 excluding those
where objects matched to X-ray point sources were preferentially
targeted. We confirmed that this sampling is representative of the
entire galaxy population by recreating composite spectra using only
the first two spectral masks. Since there are no large differences
between the average spectral features, the method should be
accurate. The results of this comparison are displayed in the second
column of Table \ref{bftab}. However, we did not calculate a blue
fraction for Cl1604 in this way, for two reasons. First, there are a
total of 24 spectroscopic masks for Cl1604, from several different
telescopes. Choosing which ones to include is difficult and it may be
impossible to select a population congruous with any of the other
fields in this manner. Second, our spectroscopy for Cl1604 is
relatively complete, so we are confident in the blue fraction measured
down to ${\rm F814W} = 23.5$.

Examining the results, we can see the same color hierarchy in the
structures for both methods of measuring the blue fraction, with
RXJ1757 having the smallest fraction and Cl1604 and Cl0023 the
largest. These large blue fractions, in particular those for Cl1604
and Cl0023, are consistent with the Butcher-Oemler effect
\citep{BO84}. We also note that the isolated X-ray-selected clusters,
RXJ1757 and RXJ1821, are the reddest structures, suggestive of a more
advanced, dynamically relaxed evolutionary state. The color hierarchy
suggests a similar ranking of galactic star formation in the five
structures, which we can explore with our spectroscopic data.

\subsection{Spectral Properties}
\label{sec:specprop}

\begin{deluxetable}{lccc}
\tablecaption{Composite Spectral Properties}
\tablehead{
\colhead{\footnotesize{Structure}}
 & \colhead{\footnotesize{EW(\OII)\tablenotemark{a}}}
 & \colhead{\footnotesize{EW(H$\delta$)\tablenotemark{a}}}
 & \colhead{\footnotesize{D$_{\rm n}$(4000)\tablenotemark{b}}} \\
 \colhead{\footnotesize{}} 
& \colhead{\footnotesize{(\AA)}}
& \colhead{\footnotesize{(\AA)}}
& \colhead{\footnotesize{}}
}
\startdata
Cl0023 & $-8.33\pm0.07$ & $3.17\pm0.05$ & $1.527 \pm 0.002$\\
Cl1604 & $-10.77\pm0.09$ & $2.99\pm0.07$ & $1.377 \pm 0.002$\\
Cl1324 & $-6.19\pm0.05$ & $2.40\pm0.03$ & $1.473 \pm 0.001$\\
RXJ1821 & $-4.20\pm0.10$ & $1.39\pm0.08$ & $1.631 \pm 0.004$\\
RXJ1757 & $-1.87\pm0.18$ & $1.77\pm0.10$ & $1.591 \pm 0.005$\\
\enddata
\tablenotetext{a}{\footnotesize{Measured using bandpasses from \citet{fish98}.}}
\tablenotetext{b}{\footnotesize{Measured using bandpasses from \citet{balogh99}.}}
\label{globchartab}
\end{deluxetable}

Using our spectroscopic data, we examine the typical star formation
history of the galaxies in each structure. We formed composite spectra
by co-adding the individual spectra of galaxies within each structure,
according to the method of \citet{lemaux09} and \citet{lemaux11}. We analyze these spectra in terms of two important
features relevant to star formation: the \OII\ and H$\delta$
lines. The H$\delta$ absorption is indicative of a population of A and
B stars, which disappears $\sim1$ Gyr after the cessation of star
formation within a galaxy, due to the lifetime of A stars
\citep{pogg97}. If star formation is ongoing, a population of O stars,
which have weaker hydrogen features, can dominate the continuum and
wash out this absorption line. Infilling can also occur from Balmer
emission from \ion{H}{2} regions. The \OII\ emission line has been
used as an indicator of star formation, especially as a proxy for the
H$\alpha$ emission line at higher redshifts when H$\alpha$ has shifted
out of the optical range \citep{pogg99}. However, recent analysis
using near-IR spectroscopy of sources from the Cl1604 supercluster and
RXJ1821 has compared \OII\ and H$\alpha$ emission, finding that a
significant portion of \OII\ emission can come from LINER-- or
Seyfert--related processes \citep{lemaux10,koc11b}. These results are
supported by those of \citet{yan06}, albeit using a lower redshift
sample. In light of this, caution must be exercised when interpreting
\OII\ measurements.  For an additional diagnostic, we measure the
D$_{\rm n}$(4000) strength which is an indicator of mean stellar age
\citep{kauf03}.

\begin{figure*}
\begin{center}
\plotone{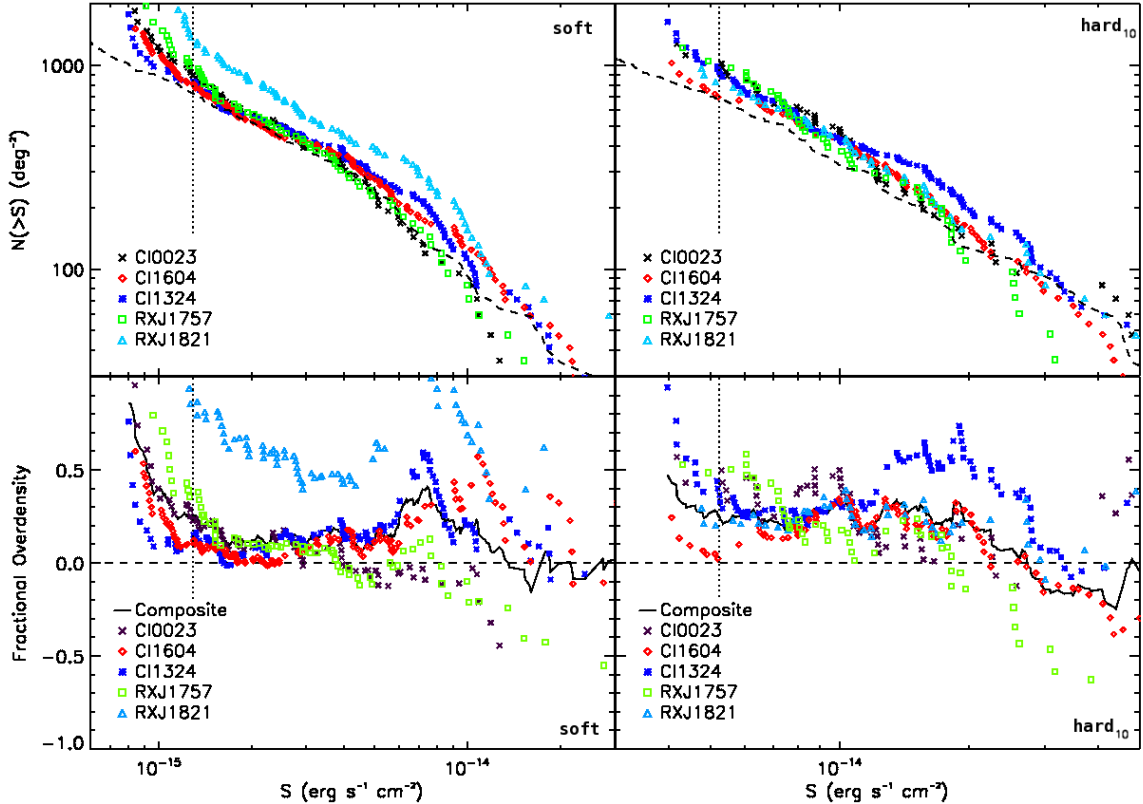}
\end{center}
\caption{
   \footnotesize{Cumulative source counts for all five structures, in the soft (0.5-2.0 keV) band (top left) and hard$_{10}$ (2.0-10 keV) band (top right). The dashed line corresponds to source counts from two pointings in the CDFN and CDFS, which we reduced and analyzed ourselves. The dotted line indicates flux corresponding to 20$\%$ sky coverage for NEP5281 in the soft band and RXJ1757 in the hard band. Note that this value is different for each field and the specific choice of field here is meant to be representative. In the lower plots, we have plotted the fractional overdensities relative to the CDF deep field measurements. Along with the five fields is a composite measurement, explained in Section \ref{sec:CSC}. }
}
\label{logNlogSall}
\end{figure*}

Figure \ref{OIIvsHd}(a) shows \OII\ vs.\ H$\delta$ equivalent widths for
the composite spectra of members of the five structures. The dotted
line represents the average spectral properties for a cluster
population composed of various fractions of ``normal'' star-forming
and quiescent galaxies \citep{dress04}, based on data from the Two-Degree Field (2dF)
Galaxy Redshift Survey \citep{coll01}.  Asterisks on this line
represent a cluster population composed of (from left to right) 20\%, 40\%, 60\%,
80\%, and 100\%\ star-forming galaxies. For a cluster whose average galaxy lies above this line,
the H$\delta$ line is too strong to be produced by normal star
formation, requiring some contribution from starbursting or
post-starbursting galaxies. The dashed lines enclose 95\% of normal
star-forming galaxies \citep{oemler09}. The shaded regions, which are
based on the spectral types of \citet{dress99} and \citet{pogg99},
denote the region of this phase space inhabited by (starting from the
upper right and moving counter-clockwise) starburst (dark blue),
post-starburst (green), quiescent (red), and normal star-forming
galaxies (light blue). Examining the positions of the structures in
our sample on this plot, we can see that RXJ1757 and RXJ1821 have
mostly quiescent populations, with each cluster having $\lesssim20\%$
normal star-forming galaxies. Because RXJ1757 is offset from the 2dF
line, there may be some contribution from post-starburst galaxies, but
the fractional contribution is low. The stronger \OII\ emission in the
Cl0023 and Cl1604 composite spectra suggests that these structures
have a higher fraction ($\gtrsim40\%$) of continuously star-forming
galaxies. These structures have larger blue fractions than the others,
so it is unlikely that the increased EW(\OII) in their average spectra
is due to LINER processes, which are primarily associated with red
sequence galaxies. The EW(H$\delta$) measured from the Cl0023 and
Cl1604 composite spectra are significantly in excess of the 2dF line,
suggesting a substantial contribution from starbursting or
post-starburst galaxies. Cl1324 is in an intermediate range, with
$\sim30\%$ normal star-forming galaxies and an observed EW(H$\delta$)
for its galaxy population smaller than that of Cl0023 and Cl1604.

Our conclusions based on the \OII\ and H$\delta$ lines are supported by
the corresponding D$_{\rm n}$(4000) measurements (see Table
\ref{globchartab}). These results are illustrated in Figure \ref{OIIvsHd}(b), where average measurements of EW(H$\delta$) and D$_{\rm n}$(4000) are plotted for the five structures. For comparison, we also indicate ranges of EW(H$\delta$)--D$_{\rm n}$(4000) phase space spanned by four different Bruzual \& Charlot
\citep{bc07} models for various times after the starburst. The four
models include a single burst ($\tau = 0.01$) and a secondary burst of
20\%, 10\%, and 5\% by mass which occurs 2 Gyr after the initial
burst. All models are solar metallicity and are corrected for
extinction using E(B-V) $= 0.25$ and a \citet{cal00} extinction
law. Because Bruzual \& Charlot models only incorporate stellar light,
emission infill corrections were made for all EW(H$\delta$)
measurements using relationships between H$\alpha$ and \OII\ taken
from \citet{yan06} and H$\alpha$ and H$\delta$ from \citet{sch06}.
Although these corrections are not perfect (e.g., see the measurement
for Region 1), these values should be accurate to within $\pm 0.5$
\AA. We can see that RXJ1757 and RXJ1821 have the
largest continuum break strengths, indicating that they possess the oldest average
stellar populations of the five structures. The average galaxy in this structure has had 1-3 Gyr since its last starburst, according to the \citet{bc07} model. The other structures
have smaller average D$_{\rm n}$(4000) measurements, consistent
with the results of Figure \ref{OIIvsHd}(a) showing larger fractions of
star-forming galaxies and younger galaxy populations. According to the \citet{bc07} model, the average galaxy in Cl1324, Cl0023, and Cl1604 has had a progressively shorter time since the last starburst. Altogether, these spectral results parallel what was found with the blue fractions, in that the reddest structures are also the
ones with the most evolved stellar populations and the lowest fraction
of star-forming galaxies.

\subsection{Summary of Global Characteristics and Grouping of the Sample}

Despite our extensive spectroscopic sample, there are too few X-ray
AGNs in any individual cluster (and even supercluster) to draw
statistical conclusions.  Hence, we divide our structure sample into
two categories. The first contains Cl0023 and Cl1604 which have the
highest level of ongoing star formation and starburst activity, as
shown in the preceding sections. The second category, consisting of
Cl1324, RXJ1757, and RXJ1821, contains structures whose member
galaxies are typically quiescent or forming stars at a lower rate. We
refer to these two categories as ``unevolved'' and ``evolved'',
respectively, as a means of describing their typical galaxy
populations. While we acknowledge that the terms ``more evolved'' and
``less evolved'' would be more appropriate, we adopt the less accurate
denominations for brevity.  In addition, these terms do not
necessarily imply differing levels of cluster dynamical evolution or
even a clear temporal sequence from one category to the other. We note
that, while some clusters or groups may not fit well with the global
characteristics of their parent supercluster (i.e. Cluster A of
Cl1604), we cannot examine all of the AGNs on a cluster-by-cluster
basis. As we will discuss in Section \ref{sec:spatdist}, many of the
AGNs in the superclusters and the supergroup are not associated with
any one cluster or group. So, when analyzing these AGNs, we take the
parent supercluster or supergroup as a whole.

The particular segregation of our structure sample is motivated by the
clear distinction between the structures shown in Figure
\ref{OIIvsHd}. The abundant star formation in the unevolved structures
suggests the presence of a large gas reservoir in many of their member
galaxies. We might expect that this same gas is available for AGN
fueling. Conversely, the typical galaxy in the evolved sample has
likely consumed most of the available gas in prior star formation
episodes, leaving less fuel for the AGNs. In the following sections, we
examine whether the properties of the AGN sin the two categories
reflects this distinction, and what we can can learn about the
relationship of star formation to AGN activity in LSSs.

\begin{figure*}
\begin{center}
\plotone{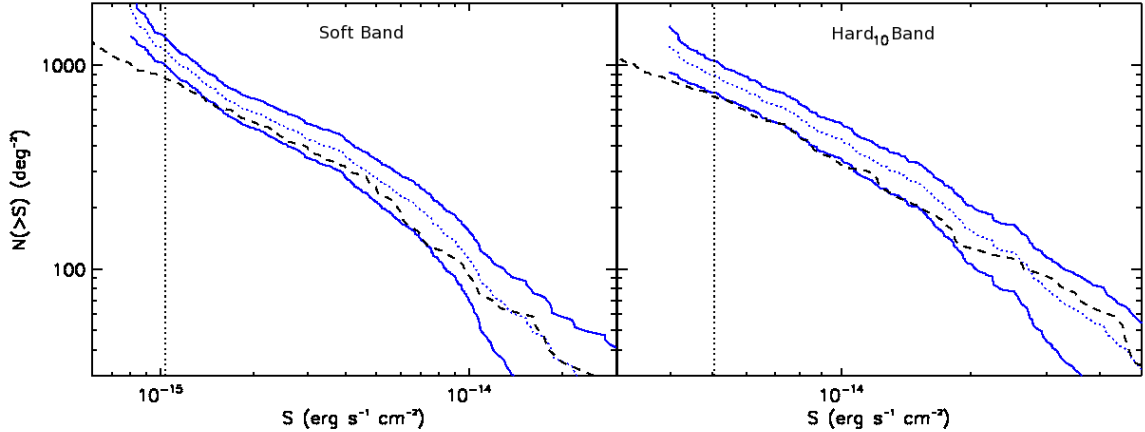}
\end{center}
\caption{
   \footnotesize{Co-added source counts created from all five structures, in the soft (0.5-2.0 keV) band (left) and hard$_{10}$ (2.0-10 keV) band (right). Composite point source lists which combined all the sources from every field were used in these plots' construction. Also, the combined background from all five fields was used for the calculation of the weights, $\Omega_i^{-1}$. The solid lines flanking the dotted line indicate the 3$\sigma$ errors on the composite cumulative source counts measurement. The dashed line corresponds to source count measurements from a control field, comprised of two pointings in the CDFN and CDFS. The vertical dotted line indicates flux corresponding to 20$\%$ sky coverage of the combined fields. }
}
\label{logNlogScomp}
\end{figure*}

\section{Cumulative X-ray Source Counts}
\label{sec:CSC}

In this section, we examine the frequency of AGN activity within the five structures to determine if there are associated excesses of X-ray point sources. X-ray point source photometry was used to calculate cumulative X-ray source number counts, $N(>S)$, using the method of \citet{gioia90}: \begin{equation} N\left(>S\right) = \sum_{i=0}^N \Omega_i^{-1}\ {deg}^{-2}\end{equation}
where $N$ is the number of point sources with fluxes greater than $S$ and $\Omega_i$ is the area of the sky in which the {\it i}th point source could have been detected at a 3$\sigma$ level or higher. Variance in $N(>S)$ was calculated using $\sigma_i^2 = \sum_{i=0}^N {\Omega}_i^{-2}$. 

To calculate $\Omega_i$, we used the method of \citet{koc09a}, which is similar to those used in other literature \citep{johnson03,cap05}. All point sources detected by {\it wavdetect}, in all bands and without significance cuts, were removed and replaced with an estimate of the background using the CIAO tool {\it dmfilth}. To create a map of the background emission, these images were binned into 32$\arcsec$ bins. According to Equation (\ref{eq:detsig}), the flux limit corresponding to a 3$\sigma$ detection\footnote{Note that we only include X-ray sources with detection significances $> 3\sigma$ in this analysis to be consistent with previous work. The inclusion of sources with detection significances between $2\sigma$ and $3\sigma$ has a significant effect only on fluxes where sky coverage is below $20\%$, and the results are generally considered unreliable.} occuring in one of these binned pixels is given by \begin{equation}
S_{lim} = 3\frac{k}{t}\left(1.0+\sqrt{0.75+B\pi R_{95}^2 A^{-1}}\right)
\end{equation}
where $B$ is the net counts in a pixel, $A$ is the area of a pixel, $R_{95}$ is the radius of the aperture enclosing 95$\%$ of an X-ray source's flux, as described in Section \ref{sec:datared}, $k$ is the conversion factor between photon count rate and X-ray flux, also described in Section \ref{sec:datared}, and $t$ is the exposure time of the image. For a given source with flux, $S$, $\Omega_i$ is then equal to the total number of binned pixels for which $S$ is greater than $S_{lim}$, multiplied by the area of a pixel. 

The cumulative source counts for all the fields in the soft band and the 2-10 keV band, hereafter the hard$_{10}$ band, are shown in Figure \ref{logNlogSall}. The latter was extrapolated, field by field, from the 2-8 keV band by fitting a power law spectral model with exponent $\gamma = 1.4\ $to each detected point source. For Cl1604 and Cl1324, the source counts for the two pointings were combined. Also shown are cumulative source counts measurements from the Chandra Deep Field North and South \citep{brandt01,rosati02}. One Chandra observation from each deep field was used (observation ID 582 and 2232). Both have exposure times of roughly 130 ks. We re-analyzed these observations using the same reduction pipeline that we used for the ORELSE fields. We used the combined source counts from these two fields (hereafter CDF) to estimate the blank-field counts for comparison with our data. 

We can see that, in the soft band, Cl0023 and RXJ1757 appear to be consistent with no overdensity compared to CDF. The other three fields are all overdense to some degree. In the range 3$\times{10}^{-15}$ to 10$^{-14}$ \ergscm, RXJ1821, Cl1324, and Cl1604 have average overdensities of 0.5$\sigma$, 1.0$\sigma$, and 1.5 $\sigma$, respectively, with $\sigma$ calculated using the cumulative source count errors from our data and those of CDF. In the hard$_{10}$ band, RXJ1821, RXJ1757, and Cl0023 all have approximate overdensities of 0.5$\sigma$ in the range 7$\times {10}^{-15}$ to $2\times {10}^{-14}$ \ergscm. RXJ1757 appears to be consistent with no overdensity. Cl1604 and Cl1324 also appear to be overdense, with average overdensities in the flux range $7\times {10}^{-15}$ to $2\times {10}^{-14}$ \ergscm\ of 1$\sigma$ and 1.5 $\sigma$, respectively. 

It should be noted that the results for Cl0023 differ from those
presented in the previous work of \citet{koc09c}. The 2-8 keV band
cumulative source counts were unintentionally presented as the 2-10
keV band counts in that paper. However, we have also found changes due
to the different versions of the CIAO software used between earlier
papers \citep{koc09a,koc09b,koc09c} and this paper, version 3.3
compared to 4.2. We believe this to be an effect of updated response
functions at off-axis angles. The net result for our data is mainly an
approximate 5\% increase in flux, with some very minor differences in
point source detection, the latter being almost entirely below the
3$\sigma$ level.

In Figure \ref{logNlogScomp}, co-added cumulative source counts for all five fields are shown. To accomplish this co-addition, all point source lists were first combined into one composite source list. For the calculation of the effective sky area weighting factors, $\Omega_i$, the combined background of every field was used. With $\sigma_i^2 = \sum_{i=0}^N 1/{\Omega}_i^2$, we can see that the error in $N(>S)$ will be reduced from its value in any individual field. In the soft band, the composite source counts have an approximate average overdensity of 1$\sigma$ in the range $2 \times 10^{-15}$ to $8 \times 10^{-15}$ \ergscm, which falls off to zero brighter than 10$^{-14}$ \ergscm. In the hard$_{10}$ band, the composite counts have an approximate average overdensity of 1$\sigma$ between $7 \times 10^{-15}$ and $2 \times 10^{-14}$ \ergscm, but an approximate underdensity of 0.5$\sigma$ between $3 \times 10^{-14}$ and $4 \times 10^{-14}$ \ergscm. While there is substantial variation in the five cumulative source count measurements for the individual fields, we can see from the composite measurement that, on average, the fields of the five structures studied here have a density of X-ray point sources in excess of the control field, though not significantly so.

\begin{figure}[!b]
\epsscale{1.1}
\plotone{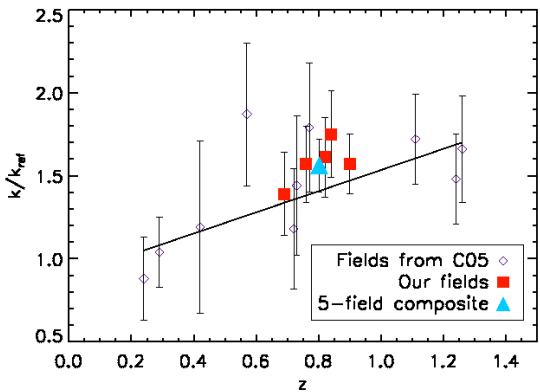}
\caption{ \footnotesize{Overdensity amplitude versus redshift in the
    hard$_{10}$ band, as defined by \citet{cap05}. $k$ is defined as
    the value of $N(>S)$ at $S_0 = {10}^{-14}$ \ergscm. Shown are the
    ten fields analyzed by C05 and their fit to this data set. Our
    fields are overplotted as squares, along with a composite
    measurement made by co-adding all five fields. (Adapted from
    Figure 6 of C05). } }
\label{capplot}
\end{figure}

\citet[][, hereafter C05]{cap05} have found a positive correlation
between cluster redshift and source count overdensity. To compare to
this, and our earlier results of \citet{koc09a}, we first fit the
cumulative source counts to a power law of the form $N(>S) =
k{(\frac{S}{S_0})}^{-\alpha}{deg}^{-2}$, using the maximum likelihood
method of \citet{crawford70} and \citet{murdoch73}. This method fixes
the dimensionless variable $k$ while fitting for $\alpha$. $k$ is
calculated by requiring consistency between the model and the data at
the flux point $S = S_0$. We use $2\times {10}^{-15}$ \ergscm\ and
$1\times 10^{-14}$ \ergscm\ for $S_0$ in the soft and hard$_{10}$
bands, respectively, which are the values used by
\citet{crawford70}. C05 measured overdensities using the ratio of $k$
in a given field to $k$ in a reference set of five blank fields. For
these fields, they found $k_{soft} = {401}_{-13}^{+31}$ and $k_{hard}
= {275}_{-9}^{+22}$. We calculated $k$ for the five fields
individually, as well as the composite of all five fields. It should
be noted that the composite measurement is not a simple average of the
five fields (see above for details on its creation) and that
significant variation in the individual $k$ values come from
calculating $k$ at a single flux point.

Our results are shown in Figure \ref{capplot} for the hard$_{10}$
band, where our data are overplotted on that of C05 and the linear fit
to their data is shown. Just as in \citet{koc09a}, our data are
consistent with their fit. Once again (refer to earlier explanation),
it is posssible that the issue with the version of the CIAO software
may have created a systematic offset from C05's data set. We would
expect this offset to increase the value of $k$ for the CO5 data by
about 5$\%$. Even without this correction, our results are still
consistent with C05 within our errors. However, we note that it is
difficult to use these overdensities to interpret the actual AGN
activity in an individual structure, even with our large spectroscopic
sample (see Section \ref{sec:AGN}).

\section{AGN and Host Properties}
\label{sec:AGN}

Using optical sources with redshifts with quality flags of $Q=3$ or 4
and the results of our optical matching, we were able to identify
X-ray sources that are members of the clusters or superclusters in our
sample (see Table \ref{srcsum}). In
summary, we found ten confirmed AGNs in the bounds of Cl1604, seven in
Cl0023, six in Cl1324, three in RXJ1821, and one in RXJ1757. Note that
these numbers include four sources, one in each structure except
RXJ1757, that were detected at a $<$3$\sigma$ (but at a $>$2$\sigma$)
level in at least one of the three X-ray passbands (see Table
\ref{AGNtab}). We show in the following that these low-significance
detections do not bias our results.
\newline
\newline
\subsection{Spatial Distribution}
\label{sec:spatdist}

\begin{figure*}
\begin{center}
\includegraphics[width=0.95\textwidth]{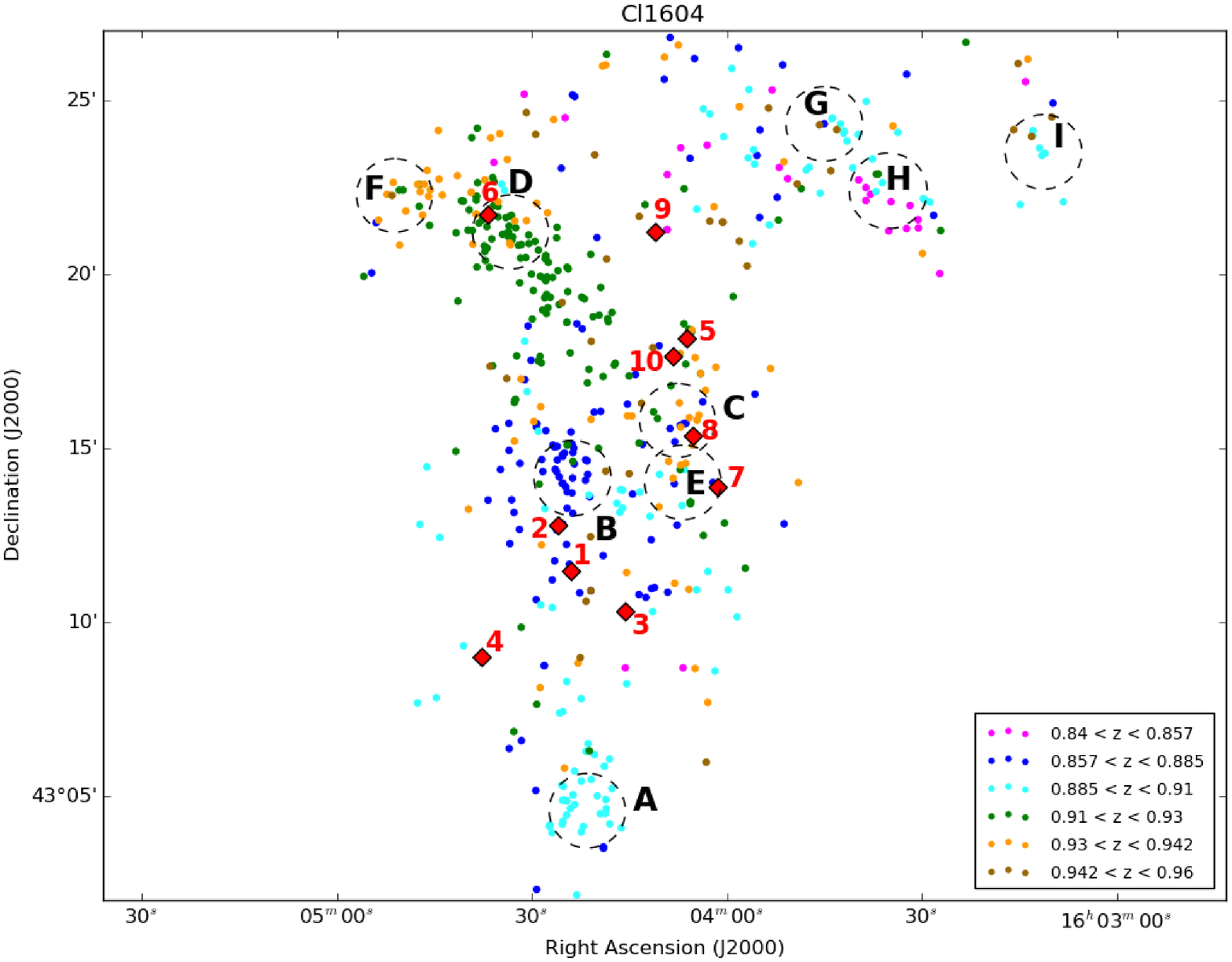}\\
\includegraphics[width=0.96\textwidth]{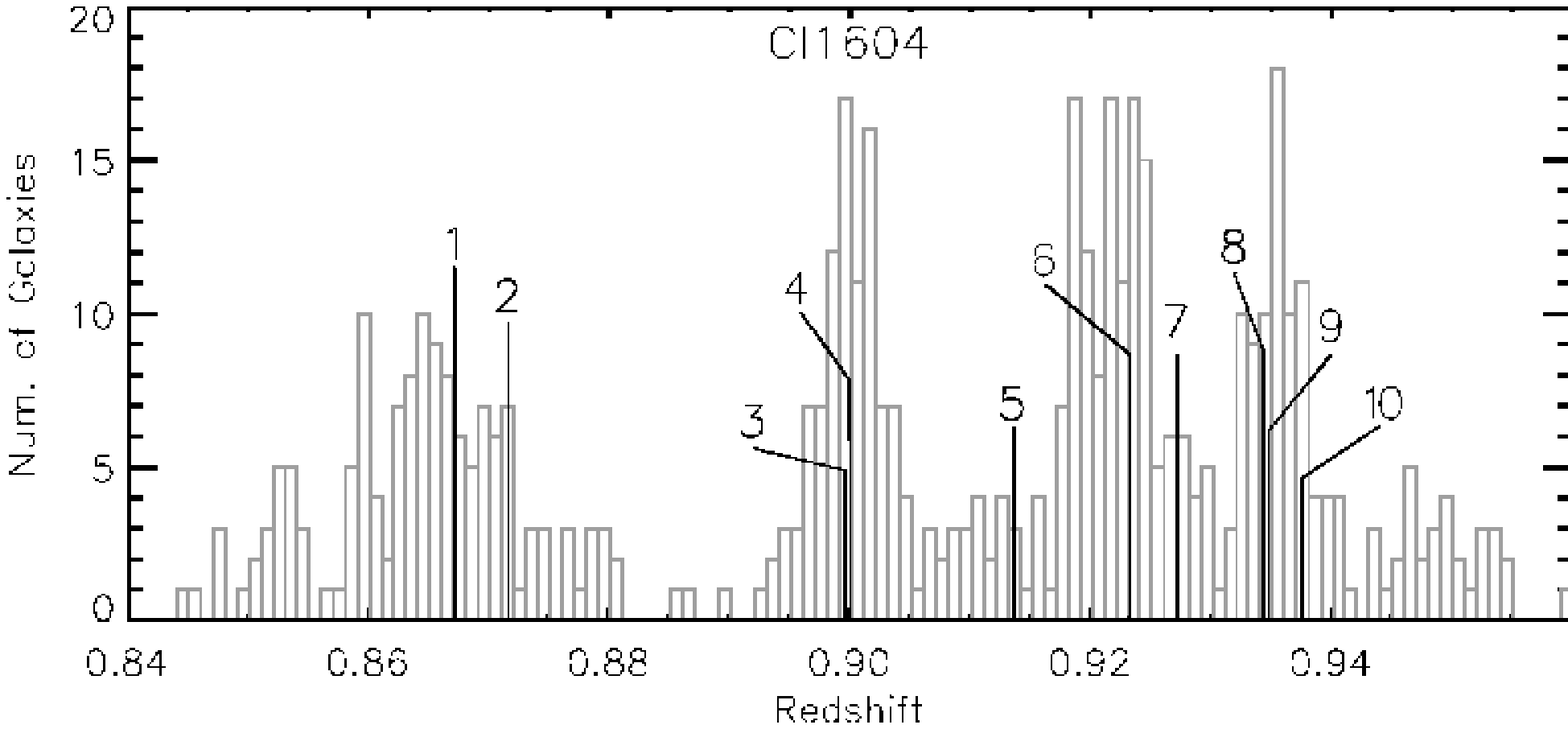}
\end{center}
\caption{\footnotesize{ Spatial map and redshift distribution for Cl1604, providing a three-dimensional map of the structure. All optical sources within the redshift bounds of the supercluster are shown, and are color-coded according to redshift. Confirmed AGNs are plotted on the spatial maps in red, and are also shown in the redshift distributions. Dashed circles show all clusters and groups in the structure, although E is not confirmed, and have radii of 0.5 $h_{70}^{-1}\ $Mpc at their respective redshifts.}}
\label{allspats0}
\end{figure*}
\begin{figure*}
\begin{center}
\includegraphics[width=0.6\textwidth]{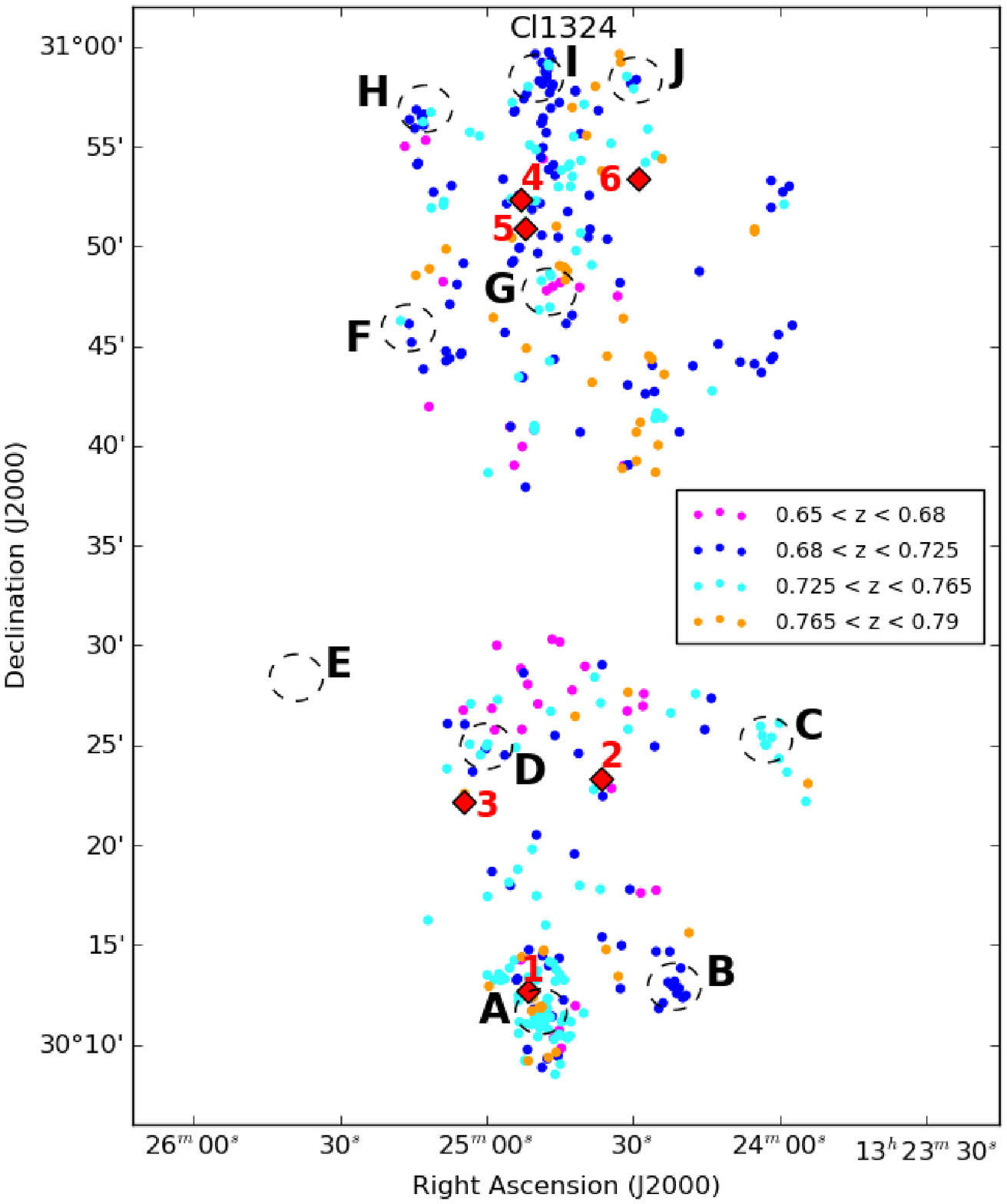} \\
\includegraphics[width=0.9\textwidth]{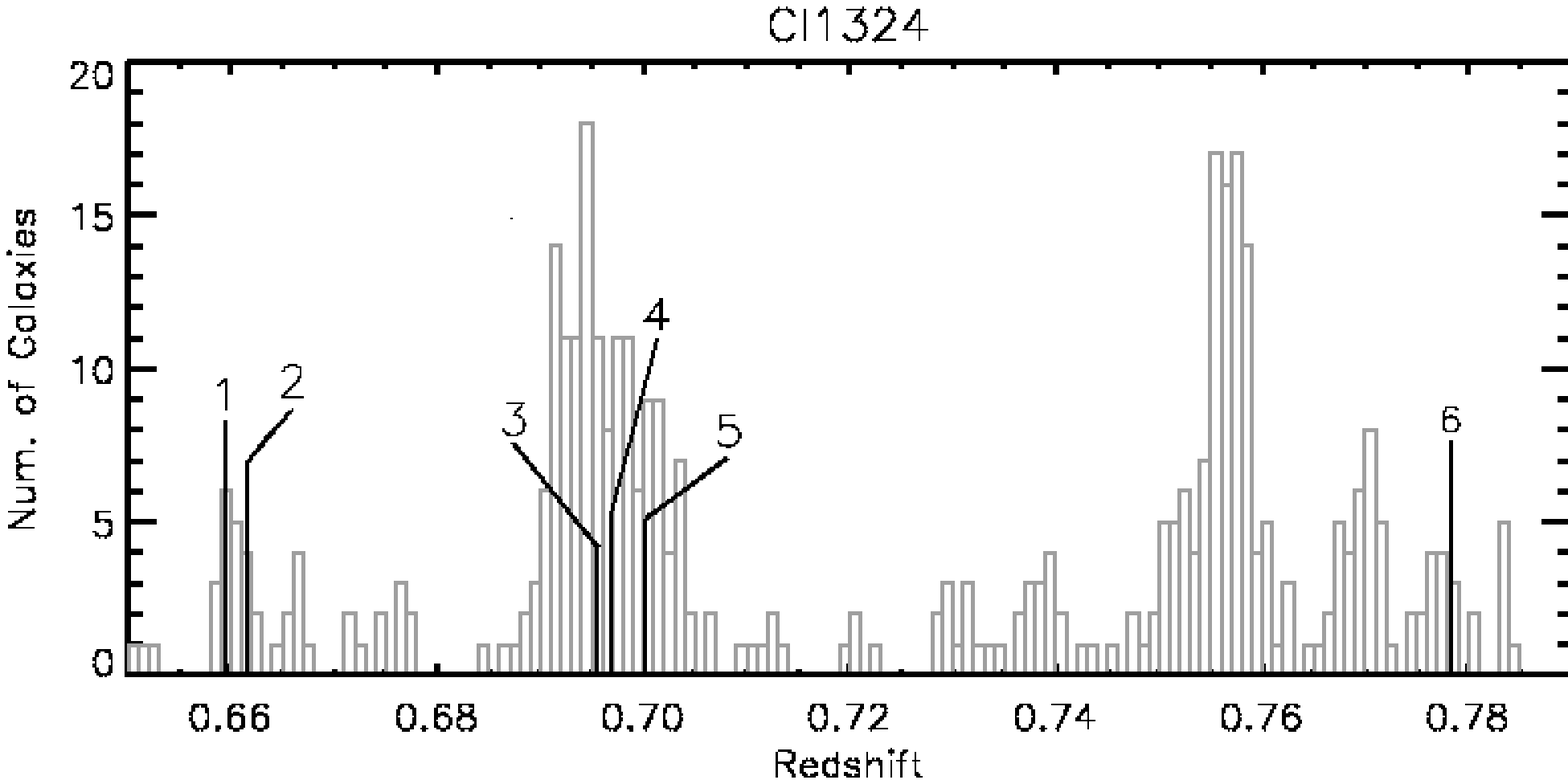}
\end{center}
\caption{\footnotesize{Spatial map and redshift distribution for
    Cl1324. All optical sources within the redshift bounds of the
    supercluster are shown, and are color-coded according to
    redshift. Confirmed AGNs are plotted on the spatial maps in red,
    and are also shown in the redshift distributions. Dashed circles
    show all clusters and groups in the structure and have radii of
    0.5 $h_{70}^{-1}\ $Mpc at their respective redshifts. These are
    derived from red galaxy overdensities, and only Clusters A, B, D,
    and I have been confirmed (see R. R. Gal et al.\ 2012, in preparation). }}
\label{allspats}
\end{figure*}
\begin{figure*}
\begin{center}
\epsscale{1.16}
\plotone{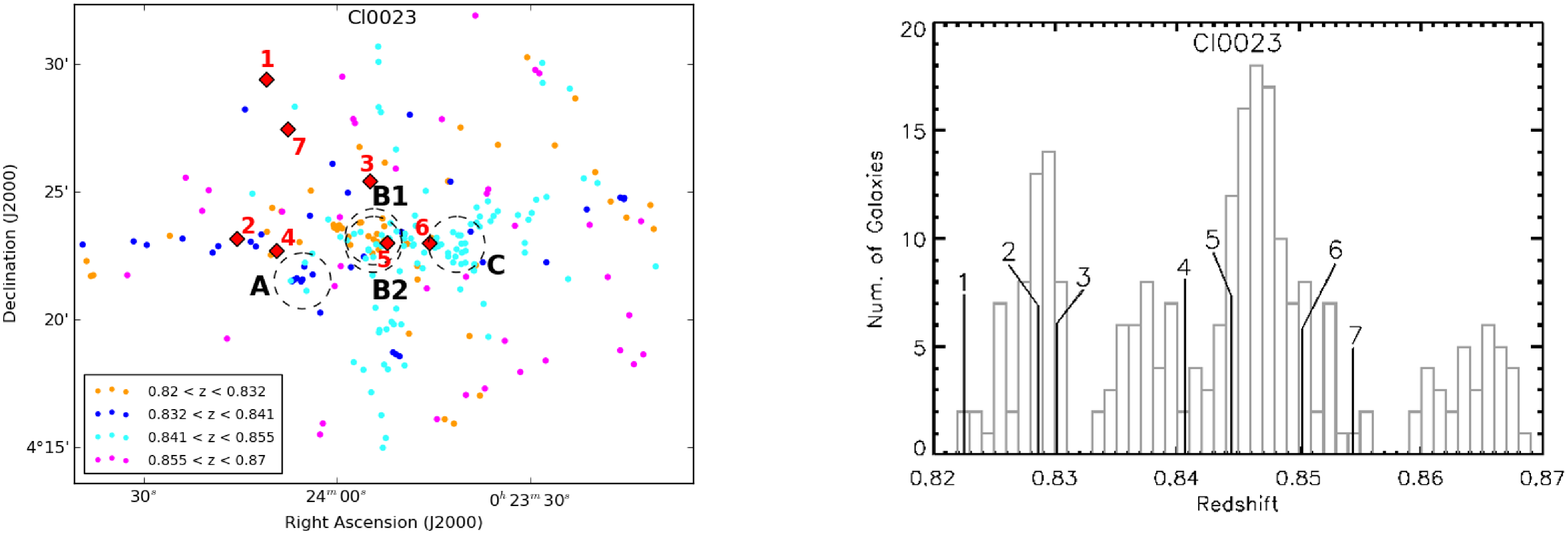}
\plotone{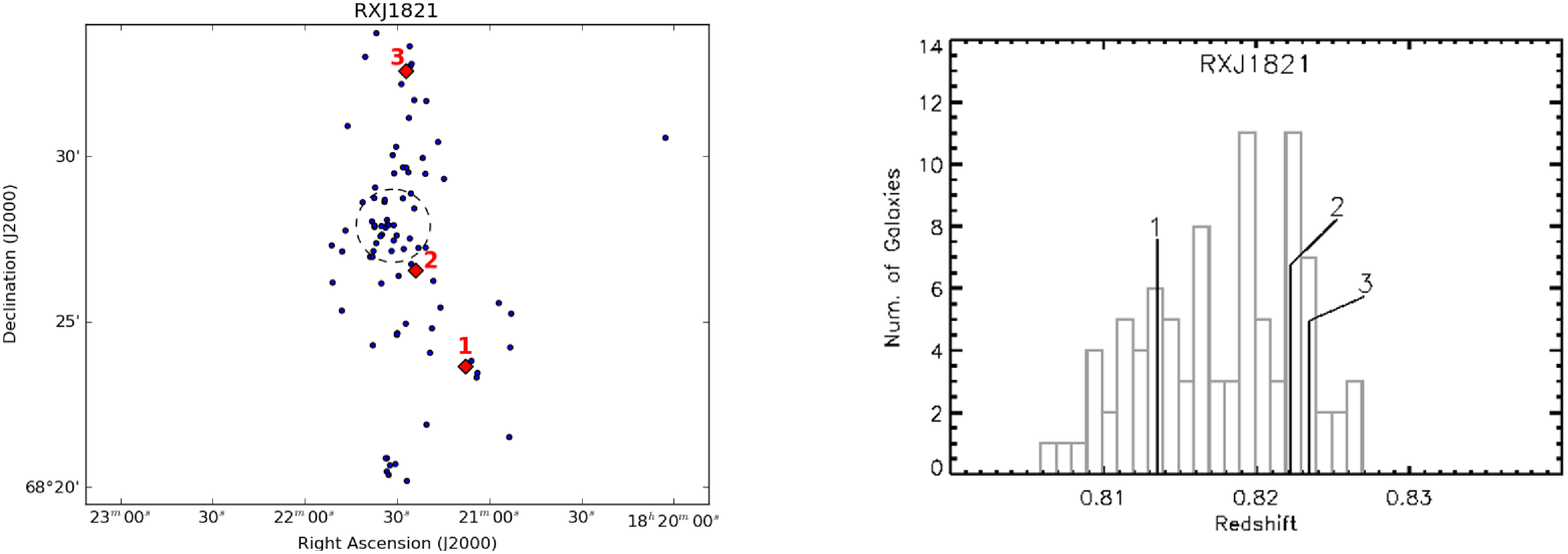}
\plotone{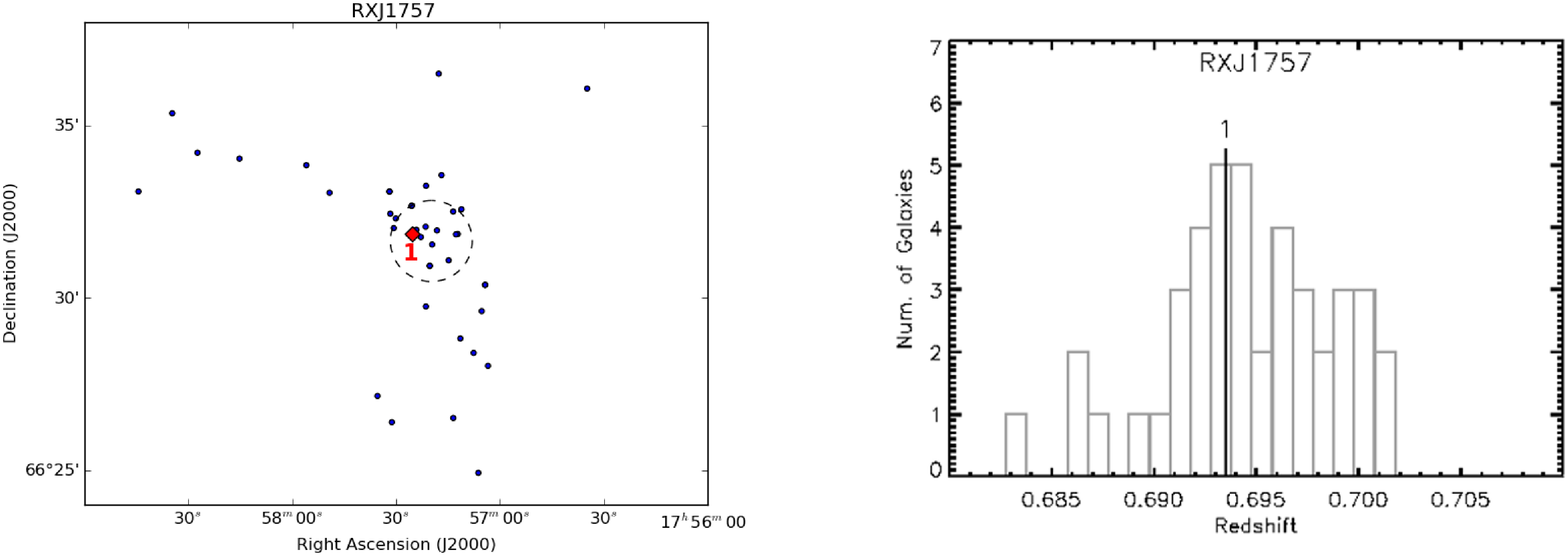}
\end{center}
\caption{\footnotesize{Spatial maps and redshift distributions for Cl0023, RXJ1757, and RXJ1821. All optical sources within the redshift bounds of the clusters and the supergroup are shown, and are color-coded according to redshift. Confirmed AGNs are plotted on the spatial maps in red, and are also shown in the redshift distributions. Dashed circles show the clusters and groups in the structures and have radii of 0.5 $h_{70}^{-1}\ $Mpc at their respective redshifts.}}
\label{allspats2}
\end{figure*}

\begin{deluxetable*}{lcccccccccc}
\tablecaption{AGN Summary}
\tablehead{
\colhead{\footnotesize{Structure}}
 & \colhead{\footnotesize{Num.}}
 & \colhead{\footnotesize{R.A.}}
 & \colhead{\footnotesize{Dec.}} 
 & \colhead{\footnotesize{$z$}}
 & \colhead{\footnotesize{}}
 & \colhead{\footnotesize{$L_x$\tablenotemark{a}}}
 & \colhead{\footnotesize{}}
 & \colhead{\footnotesize{Det.}}
 & \colhead{\footnotesize{Nearest}}
 & \colhead{\footnotesize{RS}}
 \\
   \colhead{{ }}
 & \colhead{\footnotesize{}}
 & \colhead{\footnotesize{(J2000)}}
 & \colhead{\footnotesize{(J2000)}}
 & \colhead{\footnotesize{}}
 & \colhead{\footnotesize{soft}}
 & \colhead{\footnotesize{hard}}
 & \colhead{\footnotesize{full}}
 & \colhead{\footnotesize{Sig.\tablenotemark{b}}}
 & \colhead{\footnotesize{Clus. (Mpc)\tablenotemark{c}}}
 & \colhead{\footnotesize{Offset\tablenotemark{d}}}
}
\startdata
Cl0023 &  1 & 00\ 24\ 10.9 & +04\ 29\ 23 & 0.823 & \ 5.6 & 46.7 & \ 52.3 & 12.2 & ... & -1.86 \\ 
Cl0023 &  2 & 00\ 24\ 15.5 & +04\ 23\ 09 & 0.829 & \ 7.8 & \tablenotemark{e} & \ \ 7.8 & 10.1 & 1.38 & 0.67 \\ 
Cl0023 &  3\tablenotemark{f} & 00\ 23\ 54.9 & +04\ 25\ 24 & 0.830 & \ 0.5 & \ 5.1 & \ \ 5.6 & \ 2.6 & 1.00 & 0.33\\ 
Cl0023 &  4 & 00\ 24\ 09.4 & +04\ 22\ 41 & 0.841 & 24.3 & 35.8 & \ 60.1 & 39.9 & 0.71 & -6.08 \\ 
Cl0023 &  5 & 00\ 23\ 52.2 & +04\ 22\ 59 & 0.844 & 24.8 & 83.8 & 108.6 & 63.7 & 0.25 & -2.85 \\ 
Cl0023 &  6\tablenotemark{f} & 00\ 23\ 45.6 & +04\ 22\ 59 & 0.850 & 10.8 & 17.0 & \ 27.8 & 23.2 & 0.48 & -4.70 \\ 
Cl0023 &  7 & 00\ 24\ 07.6 & +04\ 27\ 26 & 0.854 & \ 3.1 & \ 3.4 & \ \ 6.5 & \ 3.9 & 2.45$^{\ast}$ & -1.72 \\ 
Cl1604 &  1\tablenotemark{f} & 16\ 04\ 23.9 & +43\ 11\ 26 & 0.867 & 14.5 & 22.3 & \ 36.7 & 27.8 & 1.23 & -4.43 \\ 
Cl1604 &  2\tablenotemark{f} & 16\ 04\ 25.9 & +43\ 12\ 45 & 0.871 & \ 5.5 & \ 8.7 & \ 14.2 & 10.4 & 0.66 & -1.48 \\ 
Cl1604 &  3\tablenotemark{f,g} & 16\ 04\ 15.6 & +43\ 10\ 16 & 0.900 & 17.0 & 31.4 & \ 48.3 & 31.2 & 1.90$^{\ast}$ & -1.84 \\ 
Cl1604 &  4 & 16\ 04\ 37.6 & +43\ 08\ 58 & 0.900 & \ 0.9 & \ 4.4 & \ \ 5.3 & \ 2.4 & 2.44 & ... \\ 
Cl1604 &  5 & 16\ 04\ 06.1 & +43\ 18\ 07 & 0.913 & 18.3 & 29.1 & \ 47.4 & 19.3 & 1.08$^{\ast}$ & -1.02 \\ 
Cl1604 &  6\tablenotemark{f} & 16\ 04\ 36.7 & +43\ 21\ 41 & 0.923 & \ 6.3 & 18.1 & \ 24.4 & \ 9.8 & 0.36 & -3.64 \\ 
Cl1604 &  7 & 16\ 04\ 01.3 & +43\ 13\ 51 & 0.927 & 12.3 & 25.0 & \ 37.3 & 18.6 & 1.03 & ... \\ 
Cl1604 &  8 & 16\ 04\ 05.1 & +43\ 15\ 19 & 0.934 & \ 3.6 & \ 4.7 & \ \ 8.4 & \ 4.3 & 0.29 & -1.07\\ 
Cl1604 &  9 & 16\ 04\ 10.9 & +43\ 21\ 11 & 0.935 & \ 0.9 & 10.7 & \ 11.6 & \ 4.0 & 1.87$^{\ast}$ & 0.97 \\ 
Cl1604 & 10\tablenotemark{f} & 16\ 04\ 08.2 & +43\ 17\ 36 & 0.937 & \ 5.0 & 24.9 & \ 30.0 & \ 7.1 & 0.84 & -2.47 \\ 
Cl1324 &  1\tablenotemark{f} & 13\ 24\ 51.4 & +30\ 12\ 39 & 0.660 & \ 1.6 & \ 0.7 & \ \ 2.3 & \ 4.1 & 0.44$^{\ast}$ & -2.85 \\
Cl1324 &  2 & 13\ 24\ 36.4 & +30\ 23\ 16 & 0.662 & \ 1.0 & \ 6.8 & \ \ 7.8 & \ 4.2 & ... & -2.66 \\ 
Cl1324 &  3 & 13\ 25\ 04.5 & +30\ 22\ 07 & 0.696 & \ 1.2 & \ 0.6 & \ \ 1.8 & \ 2.1 & ... & -2.53 \\ 
Cl1324 &  4 & 13\ 24\ 52.9 & +30\ 52\ 18 & 0.697 & \ 1.4 & \tablenotemark{e} & \ \ 1.4 & \ 3.3 & 2.60 & -1.79 \\ 
Cl1324 &  5 & 13\ 24\ 52.0 & +30\ 50\ 51 & 0.700 & \tablenotemark{e} & 10.9 & \ 10.9 & \ 6.2 & ... & -1.23 \\ 
Cl1324 &  6 & 13\ 24\ 28.8 & +30\ 53\ 20 & 0.778 & \ 1.6 & \ 0.8 & \ \ 2.4 & \ 3.0 & 2.90$^{\ast}$ & -0.24 \\ 
RXJ1821 &  1 & 18\ 21\ 07.7 & +68\ 23\ 38 & 0.813 & \ 1.9 & \ 2.5 & \ \ 4.4 & \ 2.7 & 2.17 & -0.51 \\ 
RXJ1821 &  2\tablenotemark{f} & 18\ 21\ 23.9 & +68\ 26\ 33 & 0.822 & \ 3.4 & \ 4.1 & \ \ 7.5 & \ 5.5 & 0.67 & -1.20 \\ 
RXJ1821 &  3 & 18\ 21\ 27.0 & +68\ 32\ 34 & 0.824 & \ 8.6 & 10.6 & \ 19.3 & 10.3 & 2.13 & -0.66 \\ 
RXJ1757 &  1 & 17\ 57\ 25.2 & +66\ 31\ 50 & 0.693 & \ 1.9 & \ 5.1 & \ \ 7.1 & \ 6.1 & 0.24 & -2.11 \\ 
\enddata
\tablenotetext{a}{\footnotesize{Rest frame X-ray luminosity in units of 10$^{42}$ ergs s$^{-1}$. Soft, hard, and full bands are defined as 0.5-2.0, 2.0-8.0, and 0.5-8.0 keV, respectively.}}
\tablenotetext{b}{\footnotesize{Largest detection significance (in $\sigma$'s) in the three X-ray bands.}}
\tablenotetext{c}{\footnotesize{Projected two-dimensional distance to the nearest confirmed cluster or group. An asterisk indicates that the AGN host is not believed to be associated with the nearest cluster, which is defined here as a redshift separation of more than 0.01. For the two Cl0023 groups that are overlapping on the plane of the sky, distance was only calculated to the closer group in redshift space. Distances were not calculated to Cl1604 cluster E, for which no redshift maximum could be determined. Note that only Clusters A, B, D, and I are confirmed in Cl1324. Only distances less than 3 Mpc are listed.}}
\tablenotetext{d}{\footnotesize{Scaled offset from the center of the red sequence fit in units of RS widths, $W_{RS}$. Refer to Sections \ref{TRSATBP} and \ref{sec:hgalcolan}. Only Cl1604 AGNs with ACS data available for hosts are included.}}
\tablenotetext{e}{\footnotesize{Undetected in respective band.}}
\tablenotetext{f}{\footnotesize{Member of a close kinematic pair (see Section \ref{sec:hgalcolan}).}}
\tablenotetext{g}{\footnotesize{Has two close kinematically associated companions.}}
\label{AGNtab}
\end{deluxetable*}

Examining the spatial distribution of AGNs located within each cluster
can give insight into what processes triggered their nuclear
activity. In Figures \ref{allspats0}, \ref{allspats}, and
\ref{allspats2}, we show the spatial distributions on the sky and
redshift distributions of the five structures studied here. The AGNs
are marked in red, and their positions and characteristics are given
in Table \ref{AGNtab}.

Of particular note is the lack of AGNs in dense cluster
centers. Indeed, we find $19\%$ of AGNs in cluster cores, defined as
being within a projected distance of 0.5 Mpc to the nearest cluster or
group\footnote{Redshift is taken into account when determining
  proximity of AGNs to clusters or groups in the superclusters and the
  supergroup. Projected two-dimensional distance is only calculated to
  clusters of similar redshift.}. An additional $27\%$ lie on the
outskirts of clusters (projected distances between 0.5 and 1.5 Mpc),
and a majority ($>50\%$) of AGN host galaxies lie more than 1.5 Mpc in
projected distance from the nearest cluster or group.

\begin{figure}[!h]
\plotone{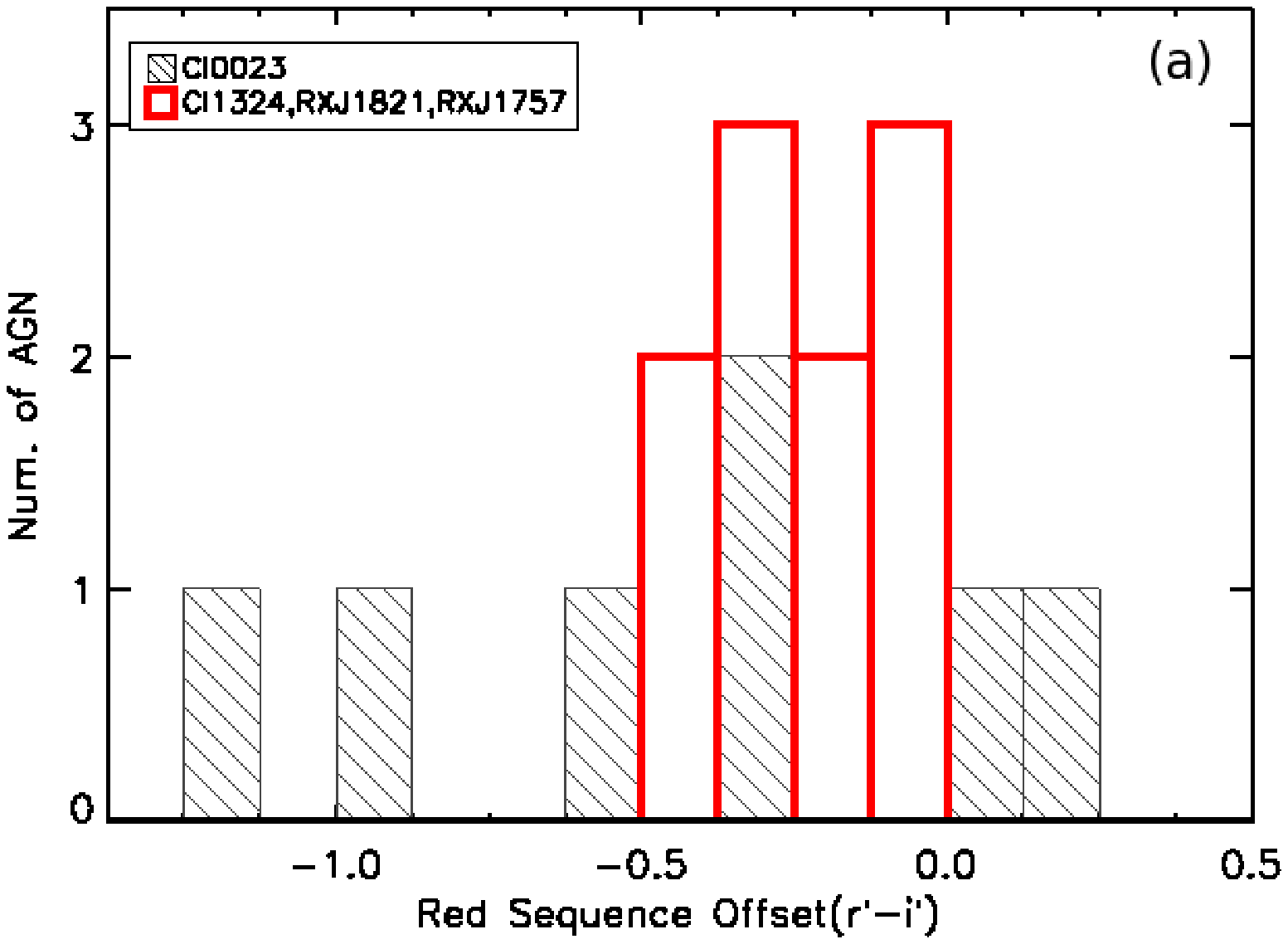}
\plotone{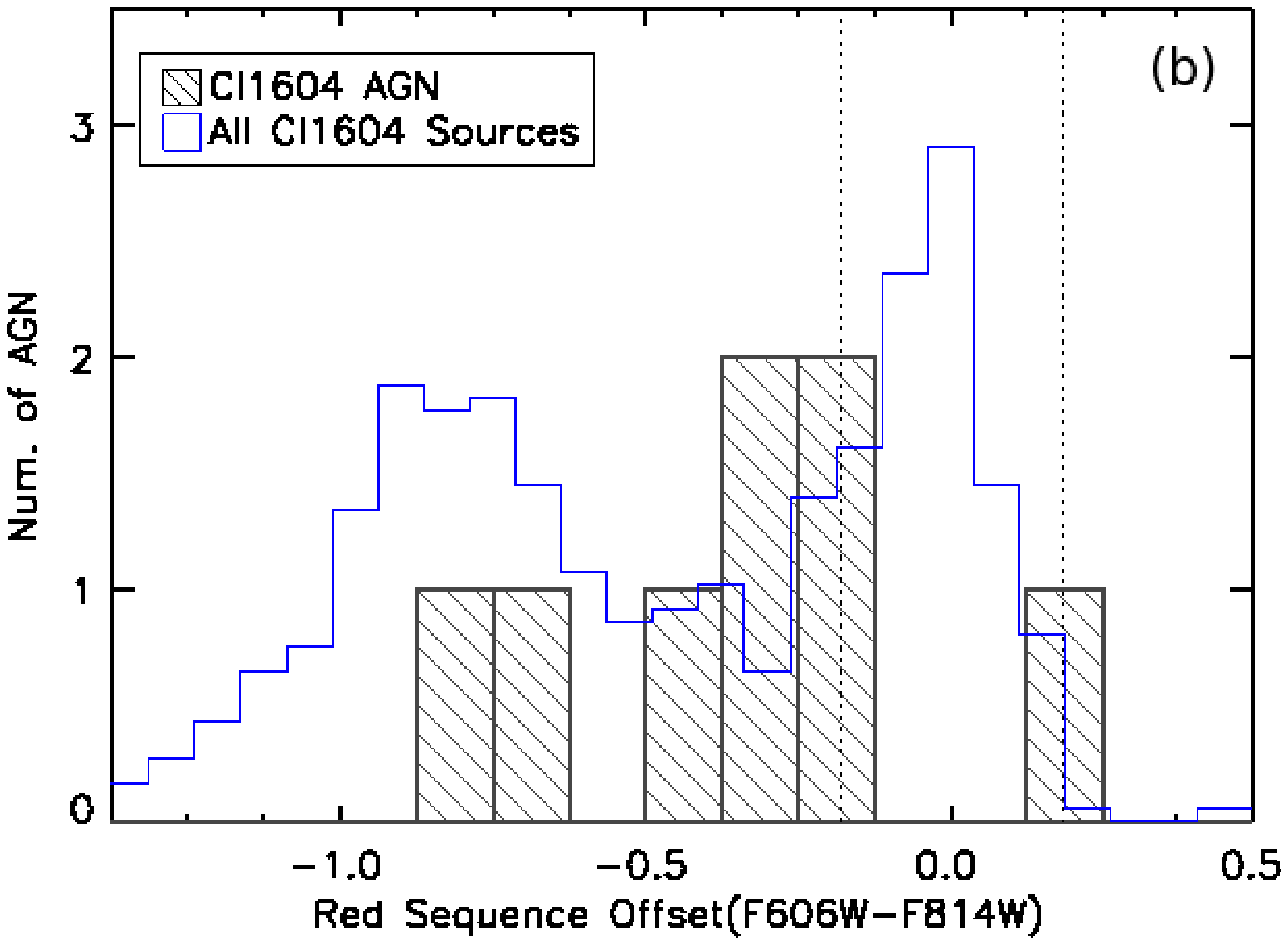}
\plotone{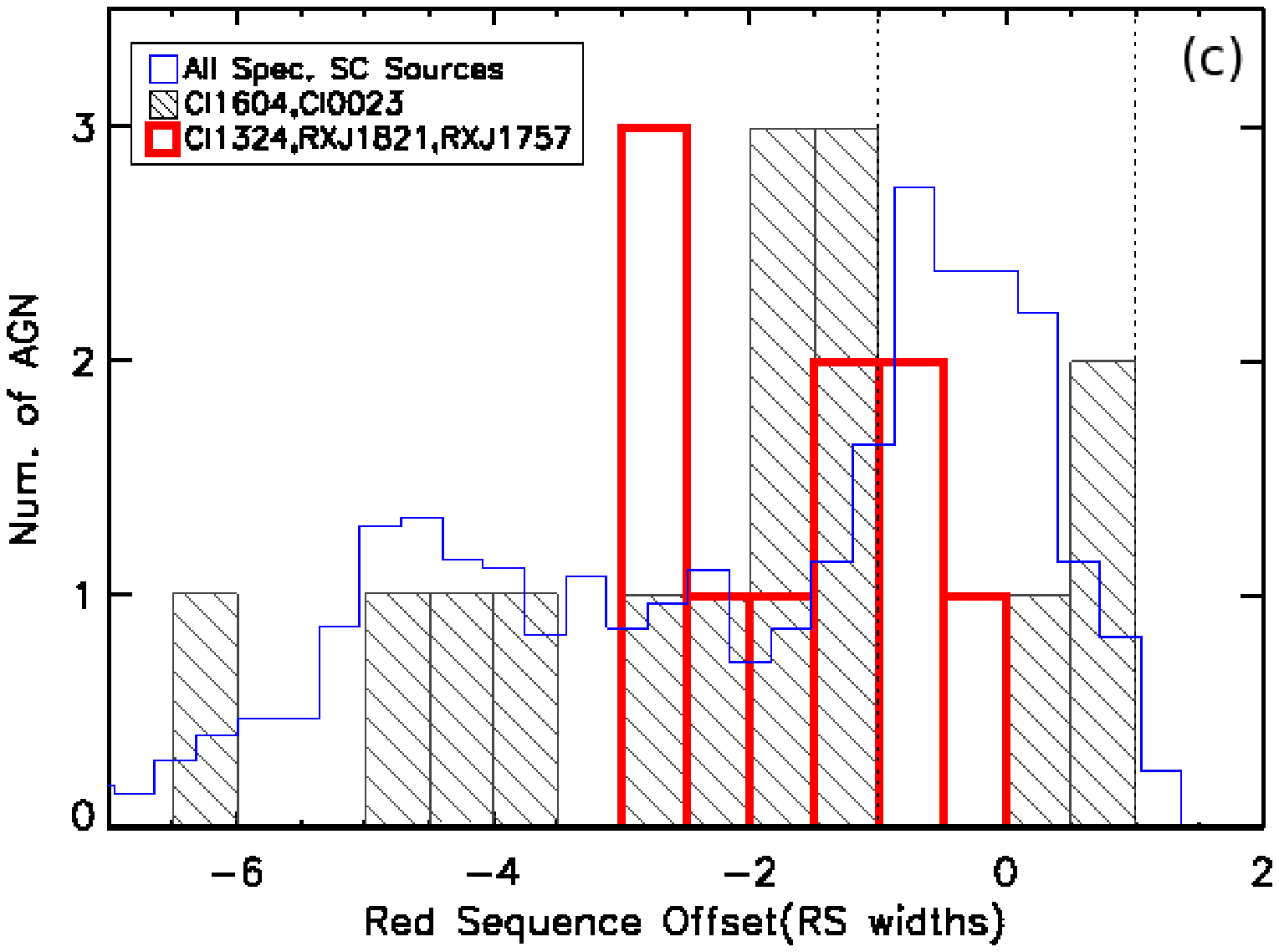}
\caption{ \footnotesize{Offsets from the red sequence (RS) for AGN
    host galaxies. Panel (a) shows only those offsets calculated
    using the LFC $r'-i'$ colors, which includes all fields except for
    Cl1604. Panel (b) shows Cl1604 AGN host galaxies where the
    ACS F606W-F814W colors were used (hashed histogram). Note that
    these offsets are calculated from the center of the red sequence,
    and the RS boundaries differ from field to field. In panel (c), all five fields are included and are scaled by the red
    sequence widths. In panels (b) and (c), scaled
    distributions of all spectroscopically confirmed supercluster
    members are also shown for reference (light-outlined solid
    histograms). In all panels, evolved structures members are
    indicated with thick-outlined histograms, while members of the
    unevolved structures are indicated by hashed histograms.} }
\label{RSoffsets}
\end{figure}

These results are consistent with previous work studying the spatial
distribution of AGNs, in that the AGNs tend to be located outside of
clusters or in their outskirts. First, the distribution of AGNs in
Cl1604, previously studied by \cite{koc09a}, is consistent with the
other four structures here. Our results are also consistent with those
of \cite{gil07}, who found that AGNs in the A901/902 supercluster at
$z\approx 0.17$ tend to avoid the densest areas. Several other studies
have also found that X-ray AGNs tend to reside in regions of moderate
density similar to group environments, up to $z\sim1$ \citep{hickox09,silver09}. It is thought that
regions of intermediate density, such as the outskirts of clusters,
are the most conducive to galaxy-galaxy interactions because of the
elevated densities, compared to the field, but relatively low
velocities \citep{cav92}. Since we find more of the AGNs in these
areas, this lends support to the theory that mergers or tidal
interactions are one of the main instigators of AGN activity. Since
only $\sim20\%\ $ inhabit dense cluster cores, processes which
preferentially occur in these regions, such as ram pressure stripping,
are probably not responsible for triggering AGNs in cluster galaxies. However, the association between AGNs and these regions could also be related to higher gas availability in galaxies farther from cluster cores. 

Although we find that most of the X-ray AGNs do not reside in the cluster cores, a number of studies have measured the fraction of cluster galaxies that host AGN \citep[e.g.,][]{mart07}. Therefore, we attempt to measure the AGN fraction for the individual clusters within the five structures in our sample. Because we are limited by the spectral completeness of our sample, we make a composite measurement of the most massive, well-sampled clusters: Cl1604A, Cl1604B, Cl1604C, Cl1604D, Cl1324A, Cl1324I, and RXJ1821. We compare to the results of \citet{mart07} for low-redshift clusters ($0.06 < z < 0.31$), who used galaxies with $M_R < -20$, within approximately 2000 \kms\ of the mean redshift of cluster members for each system, and within the field of view of their {\it Chandra} observations, which ranged from 1.2 to 4.5 Mpc in width. To approximate these criteria, we adopt a magnitude cutoff that roughly corresponds to $M_{i'/F814W}=-19$ and use galaxies within 1 Mpc and $\Delta z = 0.01$ of the cluster centers and redshifts. We measure the combined AGN fraction for the seven clusters listed above to be 0.012$\pm0.007$. This is consistent with the results of \citet{mart07}, who measured $f_A\left(M_R < -20;L_x > 10^{42}\ h_{70}^{-2}\ {\rm erg\ s}^{-1} \right)\approx1\%\ $. We note, however, that a number of arbitrary definitions went into our measurement. In addition, our spectroscopy of optical counterparts to X-ray sources is significantly incomplete, in contrast to \citet{mart07}. Correcting for this incompleteness contributes the largest source of error to our measurement. Because of the large uncertainties, we refrain from drawing any conclusions from our measurement. 
\newline
\subsection{AGN Host Galaxy Colors}
\label{sec:hgalcolan}

Figure \ref{CMDS} presents CMDs, which are
described in Section \ref{sec:globchar}, for all five
fields. Confirmed AGN members of the structures are shown with blue squares. While the AGNs in Cl1324, RXJ1821, and RXJ1757 preferentially
reside within the bounds of, or very close to, the red sequence, the AGN hosts in Cl0023 and Cl1604 are more spread out. Previous work, including \citet{koc09a} and a
number of wide-field surveys
\citep{sanchez04,nandra07,georg08,silver08}, has found an association
between AGN activity in galaxies and the transition onto the red
sequence (possibly for a second time) in the green valley. While these
galaxies could be evolving from the blue cloud onto the red sequence,
it is also possible that they could have moved down off the red
sequence after a tidal interaction or merger and are evolving back
\citep{men01,koc09a}. We note, however, that some studies using mass-selected samples have found that AGN hosts have a color distribution more similar to that of normal galaxies \citep{silver09b,xue10}. 

In addition, \citet{card10}, using a sample of galaxies with redshifts $0.8 < z < 1.2$, have found that many green valley AGN hosts are dust-reddened blue cloud members, so that AGN host colors acquire the bimodality apparent in the general galaxy population. However, \citet{ros11} have also examined the impact of extinction on the colors of AGN hosts and did not find a significant impact for galaxies in the redshift range $0.8 < z < 1.2$, although bimodality may be introduced at higher redshifts. To address this issue for our study, we are planning to implement spectral energy distribution (SED) fitting to evaluate the impact of extinction on the broadband colors of our sample. Preliminary results from SED fitting of the Cl1604 hosts suggest that extinction levels in our sample are not as drastic as those presented by \citet{card10}. It is also possible that AGN host colors are contaminated by the AGNs themselves. However, this is unlikely because (1) almost all AGN hosts in our sample have rest-frame X-ray luminosities below the quasi-stellar object (QSO) level of $10^{44}$ erg s$^{-1}$ (see Sec \ref{sec:xlum}) and (2) \citet{koc09b} found that, in Cl1604, AGN hosts with blue cores did not have a rising blue continuum indicative of QSO activity. Therefore, we proceed to investigate the AGN
association with the transition zone and to explore differences in the
evolutionary states of their host galaxies in each field by examining
color offsets of the AGN hosts from the red sequence.

Histograms of offsets from the center of the red sequence are shown in
Figure \ref{RSoffsets}. The first two panels present offsets in terms
of color. In the top panel, only the structures where LFC data were
used are shown, which is every field except Cl1604. The middle panel
shows only Cl1604, for which we used ACS colors. In the bottom panel,
all five structures are shown, with normalized offsets. In order to
compare the ACS and LFC data, we scale by the red sequence width,
$W_{RS}$. We define $W_{RS}$ as the distance from the center of the
red sequence fit to its boundary (see Section \ref{TRSATBP} and Figure
\ref{CMDS}). On this plot, AGN hosts on the red sequence will then be
located between -1.0 and 1.0.

With red sequence offsets, we can quantitatively examine the green
valley. In Figure \ref{RSoffsets}(b), we plot a histogram of RS offsets,
measured from the ACS data, for the AGNs in the Cl1604
supercluster. For comparison, we overplot a scaled distribution of RS
offsets for all spectroscopically confirmed supercluster members with
ACS photometry. In the scaled histogram, we can clearly see an area
of reduced number density between the red sequence and blue cloud. For
Cl1604, the green valley can be approximated as the region $-3W_{RS} <
\Delta C < -W_{RS}$, where $\Delta C$ is the offset of an AGN host
from the center of the red sequence. Only $\sim 17\%\ $of all confirmed supercluster members with ACS data fall within this region. However, five out of eight
of the Cl1604 AGN hosts with ACS data reside within it. While it is
unclear how well this definition of the green valley extends to LFC
data, because of larger photometric errors, we can see in Figure
\ref{RSoffsets}c that $36\%$ of all host galaxies have $-2W_{RS} <
\Delta C < -W_{RS}$ and $60\%$ are in the range $-3W_{RS} < \Delta C <
-W_{RS}$.

While many galaxies in both the evolved and unevolved structures lie
in the `green valley' region, the percentage of AGNs on the red
sequence is somewhat higher in the evolved structures compared to the
unevolved structures, with 30$\%$ and $20\%$, respectively. Examining
Figure \ref{RSoffsets}, we can see the distribution of AGN hosts in
evolved structures is clustered closer to the red sequence, while in
the unevolved structures, this distribution has a large tail extending
into the blue cloud. Indeed, none of the AGN host galaxies in the
evolved structures have $\Delta$C $<-3W_{RS}$, whereas four of the X-ray
AGN hosts in the unevolved structures have red-sequence offsets below
this limit. Although these results are suggestive (and unaffected by
our inclusion of the $<$ 3$\sigma$ X-ray sources), the two
distributions are not statistically different based on the K-S test.

Morphological analysis by \citet{koc09a} has shown that $\sim67\%$ of
the X-ray AGNs in Cl1604 have had recent mergers or tidal interactions,
which could fuel star formation through starburst events. More recent
mergers or interactions are one possible explanation for some of the
color differences that we see between the AGN host galaxies in the
evolved and unevolved structures. In particular, we find that nine of
the AGN host galaxies are members of a kinematic close pair with a
relative line-of-sight velocity of $\Delta v \le 350$ km s$^{-1}$ and
projected physical separation (on the plane of the sky) of $\Delta r_p
\le 70$ h$^{-1}_{70}$ kpc \citep[e.g.,][]{lin07}. Two are in Cl0023,
five in Cl1604, one in Cl1324, and one in RXJ1821 (see Table
\ref{AGNtab}). Those AGN hosts in pairs include three out of the four
galaxies with the largest red sequence offsets (i.e., the bluest), all
of which are members of the unevolved structures. Based on their $z'$ magnitudes or measured stellar masses (in the case of the Cl1604 members; see \citet{lemaux11}), seven out of the nine kinematic pairs have flux or mass ratios of $\gtrsim 50\%$, implying a major merger scenario.

The differences in color and, perhaps, merger activity are likely
related to the increased level of star formation and starburst
activity in the unevolved compared to evolved structures (see Section
\ref{sec:specprop}).  To explore the connection between the AGN and
star formation history, we can use our high-resolution spectroscopy to
examine the average spectral properties of their host galaxies.

\begin{figure}[!b]
\epsscale{1.2}
\plotone{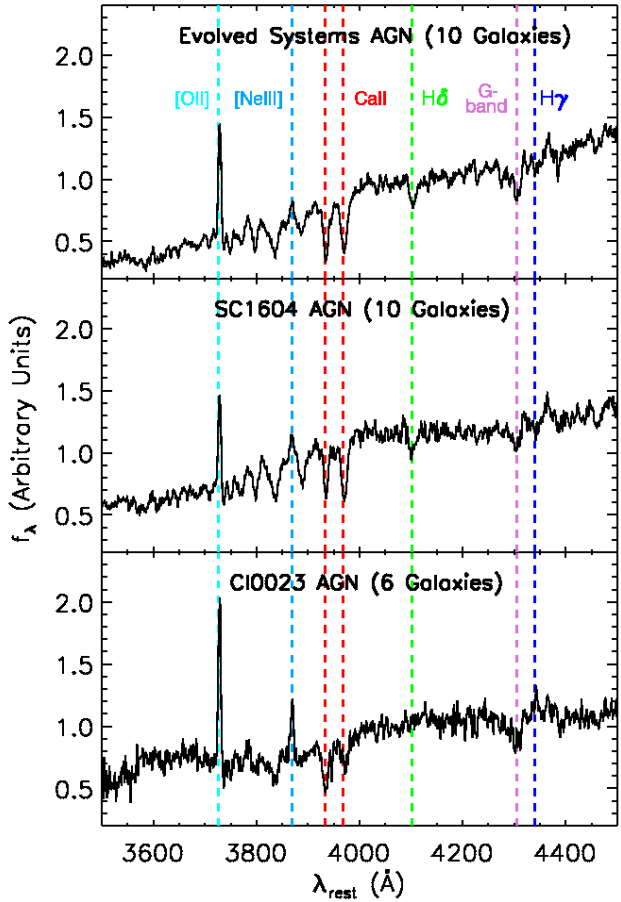}
\caption{ \footnotesize{Composite spectra for three bins of AGNs from
    the five fields studied here. The top bin includes AGN hosts from
    Cl1324, RXJ1821, and RXJ1757. Dashed lines indicate spectral
    features of interest. Notice the increasing depth of the H$\delta$
    absorption feature, which tracks recent star formation, going from
    the least evolved structure in our survey, Cl0023, to the top bin,
    made up of the most evolved structures. While all three spectra
    have significant differences, Cl0023 seems to be the only
    structure with substantial Balmer line emission and ongoing star
    formation.}}
\label{CSplot}
\end{figure}

\subsection{Average Spectral Properties of AGN Hosts}
\label{sec:compspec}

We measure the average spectral properties of the AGN host galaxies in
the five ORELSE structures using three composite spectra: one
comprised of six AGNs from Cl0023\footnote{While we have seven AGNs in
  our Cl0023 sample, one was not included in the composite
  spectrum. Our Cl0023 sample contains two AGNs which are not type-2
  (Cl0023 is the only structure with any such AGN). One of these AGNs
  has very broad-line features (rotational velocities $\gtrsim
  1000\ $\kms) which would dominate a composite spectrum and was
  therefore not included. While the other has a very broad \ion{Ne}{5}
  line, the other features are narrow and do not dominate the
  composite spectrum. As a result, it was included.}, one comprised of
all ten AGNs from Cl1604, and one comprised of the ten AGNs from the
combined fields of Cl1324, RXJ1821, and RXJ1757. The last spectrum
combines all the evolved structures, necessitated by the low number of
AGNs in RXJ1821 and RXJ1757 compared with the other structures. These
three composite spectra are shown in Figure \ref{CSplot}, and
measurements of spectral features are listed in Table \ref{co-addtab}.

First, we can see that the AGN hosts in all three groupings have
substantial \OII\ emission. In Cl1604 and the evolved structures, most
of this emission is probably from the AGNs, rather than star
formation. Six of the ten AGN host galaxies in our Cl1604 sample were
analyzed by \citet{koc11b} using the Keck II Near-Infrared Echelle
Spectrograph \citep[NIRSPEC;][]{mclean98}. For five out of the six
targets, \citet{koc11b} found that the [\ion{N}{2}]/H$\alpha$ flux
ratio was too large for a normal star-forming galaxy, which implies
that AGNs are the dominant contributor to \OII\ emission
\citep{kauf03}. The Cl1324, RXJ1821, and RXJ1757 structures have AGNs
mostly near or on the red sequence. In addition, we found in Section
\ref{sec:globchar} that star formation is low in all three structures
($\lesssim30\%\ $of all galaxies are star-forming). Because of this
low star formation rate, and because \citet{lemaux10} found that $\sim
90\%$ of red \OII\ emitters are dominated by LINER/Seyfert emission,
it is likely that most of the \OII\ emission from the AGN host galaxies in
the evolved structures comes from the AGNs as well.

Deciphering the origin of the \OII\ emission in the AGN host galaxies
in Cl0023 is not as straightforward. While some AGN host galaxies in
Cl0023 are on the red sequence, the structure also has the bluest host
galaxies in our sample. Also, as discussed in Section
\ref{sec:globchar}, there is significant star formation in the general
population of Cl0023. While this is also true of Cl1604, the NIRSPEC
results of \citet{koc11b} showed most of the \OII\ emission in the AGN
host galaxies in that structure comes from the AGNs. However, we do not
have any near-IR spectroscopic data for the Cl0023 AGNs, so we must use
other means to determine the emission source. The average
[\ion{Ne}{3}]/\OII\ ratio of the Cl0023 X-ray AGN hosts is 0.429,
typical of type-2 AGN emission, emission from metal-poor star
formation, or a superposition of the two processes \citep{nagao06,tro11}. Note that Cl1604 and the evolved structures have values of the [\ion{Ne}{3}]/\OII\ ratio of 0.549 and 0.229, respectively, also typical of type-2 AGN emission. Combining this result with the blue colors of the Cl0023
hosts and the high fraction of star-forming galaxies, it is likely
that the observed \OII\ emission of the AGN hosts in Cl0023 is due to
a combination of normal star formation and type-2 AGN activity.

Examination of the Balmer features reveals further insight into the
star formation histories of the AGN host galaxies. Specifically, based
on a single stellar population model, the EW(H$\delta$) rises quickly
from zero after a starburst in a galaxy, peaks after about 300--500
Myr, and then declines back to approximately zero at $\sim 1$ Gyr
after the burst \citep{pogg97}. Infill can complicate the
interpretation of this feature, so care must be taken when measuring
the line strength. For the composite spectrum of the Cl0023 AGN hosts,
the equivalent width of H$\delta$, attempting to correct for infill,
is consistent with zero. However, we observe strong emission from
other Balmer features, EW$\left(H\beta\right) =
-7.81$\AA\footnote{Because a significant number of the AGN host spectra across all
  fields do not have spectral coverage including the H$\beta$ line,
  Figure \ref{CSplot} was not drawn out to this range. However,
  approximately two-thirds of AGNs in the Cl0023 structure do have
  spectral coverage for H$\beta$, and the measurement presented here
  represents the average value for these galaxies.} and
EW$\left(H\gamma\right) = -2.55$\AA, suggesting that emission infill
has a significant effect on the measured EW(H$\delta$). This infill
could be due to emission from \ion{H}{2} regions, emission from AGNs or
some other LINER processes, or from continuum emission produced by O
stars. Since we observe H$\beta$ and H$\gamma$ in emission, it is
unlikely that O stars are solely responsible for the observed
EW(H$\delta$). The Balmer emission lines observed in the Cl0023 AGN
hosts are not broad\footnote{With the exception of the type-1
  AGN excluded from the composite spectrum, rotational velocities are $\lesssim 200\ $\kms.}
and are quite strong. In the average type-2 AGN, a large fraction of
the Balmer emission originates from star formation \citep{kauf03},
which suggests ongoing star formation in the Cl0023 hosts. Compared to Cl0023, H$\gamma$ emission from the Cl1604 hosts is low, with EW(H$\gamma$) = -0.69$\pm0.17$, which is consistent with a lower level of star formation in the supercluster\footnote{EW(H$\beta$) could not be measured for a majority of galaxies in the Cl1604 host sample.}. Balmer lines for the evolved structures are in absorption, EW(H$\beta$) = 0.48$\pm0.12$ and EW(H$\gamma$) = 1.41$\pm0.16$, consistent with an even lower level of star formation. These results
combined with the earlier result analyzing the average
[\ion{Ne}{3}]/\OII\ ratios of the AGN hosts strongly indicates that
star formation is occurring in the Cl0023 galaxies coevally with AGN
activity, while less star formation activity is occuring in the other structures.

\begin{deluxetable*}{lcrccc}
\tablecaption{Spectral Measurements of AGN Hosts}
\tablehead{
\colhead{\footnotesize{Bin}}
 & \colhead{\footnotesize{Num. of}}
 & \colhead{\footnotesize{EW(\OII)\tablenotemark{a}}}
 & \colhead{\footnotesize{EW(H$\delta$)\tablenotemark{a,b}}}
 & \colhead{\footnotesize{D$_{\rm n}$(4000)\tablenotemark{c}}}
 & \colhead{\footnotesize{\ion{Ca}{2} H+H$\epsilon$/\ion{Ca}{2} K\tablenotemark{d}}} \\
\colhead{\footnotesize{}} 
 & \colhead{\footnotesize{Galaxies}}
 & \colhead{\footnotesize{(\AA)}}
 & \colhead{\footnotesize{(\AA)}}
 & \colhead{\footnotesize{}}
 & \colhead{\footnotesize{}}
}
\startdata
Cl0023 & \ 6 & $-11.20\pm0.21$ & $0.36\pm0.39$ & $1.05\pm0.003$ & $0.42\pm0.07$\\
Cl1604 & 10 & $-7.09\pm0.17$ & $2.14\pm0.19$ & $1.22\pm0.005$ & $1.58\pm0.13$\\
Evolved Sys.\tablenotemark{e} & 10 & $-12.92\pm0.21$ & $4.51\pm0.15$ & $1.44\pm0.006$ & $1.02\pm0.02$
\enddata
\tablenotetext{a}{\footnotesize{Measured using bandpasses from \citet{fish98}.}}
\tablenotetext{b}{\footnotesize{Corrected for infill from emission, based on EW(\OII). We use the method described in \citet{koc11b}, although we made a slightly different choice for the relationship between EW(H$\alpha$) and EW(\OII) appropriate for Seyfert galaxies.}}
\tablenotetext{c}{\footnotesize{Measured using bandpasses from \citet{balogh99}.}}
\tablenotetext{d}{\footnotesize{The equivalent widths are measured by fitting a line to the background and a Gaussian to the absorption feature.}}
\tablenotetext{e}{\footnotesize{Includes Cl1324, RXJ1821, and RXJ1757.}}
\label{co-addtab}
\end{deluxetable*}

The H$\delta$ equivalent widths, combined with measurements of the
4000\AA\ break, suggest that starbursts have occurred more recently in
the average Cl1604 AGN host compared to those in the evolved
structures. Larger values of D$_{\rm n}$(4000) indicate a more passive
galaxy, with an older average stellar population, which could mean that more
time has passed since the cessation of star formation
\citep{balogh99,kauf03}. The EW(H$\delta$) measured from the Cl1604
composite is $2.14\pm0.19\ $\AA, roughly half of that from the evolved
structures composite. Similarly, the Cl1604 composite has a lower
value for D$_{\rm n}$(4000) than those in the evolved structures,
indicating that the average Cl1604 host is more actively star-forming
or has a younger stellar population on average. This result is
supported by the bluer colors of the Cl1604 AGN hosts compared to
those in the evolved structures. All of these results imply the average
Cl1604 host has had a starburst more recently than the average AGN
host in the evolved structures.

The AGN host composite spectra for Cl0023 has a particularly low
D$_{\rm n}$(4000) measurement (see Table \ref{co-addtab}). Since we found that
this composite spectra (which excluded one broadline source) was
consistent with type-2 AGNs, the AGNs themselves should not contribute
most of the blue continuum. This points to a stellar source,
particulary O and B stars. We would then expect significant star
formation in the Cl0023 hosts within the last 10-100 Myr, as indicated
by the other spectral features as well.

Related to the D$_{\rm n}$(4000) measurement, the \ion{Ca}{2}
H+H$\epsilon$ and the \ion{Ca}{2} K lines also provide information on
star formation. For F, G, and K stars, the ratio of these lines is
constant, while the \ion{Ca}{2} H+H$\epsilon$/\ion{Ca}{2} K ratio
increases for A and B stars as the overall \ion{Ca}{2} strength
decreases and the H$\epsilon$ strength increases \citep{rose85}.  The
\ion{Ca}{2} H+H$\epsilon$/\ion{Ca}{2} K ratio is $1.58\pm0.13$ for
Cl1604 and $1.02\pm0.02$ for the evolved structures (see Table
\ref{co-addtab}), consistent with the evolved structures having (on
average) older stellar populations. We do see a decrease in the
overall strengths of both \ion{Ca}{2} lines in the average spectrum of
the Cl0023 host galaxies relative to the other structures; however, we
actually measure a dramatic decrease in the ratio for Cl0023
($0.42\pm0.07$), the opposite of what is expected from a population of
A and B stars. The most likely explanation is significant H$\epsilon$
emission, which would be in concert with the other observed Balmer
emission. The H$\epsilon$ emission could be coming from some
combination of AGNs and \ion{H}{2} regions, which would support previous
conclusions about the level of activity in the Cl0023 hosts.

Altogether, the composite spectra of the AGN hosts in all three bins
of Figure \ref{CSplot} show that the average host galaxy has ongoing
star formation or has had star formation within the last $\sim1$
Gyr. However, the hosts in Cl1604 and the evolved structures each
have, on average, less ongoing star formation than Cl0023, as
evidenced by larger values of D$_{\rm n}$(4000) and the absence of the
Balmer emission that is observed in Cl0023. These differences suggest a
progression in the temporal proximity of the last starburst event,
with the hosts in Cl0023 having significant ongoing star formation
characteristic of a current starburst, to those in Cl1604 and the
evolved structures each having successively more time since the last
significant starburst event. We have confirmed that none of our
results based on the composite spectra are changed by removing the
four lowest significance ($<$3 $\sigma$) X-ray sources from our sample, with
most spectral measurements remaining the same within the errors.

With our analysis of the composite spectra of AGN host galaxies in the
different structures, we can compare the average properties of these
hosts with the average properties of all spectroscopically confirmed
galaxies within the same structures. In Section \ref{sec:globchar}, we
found that the galaxy populations in the evolved structures were
largely quiescent, with little or no contribution from starburst or
post-starburst galaxies. In contrast, the populations in the unevolved
structures were comprised of large fractions of star-forming galaxies,
with a more significant contribution from starburst or post-starburst
galaxies. When comparing these results to the average spectral
properties of the AGN hosts, we find that, in {\it all} cases, the
average AGN host galaxy has a {\it younger} stellar population than
the average galaxy in the parent structure, irregardless of the
evolved or unevolved classification. This result holds even when comparing to member galaxies outside the dense cluster cores ($> 0.5\ $Mpc), where the vast majority of X-ray AGNs reside. 

The most prominent difference comes from the evolved structures where their average AGN host
galaxy has significantly larger EW(H$\delta$) and smaller D$_{\rm n}$(4000) than the average structure member, indicative of a post-starburst galaxy with a substantial
star-formation event within the last $\sim 1\ $Gyr.  Such galaxies
make a small contribution to the overall population in the evolved
structures, which is largely quiescent. While we do observe clear
differences between the spectra of the AGN hosts, with those in Cl0023
having significant ongoing star formation to those in the evolved
structures having the most time since the last significant starburst
event, these differences are not nearly as pronounced as the
differences between the average galaxy in each structure. This
suggests that AGN activity has a common origin associated with current
or recent star-formation.

\begin{figure*}
\begin{center}
\epsscale{1.2}
\plotone{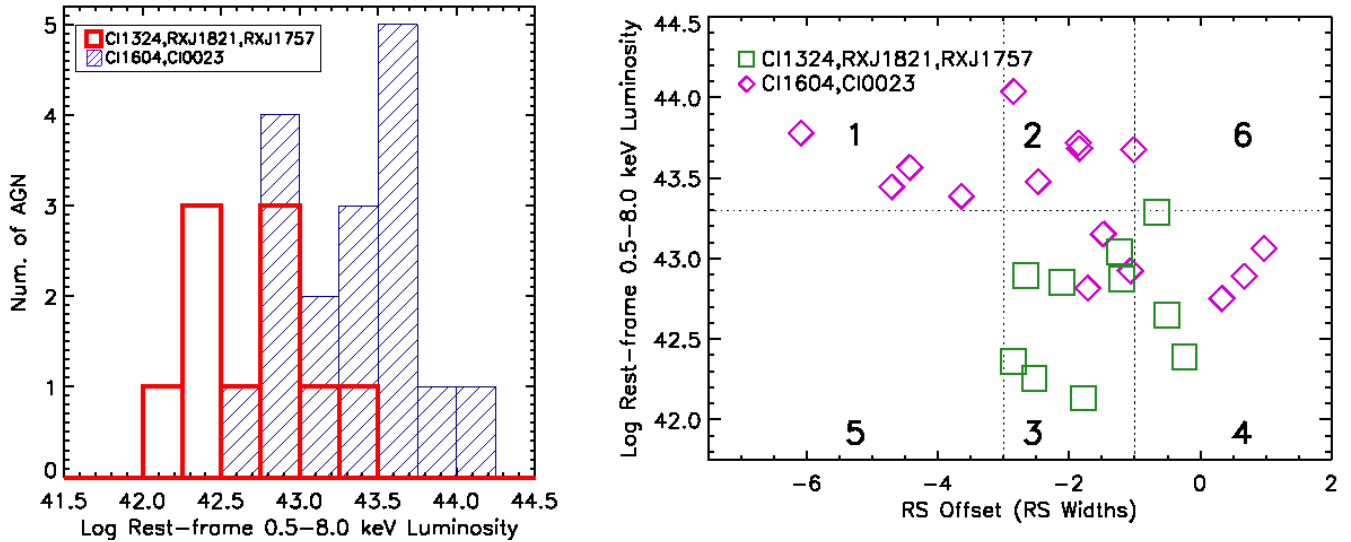}
\end{center}
\caption{ \footnotesize{(Left: rest-frame X-ray luminosities of
    AGNs in the full band (0.5--8 keV). Thick-outlined histograms
    indicate members of Cl1324, RXJ1821, and RXJ1757, while members of
    Cl1604 and Cl0023 are shown with hashed histograms. The former
    make up a bin of more evolved structures, and the distribution
    demonstrates that AGNs in these clusters tend to have lower
    luminosities than the less evolved structures. Right: AGN
  rest-frame full-band X-ray luminosities plotted against host galaxy
  color offsets from the red sequence, scaled by the red sequence
  width. AGN hosts from Cl1604 and Cl0023 are shown with diamonds,
  while those from the three evolved structures are shown with
  squares. The horizontal line delineates between regions of high and
  low X-ray luminosity (at $L_x = 10^{43.3}$ erg s$^{-1}$), and the
  vertical lines delineate the color regions of the blue cloud, green
  valley, and red sequence, respectively. We observe AGNs in Regions
  1--4 but none in Regions 5--6. See Section \ref{sec:xlums} for
  details.}}
\label{xlums}
\end{figure*}

\subsection{X-ray Luminosity}
\label{sec:xlum}

In this section, we explore the differences between the structures and
AGN host properties based on the X-ray luminosities of the confirmed
AGNs. We calculate rest-frame luminosities for X-ray point sources with
known redshifts. K-corrections were carried out using the power law
spectral models for sources, with a photon index of $\gamma=1.4$,
described in Section \ref{sec:red}. Luminosities are measured in the
X-ray soft, hard, and full bands. A histogram of full-band rest-frame
luminosity\footnote{Note that we are using all ten AGNs from Cl1604
  here, not just the eight in our ACS pointings.}, binned by evolved
and unevolved structures, is shown in Figure \ref{xlums}. The
luminosity distributions in the soft and hard bands are similar to the
one shown.

In the left panel of Figure \ref{xlums}, we can see that the AGNs in
the unevolved structures have higher X-ray luminosities than those in
the more evolved structures.  K-S tests show that the distributions of
the two bins are different at the 99$\%$ level in each of the three
bands. This statistically-significant result is independent of our
inclusion of the four low-significance ($<$ 3$\sigma$) X-ray
sources. 

While 10 out of 17 AGNs in the unevolved structures have full
band luminosities above $L_x = 10^{43.3}\ h^{-2}_{70}$ ergs s$^{-1}$,
there are {\it no} AGNs above this limit in Cl1324, RXJ1757, or RXJ1821.  
We follow a Bayesian approach, using Poisson statistics to calculate
the likelihoods, to estimate the probability of finding no high
$L_{x}$ AGNs in the evolved structures ($N_{\rm det}^{\rm E} = 0$)
given the detection rate in the unevolved structures. Specifically, we
calculate $P(N_{\rm det}^{\rm E} = 0~|~N_{\rm det}^{\rm UE},N_{\rm
  targ}^{\rm UE}, N_{\rm targ}^{\rm E})$.  Here, $N_{\rm det}^{\rm
  UE}$ is the number of high $L_x$ AGNs that are
spectroscopically confirmed members in the unevolved structures, and
$N_{\rm targ}^{\rm UE}$ and $N_{\rm targ}^{\rm E}$ are the total
number of high $L_x$ AGNs that were targeted for spectroscopy in the
unevolved and evolved structures, respectively. Based on this
calculation, the probability of finding no high $L_{x}$ AGNs in the
evolved structures is only 0.25\%.
This result is likely related to the smaller fractions of blue cloud
galaxies (and overall more quiescent populations) in Cl1324, RXJ1757,
or RXJ1821 and, thus, the unavailability of large gas reservoirs.

At the faint end, there are four sources in the evolved structures
below $10^{42.5}\ h^{-2}_{70}$ ergs s$^{-1}$, all of which are members
of Cl1324. However, the X-ray source counts, in all five fields, are
significantly incomplete at these luminosity levels with only $\sim
5\%$ of optically matched X-ray sources below this limit, most of
which are $<$ 3$\sigma$ detections. Therefore, we cannot say anything
definitive about the lack of faint sources in the unevolved
structures; however, if we remove the four low-luminosity sources, the
difference between the luminosity distributions in the evolved and
unevolved samples is still significant at a 95\% level according to
the K-S test. The reasonably high significance is clearly due to the
lack of high-luminosity sources in the evolved structures.

\subsubsection{Relation to Host Galaxy Color}
\label{sec:xlumc}

In the right panel of Figure \ref{xlums}, we plot the full-band
rest-frame luminosity versus the red-sequence offset, scaled by RS
width (see Section \ref{sec:hgalcolan}). From this figure, we can see
that all AGNs with host galaxies on the red-sequence have lower
X-ray luminosities, all below $L_x < 10^{43.3}$ ergs s$^{-1}$ (Region
4). This result is not unexpected since AGN activity should diminish
as the host galaxy moves onto the red sequence.  We can also see that
60\% of the confirmed AGNs lie in the green valley ($-3W_{RS} < \Delta
C < -W_{RS}$; Regions 2 and 3). Most interestingly, in the green
valley there are almost two orders of magnitude variation in X-ray
luminosity, with the unevolved structures having all of the highest
$L_X$ sources. We try to decipher the origin of these variations in
Section \ref{sec:xlums}.

We also do not detect any low $L_x$ blue AGNs (Region 5) or any high
$L_x$ red AGNs (Region 6), not necessarily unexpected given their
expected gas contents. However, it is difficult to say for certain if
these null results are significant. We do sample optically matched
X-ray sources in these regions. Specifically, in the five fields,
there are a total of 85 sources (at $>$ 3$\sigma$) in Region 5 ($-6.1
W_{RS} < \Delta C < -3 W_{RS}$ and $10^{42.5} < L_x < 10^{43.3}$ ergs
s$^{-1}$), of which we have targeted 24 (28\%). Here, we estimate the
X-ray luminosities by assuming that all sources in a particular field
are at the mean redshift of the structure, and we choose the lower
limit of $L_x = 10^{42.5}$ ergs s$^{-1}$ so as not to be adversely
affected by incompleteness (see above). Based on \citet{bar05}, the
X-ray luminosity function of field AGNs at similar redshifts in
``optically normal'' galaxies (comparable to the vast majority of our
AGN hosts) shows increasing number densities down to $L_x \approx
10^{42}$ ergs s$^{-1}$. As a result, we would naively expect to detect
a larger number of fainter X-ray sources in the blue cloud. Using the
Bayesian approach described in Section \ref{sec:xlum}, the probability
of finding no AGN in Region 5 is 3\%, given our success rate of
confirming cluster members for the high $L_x$ blue galaxies
(Region 1).
This formal probability may, in fact, be an {\it upper limit} as we
would expect a higher success rate given the larger number densities
at fainter X-ray luminosities. The fact that we observe no low $L_x$
blue galaxies in our sample may suggest that, in high-density
environments compared to the field, either (1) the time to reach the
highest X-ray luminosities is shorter after AGN turn-on or (2) the
host galaxies are transformed more quickly, moving to redder colors by
the time their X-ray luminosities drop to lower levels.

Similarly, there are 25 sources in Region 6 ($-1 W_{RS} > \Delta C > 2
W_{RS}$ and $L_x > 10^{43.3}$ ergs s$^{-1}$), of which we have targeted 7 (28\%). If
we assume our success rate for confirming cluster members as measured
from all high $L_x$ AGNs (Regions 1 and 2), the probability of finding
no AGN in Region 6 is 18\%.
Although not a significant result, the conclusions for this region
are, of course, more obvious as we do not expect any true red-sequence
(i.e., non-dust reddened) galaxies to have enough cold gas to fuel a
luminous AGN.

\subsubsection{Relation to Average Spectral Properties}
\label{sec:xlums}

\begin{figure}[!h]
\epsscale{1.1}
\plotone{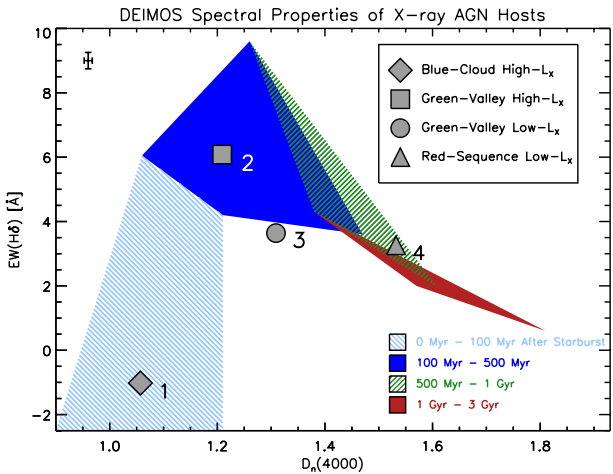}
\caption{ \footnotesize{EW(H$\delta$) vs. D$_{\rm n}$(4000) from
    the composites made by AGN host galaxies in Regions 1-4 of Figure
    \ref{xlums}. For comparison, the ranges of EW(H$\delta$)--D$_{\rm
      n}$(4000) phase space spanned by four different Bruzual \&
    Charlot \citep{bc07} models for various times after the starburst
    are shown. Because Bruzual \& Charlot models only incorporate
    stellar light, emission infill corrections were made for all
    EW(H$\delta$) measurements (see Section \ref{sec:specprop} for
    details). The average host galaxy in each region either is in the
    process of having a starburst or has had one within the last
    $\lesssim 1$ Gyr. The hosts of AGNs with higher X-ray luminosities,
    even when comparing green valley galaxies, have younger stellar
    populations, with shorter times since the last burst. A typical
    error bar is shown in the upper left corner.}}
\label{specreg}
\end{figure}

To explore the origin of the variations observed in Figure
\ref{xlums}, we examine the average spectral properties of the host
galaxies within the four distinct regions. Specifically, the sample is
split by X-ray luminosity at $L_x = 10^{43.3}$ and red-sequence
offset, scaled by RS width, to delineate regions containing high $L_x$
blue cloud (Region 1), high $L_x$ (Region 2) and low $L_x$ (Region 3)
green valley, and low $L_x$ red-sequence (Region 4) host galaxies.  In
Figure \ref{specreg}, we plot the measured EW(H$\delta$) versus
D$_{\rm n}$(4000) from the spectral composites in the four regions. For comparison, we also plot post-starburst temporal regimes derived from four \citet{bc07} models, described in Section \ref{sec:specprop}.

The small D$_n$(4000) and H$\delta$ in emission indicate that the high
$L_x$ blue hosts are coeval with the starburst or ongoing star
formation. As we examine galaxies in Regions 2 to 4 going from
the high to low $L_x$ hosts in the green valley to the low $L_x$ hosts
in the red sequence, the time since the burst gets progressively
larger. While there is some degeneracy between time since burst and
burst strength, it is clear that the low $L_x$ green valley hosts are
either (1) further along since the burst than their high $L_x$
counterparts or (2) had a weaker initial burst which could explain
their lower X-ray luminosities as less gas would likely be funneled to
the center. Our results are robust to removing the four
lowest significance ($<$ 3$\sigma$) X-ray sources, as well as the four
X-ray sources below $L_x = 10^{42.5}$ ergs s$^{-1}$, where in both
cases we are highly incomplete.

The most striking results from this spectral analysis are, first, that
the average AGN host in {\it every} region is either in the process of
having a starburst or has had one within last $\lesssim 1$ Gyr. This
global finding clearly demonstrates the close connection between
starburst and AGN activity as normal star formation does not typically
produce the H$\delta$ values seen in these host galaxies. Second, we
do not detect high X-ray luminosity, young (as indicated by time since
starburst) galaxies in the evolved structures. This result implies
that the entire galaxy population in these structures (certainly in
the isolated, X-ray selected ones) are more advanced, suggesting that
the peak of gas consumption, seen in both star formation and AGN
activity, occurred at an earlier time.

\section{Discussion}

We studied AGN activity in five high-redshift clusters and
superclusters in the redshift range $z\approx 0.7-0.9$. Before
identifying individual AGNs, we analyzed the structures using the
statistical measure of cumulative source counts. We found every
structure to have X-ray excesses of 0.5-1.5 $\sigma$ with respect to the
CDFN and CDFS control fields. The method is highly dependent on the
field used as a sky estimate, which makes comparing results between
studies difficult. \citet{cap05} measured cumulative source counts in
a number of structures in the range $0.2 \lesssim z \lesssim 1.4$ and
found a dependence on redshift. While our data are consistent within
the errors with the results of C05, our redshift range is not large
enough to evaluate the relation. We note that it is difficult to use
these overdensities to interpret the actual AGN activity in an
individual structure, even with extensive spectroscopy, as we have
attempted to do here with our large spectroscopic sample. We recommend
caution in using this technique, as its precision can easily be
overestimated.

We employed a maximum likelihood technique to match X-ray sources to
optical counterparts. With an extensive DEIMOS optical spectroscopic
campaign with $\sim 6000$ targets, accurate redshifts have been
obtained for 126 of these X-ray sources, allowing us to identify a
total of 27 AGNs within all of the structures. These results show that
significant spectroscopy is needed to confirm even small numbers of
AGN members. We find that the spatial distribution of the AGN is
largely consistent with previous work at lower redshift. Across all
five structures, we find that AGN host galaxies tend to be located
away from dense cores (within 0.5 Mpc of a cluster or group center),
with many instead located on the outskirts of clusters or poorer
groups\footnote{Only including groups with $\sigma > 300$
  \kms.}. Previous studies have found similar results up to $z\sim1$
\citep{cold02,kauf04,gil07,koc09b}. These intermediate environments
are thought to be conducive to galaxy-galaxy interactions, because of
the relatively high densities compared to the field and low velocity
dispersions compared to cluster cores \citep{cav92}. Our results would then lend
support to these interactions as the trigger of X-ray AGN activity in
the environments of LSSs. However, these may also be regions where gas availability is higher in the member galaxies, which could lead to increased AGN activity. 

With optical counterparts to X-ray sources identified, we were able to
analyze the color properties of AGN hosts. Our analysis showed that
AGN host galaxies are overrepresented in the green valley. In Cl1604,
only $\sim17\%$ of all the supercluster members with ACS data were
within $2W_{RS}$ of the lower boundary of the red sequence, where
$W_{RS}$ is the width of the red sequence, defined in Section
\ref{sec:globchar}. However, five out of the eight AGN hosts in our ACS
pointings were in this range. In fact, $36\%$ of host galaxies in all
five structures lie within $W_{RS}$ of the lower boundary of the red
sequence and $60\%$ are within $2W_{RS}$. Our results are supported by
other studies which have found an overabundance of AGN activity in the
green valley \citep{nandra07,georg08,silver08,hickox09}. Since the
green valley is thought to be a transitional region for galaxies
\citep{faber07}, these results suggest X-ray AGNs in LSS are a
transitional population between blue star-forming and red quiescent
galaxies. However, we note that, while our sample is magnitude-limited, several studies using mass-selected samples have found less overrepresentation of AGN hosts in the green valley \citep{silver09b,xue10}. Additionally, \citet{card10} have found that $\sim75\%\ $of AGN hosts in the green valley are dust-reddened blue cloud galaxies, although these results conflict with the recent studies of \citet{ros11}. Our results from preliminary SED fittings to AGN hosts suggest that the effect is not as drastic in our sample.

The five structures studied in this paper occupy a range of
evolutionary states. Based on the \OII\ and H$\delta$ features of the
composite spectra for each structure, we grouped our sample into the
least evolved structures (``unevolved''), Cl1604 and Cl0023, and the most evolved
structures (``evolved''), Cl1324, RXJ1821, and RXJ1757. This distinction is based on
the average stellar populations and the presence, or lack thereof, of
current star formation. With these two categories, we sought to
explore differences in AGN activity between structures with different
galaxy populations. We did not find any significant differences
between the five structures when examining the cumulative source
counts or the spatial distributions of the AGN hosts. However, the AGN
host galaxies in the unevolved structures were skewed more towards
bluer colors, although this was not at a statistically significant
level for this sample size.

We did, however, find significant differences between the subsets when examining
the X-ray luminosities of the AGNs and the optical spectra of their hosts. We found
that AGNs in the unevolved structures tend to have higher full band
(0.5--8 keV) X-ray luminosities relative to those in the evolved structures at a $99\%$ level, with all of the
most luminous AGNs ($L_x > 10^{43.3}$ erg s$^{-1}$) found in the unevolved
structures. While {\it all} AGN host galaxies either have on-going
star formation or have had a starburst within the last $\lesssim 1$
Gyr, the host galaxies in the unevolved structures are distinctly
younger than those in the evolved structures, with shorter times since
the last starburst as indicated by smaller average EW$(H\delta)$ and
D$_{\rm n}$(4000) in their composite spectra. The average Cl0023 host
has current star formation, and the average Cl1604 host has had a
burst within the last $\sim 100$ Myr. We do not detect any of these
young, high X-ray luminosity AGNs in the evolved structures, implying
that the peak of both star formation and AGN activity occurred at an
earlier time. We note that, regardless of whether they are members of
the evolved or unevolved structures, {\it all} AGN host galaxies are
younger than the average galaxy in their parent population.

We also find a large (two orders of magnitude) variation in X-ray
luminosity for AGNs within the green valley, while AGNs in the red
sequence have consistently lower luminosities ($L_x < 10^{43.3}$ erg
s$^{-1}$). As we move from the high to low $L_x$ green-valley hosts to
low $L_x$ red-sequence hosts, the time since starburst gets
progressively longer. Although there is some degeneracy between burst
strength and time since burst, the low $L_x$ green valley hosts are
either further along since the burst than their high $L_x$
counterparts or have had a weaker initial burst, which may explain their
lower X-ray luminosities.

The higher AGN X-ray luminosities in the unevolved structures are most
likely related to their bluer colors and, hence, larger reservoirs of
gas that could lead to higher levels of black hole accretion and the
higher X-ray luminosities. In addition, both the X-ray and spectral
results can be explained if these galaxies had more recently undergone
merger-induced, or other, starburst events. Specifically, simulations
and observations have found that AGN activity peaks soon after maximum
star formation in a starburst event ($\sim$0.1-0.25 Gyr)
\citep{davies07,scha07,scha09,wild10,hopkins11}. As the AGN X-ray
luminosity declines after reaching its peak, the star formation rate
should be declining as well. This could explain why the most X-ray
luminous AGNs are in the least evolved structures. AGN host galaxies in
Cl0023 were also found to have significant ongoing star formation,
which could mean these galaxies have had the most recent merger events,
where star formation is still near its peak.

The X-ray luminosity differences between the AGNs in the evolved and
unevolved structures could also be viewed as a transition from quasar
mode emission towards radio mode emission, as defined in
\citet{croton06} and \citet{mc07}, which is also related to the star
formation rate. Since AGN feedback deters gas from falling into the
core and depletes the cold gas in a galaxy, the fuel for both star
formation and further AGN emission is decreased. Ultimately, the
dominant fuel for black hole accretion transitions from cold gas
funneled to the galactic core from the starburst event to hot halo
gas, which leads to a much more quiescent state \citep[][; for a similar model, see also, e.g., \citet{bower06}, \citet{fan11}]{croton06,mc07}. In our sample, the Cl0023 AGNs are emitting more
similarly to the quasar mode, accompanied by substantial star
formation. The AGN host galaxies in Cl1604 and the evolved structures
have lower star formation rates than those in Cl0023, while the
evolved structures have AGNs with lower X-ray luminosities, which
suggests that they are sequentially further along the track leading
toward domination of radio mode AGN emission. If the AGNs in the
evolved structures were found to be radio emitters, it would support
these conclusions. Existing VLA B-array observations (at 1.4 Ghz) of
these five structures are currently being analyzed to explore this
connection (C. D. Fassnacht et al.\ 2012, in preparation).

Altogether, many of our results could support several potential AGN
triggering scenarios. Two possibilities are that: (1) AGNs in these structures represent a
transitional population where hosts are evolving from the blue cloud
onto the red sequence or (2) AGNs represent a population evolving
mainly in mass space, where red sequence hosts have undergone episodic
nuclear star formation induced by minor mergers. Additionally, many observable effects of the latter would appear similar to models in which recycled stellar material fuels central starbursts and nuclear activity in elliptical galaxies, which tend to be located on the red sequence. In the first case,
AGNs are triggered by mergers or strong tidal interactions which lead
to a starburst. The feedback from the AGN quenches star formation,
leading to a rapid evolution across the green valley and onto the red
sequence \citep{hopkins05,spring05,hopkins07,somer08}. In the second
case, red sequence galaxies undergo minor mergers, which funnel gas
into the galactic core, creating a burst of nuclear star formation and
fueling the AGN \citep{men01}. Alternatively, recycled stellar material could create a nuclear starburst and a central instability, leading to black hole accretion \citep{CO07}. In either case, the galaxies could evolve into the
green valley before AGN feedback brought them onto the red sequence
again. Once on the red sequence, there would not be a large net change in color, in contrast to the dramatic color evoltion of the first possibility. In the case of minor mergers, this process would mainly entail an evolution in mass. 

In support of AGNs as a transitional population, many of the host galaxies across all five
structures are located close to the red sequence, where the green
valley should lie, although it is difficult to determine the location
of the green valley in the four structures without precise ACS
data. In support of all scenarios involving significant starbursts, the average AGN host in both the
evolved structures and the Cl1604 supercluster has substantial
H$\delta$ absorption, which is a sign of recently quenched star
formation. This is expected for green valley galaxies evolving onto
the red sequence. Previous morphological analysis of part of our
sample by \citet{koc09b} found that two-thirds of Cl1604 host galaxies
studied showed signs of recent or pending mergers or tidal
interactions, which is expected in the context of both major and minor merger theories. In
addition, we find that at least nine of the 27 AGN host galaxies are
part of a kinematic close pair. In seven of these
cases, the companion galaxy has a similar stellar mass or $z'$ magnitude to the AGN host. While being far from conclusive due to sampling and selection effects, these results could point to a major merger scenario.

\citet{koc09b} examined eight of the AGNs in Cl1604 and found that half
of the hosts had blue cores in an otherwise red galaxy. Other studies have also found blue cores or blue early type galaxies
\citep{lee06,martel07}. These blue
cores are predicted by minor merger simulations \citep{MiHern96}, as well as the recycled gas models of \citet{CO07}. In addition simulations show that central black hole
accretion is most highly correlated with star formation in the
nucleus, as opposed to the entire galaxy
\citep{HQ10,diamond11}. During a major merger,
star formation peaks at later times closer to the galactic core
\citep{hopkins11}. This could potentially create bluer cores as
well. From our data, it seems our results are ambiguous with regards
to the various scenarios. However, with evidence supporting more than one possibility,
our results could indicate a combination of different triggering mechanisms. 

These different AGN triggering mechanisms could also potentially
explain some of the differences that we see between the evolved and
unevolved structures. The second AGN mode, involving episodic nuclear
activity in red sequence galaxies fueled by minor mergers, or recycled gas, is expected
to involve accretion rates at a much lower Eddington ratio than in
major merger driven AGN activity \citep{marc04,merl08,has08}. While
AGN luminosity depends on both the Eddington ratio and the black hole
mass \citep{kauff00}, lower Eddington ratios will, on average,
correspond to lower luminosities. A difference in Eddington ratios
could potentially explain the lower X-ray luminosities that we observe in
the evolved structures, if our sample of host galaxies consists
of a mix of the two AGN modes. In the unevolved structures, we
would expect to observe more major mergers between blue cloud galaxies
due to the larger fraction of blue galaxies, fueling the brightest AGNs
in our sample. Since all of our structures have a substantial number
of red-sequence galaxies we would expect all structures to have
galaxies undergoing the second mode of AGN activity, whether induced by minor mergers or, perhaps, recycled gas in ellipticals. We would then
expect the X-ray luminosity distributions of the evolved and unevolved
structures to look similar except for a tail of higher luminosity
objects in the unevolved structures, which is roughly what we observe.
While a combination of the two AGN modes considered could explain our
results, they could also be explained if major mergers were the
primary driver, and the quasar mode is less dominant for the AGNs in
the evolved structures, because of the larger red fraction
therein. With our current data, we still lack the ability to
distinguish between the different triggering mechanisms. Breaking the degeneracy will require high-resolution imaging to
examine the morphologies of the AGN hosts and the colors of their
cores in the structures other than Cl1604, as well as reliable stellar
masses and HST data, for measuring bulge-to-disk-ratios, with which we
could reliably calculate Eddington ratios.

\section{Conclusions}

In summary, we find that most X-ray AGN hosts, across all five
structures, avoid the dense cluster cores, in agreement with a number
of previous studies at a range of redshifts. We interpret this to mean
that X-ray AGN activity is preferentially triggered in intermediate-density
environments, such as the outskirts of clusters. We also find many AGN
host galaxies in or near the green valley, with 36$\%$ within one
red-sequence width of the lower boundary of the red sequence and
60$\%$ within two red-sequence widths. With numerous other studies
finding a similar connection, this implies that there is an
association between this transitional region and AGN activity.

We divided our sample of five structures into two groups: the more and
less evolved structures, which we separated using composite spectra
made of all of their spectroscopically-confirmed member galaxies.  We
define the more evolved structures as those with member galaxies that exhibit, on average, less
\OII\ emission and less H$\delta\ $absorption, where the \OII\ and H$\delta$ lines are
taken as indicators of ongoing and recent star formation,
respectively. The more evolved structures also have galaxy populations with a higher red fraction than the less evolved structures. Our spectral results indicate that the AGN hosts in the less
evolved structures have more ongoing star formation, while those in
the more evolved structures have stronger average H$\delta$
features. Stronger H$\delta$ lines are indicative of star formation
within the past $\sim$1 Gyr, and our results indicate starbursts
occurred more recently in the AGN host galaxies in the less evolved
structures. However, all of the AGN hosts, regardless of whether they
are members of the more evolved or less evolved structures, are younger than
the average galaxy in their parent population.

We also found that AGNs in the less evolved structures had more
luminous X-ray emission. This may be expected, since these
structures contain the highest fraction of blue galaxies which are likely to have larger reservoirs
of cool gas to fuel nuclear activity. If AGNs were triggered more
recently in the less evolved structures, as the spectral data
suggests, the difference in luminosity could also be related to a
transition from ``quasar mode'' emission in newly triggered AGNs to
``radio mode'' emission at later times as supplies of inflowing cool
gas are shut off.

We consider several scenarios for AGN triggering that are in agreement
with our results. AGNs are triggered by major mergers or tidal interactions between blue cloud galaxies and/or AGNs are triggered episodically in red-sequence galaxies, fueled by recycled stellar material or induced by minor mergers. Each of these scenarios could explain the
association of AGN hosts with the green valley. Also, the
H$\delta\ $absorption we observe could be indicative of the quenching
of star formation that drives galaxies across the green valley. A
previous study of the Cl1604 AGNs by \citet{koc09b} found that a
majority of hosts had recent or pending mergers. Half of the hosts had blue cores in otherwise red galaxies,
which could support the second two scenarios, although other explanations
are possible. Some of our results, such as the difference in X-ray
luminosities, could be explained if the AGNs in our sample were
triggered by a mix of these modes, with a larger fraction of AGNs in
the less evolved structures triggered by major mergers. However, we
cannot distinguish between the triggering scenarios with our data,
so the cause of the AGN activity is still ambiguous. Future work
investigating AGN host morphologies, examining for blue cores, and
calculating Eddington ratios could potentially break the degeneracy.
\newline
\newline
The authors thank Phil Marshall and Robert Lupton for useful conversations. This work is supported by the {\it Chandra} General Observing Program under
award numbers GO6-7114X, GO7-8126X, GO8-9123A, and GO9-0139A.  In
addition, we acknowledge support by the National Science Foundation
under Grant No. AST-0907858. The spectrographic data presented herein
were obtained at the W.M. Keck Observatory, which is operated as a
scientific partnership among the California Institute of Technology,
the University of California, and the National Aeronautics and Space
Administration. The Observatory was made possible by the generous
financial support of the W.M. Keck Foundation. As always, we thank the
indigenous Hawaiian community for allowing us to be guests on their
sacred mountain. We are most fortunate to be able to conduct
observations from this site.
\newline
\newline


\begin{thebibliography}{}
\bibitem[Baldry et al.(2004)]{baldry04} Baldry, I.~K., Glazebrook, K., Brinkmann, J., et al.\ 2004, \apj, 600, 681
\bibitem[Balogh et al.(1999)]{balogh99} Balogh, M.~L., Morris, S.~L., Yee, H.~K.~C., Carlberg, R.~G., \& Ellingson, E.\ 1999, \apj, 527, 54 
\bibitem[Barger et al.(2005)]{bar05} Barger, A.~J., Cowie, L.~L., Mushotzky, R.~F., et al.\ 2005, \aj, 129, 578 
\bibitem[Bluck et al.(2011)]{bluck11} Bluck, A.~F.~L., Conselice, C.~J., Almaini, O., et al.\ 2011, \mnras, 410, 1174
\bibitem[Bournaud et al.(2011)]{bour11} Bournaud, F., Chapon, D., Teyssier, R., et al.\ 2011, \apj, 730, 4
\bibitem[Bower et al.(2006)]{bower06} Bower, R.~G., Benson, A.~J., Malbon, R., et al.\ 2006, \mnras, 370, 645
\bibitem[Boyle \& Terlevich(1998)]{BoyTer98} Boyle, B.~J., \& Terlevich, R.~J.\ 1998, \mnras, 293, L49
\bibitem[Brandt et al.(2001)]{brandt01} Brandt, W.~N., Alexander, D.~M., Hornschemeier, A.~E., et al.\ 2001, \aj, 122, 2810
\bibitem[Bruzual(2007)]{bc07} Bruzual, G.\ 2007, From Stars to Galaxies: Building the Pieces to Build Up the Universe, 374, 303
\bibitem[Butcher \& Oemler(1984)]{BO84} Butcher, H., \& Oemler, A., Jr.\ 1984, \apj, 285, 426
\bibitem[Calzetti et al.(2000)]{cal00} Calzetti, D., Armus, L., Bohlin, R.~C., et al.\ 2000, \apj, 533, 682 
\bibitem[Cappelluti et al.(2005)]{cap05} Cappelluti, N., Cappi, M., Dadina, M., et al.\ 2005, \aap, 430, 39 
\bibitem[Cardamone et al.(2010)]{card10} Cardamone, C.~N., Urry, C.~M., Schawinski, K., et al.\ 2010, \apjl, 721, L38
\bibitem[Cavaliere et al.(1992)]{cav92} Cavaliere, A., Colafrancesco, S., \& Menci, N.\ 1992, \apj, 392, 41
\bibitem[Ciotti \& Ostriker(2007)]{CO07} Ciotti, L., \& Ostriker, J.~P.\ 2007, \apj, 665, 1038
\bibitem[Coldwell et al.(2002)]{cold02} Coldwell, G.~V., Mart{\'{\i}}nez, H.~J., \& Lambas, D.~G.\ 2002, \mnras, 336, 207
\bibitem[Colless et al.(2001)]{coll01} Colless, M., et al.\ 2001, \mnras, 328, 1039 
\bibitem[Crawford et al.(1970)]{crawford70} Crawford, D.~F., Jauncey, D.~L., \& Murdoch, H.~S.\ 1970, \apj, 162, 405
\bibitem[Croton et al.(2006)]{croton06} Croton, D.~J., Springel, V., White, S.~D.~M., et al.\ 2006, \mnras, 365, 11 
\bibitem[Davies et al.(2007)]{davies07} Davies, R.~I., M{\"u}ller S{\'a}nchez, F., Genzel, R., et al.\ 2007, \apj, 671, 1388 
\bibitem[Diamond-Stanic \& Rieke(2011)]{diamond11} Diamond-Stanic, A.~M., \& Rieke, G.~H.\ 2011, \apj, in press
\bibitem[Dickey \& Lockman(1990)]{DicLock90} Dickey, J.~M., \& Lockman, F.~J.\ 1990, \araa, 28, 215
\bibitem[Dressler et al.(2004)]{dress04} Dressler, A., Oemler, A., Jr., Poggianti, B.~M., et al.\ 2004, \apj, 617, 867 
\bibitem[Dressler \& Shectman(1988)]{DS88} Dressler, A., \& Shectman, S.~A.\ 1988,\aj, 95, 284
\bibitem[Dressler et al.(1999)]{dress99} Dressler, A., Smail, I., Poggianti, B.~M., et al.\ 1999, \apjs, 122, 51
\bibitem[Eastman et al.(2007)]{east07} Eastman, J., Martini, P., Sivakoff, G., et al.\ 2007, \apjl, 664, L9 
\bibitem[Faber et al.(2003)]{faber03} Faber, S.~M., Phillips, A.~C., Kibrick, R.~I., et al.\ 2003, \procspie, 4841, 1657
\bibitem[Faber et al.(2007)]{faber07} Faber, S.~M., Willmer, C.~N.~A., Wolf, C., et al.\ 2007, \apj, 665, 265
\bibitem[Fanidakis et al.(2011)]{fan11} Fanidakis, N., Baugh, C.~M., Benson, A.~J., et al.\ 2011, \mnras, 410, 53
\bibitem[Ferrarese \& Merritt(2000)]{FM00} Ferrarese, L., \& Merritt, D.\ 2000, \apjl, 539, L9 
\bibitem[Fisher et al.(1998)]{fish98} Fisher, D., Fabricant, D., Franx, M., \& van Dokkum, P.\ 1998, \apj, 498, 195
\bibitem[Ford et al.(2003)]{ford03} Ford, H.~C., Clampin, M., Hartig, G.~F., et al.\ 2003, \procspie, 4854, 81
\bibitem[Fruscione et al.(2006)]{frusc06} Fruscione, A., McDowell, J.~C., Allen, G.~E., et al.\ 2006, \procspie, 6270, 60
\bibitem[Gal et al.(2008)]{gal08} Gal, R.~R., Lemaux, B.~C., Lubin, L.~M., Kocevski, D., \& Squires, G.~K.\ 2008, \apj, 684, 933
\bibitem[Gal \& Lubin(2004)]{GalLub04} Gal, R.~R., \& Lubin, L.~M.\ 2004, \apjl, 607, L1 
\bibitem[Gal et al.(2005)]{gal05} Gal, R.~R., Lubin, L.~M., \& Squires, G.~K.\ 2005, \aj, 129, 1827 
\bibitem[Gebhardt et al.(2000)]{geb00} Gebhardt, K., Bender, R., Bower, G., et al.\ 2000, \apjl, 539, L13
\bibitem[Gehrels(1986)]{geh86} Gehrels, N.\ 1986, \apj, 303, 336 
\bibitem[Georgakakis et al.(2008)]{georg08} Georgakakis, A., Nandra, K., Yan, R., et al.\ 2008, \mnras, 385, 2049
\bibitem[Georgantopoulos et al.(2011)]{georg11} Georgantopoulos, I., Rovilos, E., \& Comastri, A.\ 2011, \aap, 526, A46
\bibitem[Gilmour et al.(2007)]{gil07} Gilmour, R., Gray, M.~E., Almaini, O., et al.\ 2007, \mnras, 380, 1467
\bibitem[Gioia et al.(2003)]{gioia03} Gioia, I.~M., Henry, J.~P., Mullis, C.~R., et al.\ 2003, \apjs, 149, 29 
\bibitem[Gioia et al.(1990)]{gioia90} Gioia, I.~M., Maccacaro, T., Schild, R.~E., et al.\ 1990, \apjs, 72, 567
\bibitem[Gioia et al.(2004)]{gioia04} Gioia, I.~M., Wolter, A., Mullis, C.~R., et al.\ 2004, \aap, 428, 867 
\bibitem[Gladders et al.(1998)]{gladders98} Gladders, M.~D., Lopez-Cruz, O., Yee, H.~K.~C., \& Kodama, T.\ 1998, \apj, 501, 571
\bibitem[Gunn et al.(1986)]{gunn86} Gunn, J.~E., Hoessel, J.~G., \& Oke, J.~B.\ 1986, \apj, 306, 30 
\bibitem[H{\"a}ring \& Rix(2004)]{HR04} H{\"a}ring, N., \& Rix, H.-W.\ 2004, \apjl, 604, L89
\bibitem[Hasinger(2008)]{has08} Hasinger, G.\ 2008, \aap, 490, 905
\bibitem[Heckman et al.(2004)]{heck04} Heckman, T.~M., Kauffmann, G., Brinchmann, J., et al.\ 2004, \apj, 613, 109
\bibitem[Henry et al.(2006)]{hen06} Henry, J.~P., Mullis, C.~R., Voges, W., et al.\ 2006, \apjs, 162, 304 
\bibitem[Hickox et al.(2009)]{hickox09} Hickox, R.~C., Jones, C., Forman, W.~R., et al.\ 2009, \apj, 696, 89
\bibitem[Hopkins(2011)]{hopkins11} Hopkins, P.~F.\ 2011, \mnras, in press
\bibitem[Hopkins et al.(2007)]{hopkins07} Hopkins, P.~F., Bundy, K., Hernquist, L., \& Ellis, R.~S.\ 2007, \apj, 659, 976
\bibitem[Hopkins \& Hernquist(2006)]{HH06} Hopkins, P.~F., \& Hernquist, L.\ 2006, \apjs, 166, 1
\bibitem[Hopkins et al.(2005)]{hopkins05} Hopkins, P.~F., Hernquist, L., Martini, P., et al.\ 2005, \apjl, 625, L71
\bibitem[Hopkins \& Quataert(2010)]{HQ10} Hopkins, P.~F., \& Quataert, E.\ 2010, \mnras, 407, 1529 
\bibitem[Johnson et al.(2003)]{johnson03} Johnson, O., Best, P.~N., \& Almaini, O.\ 2003, \mnras, 343, 924 
\bibitem[Kartaltepe et al.(2007)]{kart07} Kartaltepe, J.~S., Sanders, D.~B., Scoville, N.~Z., et al.\ 2007, \apjs, 172, 320
\bibitem[Kauffmann \& Haehnelt(2000)]{kauff00} Kauffmann, G., \& Haehnelt, M.\ 2000, \mnras, 311, 576
\bibitem[Kauffmann et al.(2003)]{kauf03} Kauffmann, G., Heckman, T.~M., Tremonti, C., et al.\ 2003, \mnras, 346, 1055
\bibitem[Kauffmann et al.(2004)]{kauf04} Kauffmann, G., White, S.~D.~M., Heckman, T.~M., et al.\ 2004, \mnras, 353, 713
\bibitem[Kim et al.(2007)]{kim07} Kim, M., Kim, D.-W., Wilkes, B.~J., et al.\ 2007, \apjs, 169, 401 
\bibitem[Kocevski et al.(2012)]{koc11a} Kocevski, D.~D., Faber, S.~M., Mozena, M., et al.\ 2012, \apj, 744, 148 
\bibitem[Kocevski et al.(2011)]{koc11b} Kocevski, D.~D., Lemaux, B.~C., Lubin, L.~M., et al.\ 2011, \apj, 736, 38
\bibitem[Kocevski et al.(2009a)]{koc09a} Kocevski, D.~D., Lubin, L.~M., Gal, R., et al.\ 2009a, \apj, 690, 295 
\bibitem[Kocevski et al.(2009b)]{koc09b} Kocevski, D.~D., Lubin, L.~M., Lemaux, B.~C., et al.\ 2009b, \apj, 700, 901
\bibitem[Kocevski et al.(2009c)]{koc09c} Kocevski, D.~D., Lubin, L.~M., Lemaux, B.~C., et al.\ 2009c, \apjl, 703, L33 
\bibitem[Kushino et al.(2002)]{kush02} Kushino, A., Ishisaki, Y., Morita, U., et al.\ 2002, \pasj, 54, 327
\bibitem[Laird et al.(2010)]{laird10} Laird, E.~S., Nandra, K., Pope, A., \& Scott, D.\ 2010, \mnras, 401, 2763
\bibitem[Lee et al.(2006)]{lee06} Lee, J.~H., Lee, M.~G., \& Hwang, H.~S.\ 2006, \apj, 650, 148
\bibitem[Lemaux et al.(2012)]{lemaux11} Lemaux, B.~C., Gal, R.~R., Lubin, L.~M., et al.\ 2012, \apj, 745, 106
\bibitem[Lemaux et al.(2009)]{lemaux09} Lemaux, B.~C., Lubin, L.~M., Sawicki, M., et al.\ 2009, \apj, 700, 20
\bibitem[Lemaux et al.(2010)]{lemaux10} Lemaux, B.~C., Lubin, L.~M., Shapley, et al.\ 2010, \apj, 716, 970
\bibitem[Lin et al.(2007)]{lin07} Lin, L., Koo, D.~C., Weiner, B.~J., et al.\ 2007, \apjl, 660, L51
\bibitem[Lubin et al.(2000)]{lubin00} Lubin, L.~M., Brunner, R., Metzger, M.~R., Postman, M., \& Oke, J.~B.\ 2000, \apjl, 531, L5
\bibitem[Lubin et al.(2009)]{lub09} Lubin, L.~M., Gal, R.~R., Lemaux, B.~C., Kocevski, D.~D., \& Squires, G.~K.\ 2009, \aj, 137, 4867
\bibitem[Lubin et al.(2004)]{lubin04} Lubin, L.~M., Mulchaey, J.~S., \& Postman, M.\ 2004, \apjl, 601, L9 
\bibitem[Lubin et al.(2002)]{lubin02} Lubin, L.~M., Oke, J.~B., \& Postman, M.\ 2002, \aj, 124, 1905 
\bibitem[Lubin et al.(1998)]{lub98} Lubin, L.~M., Postman, M., \& Oke, J.~B.\ 1998, \aj, 116, 643  
\bibitem[Mann et al.(2002)]{mann02} Mann, R.~G., Oliver, S., Carballo, R., et al.\ 2002, \mnras, 332, 549 
\bibitem[Mann et al.(1997)]{mann97} Mann, R.~G., Oliver, S.~J., Serjeant, S.~B.~G., et al.\ 1997, \mnras, 289, 482
\bibitem[Marconi \& Hunt(2003)]{MH03} Marconi, A., \& Hunt, L.~K.\ 2003, \apjl, 59, L21
\bibitem[Marconi et al.(2004)]{marc04} Marconi, A., Risaliti, G., Gilli, R., et al.\ 2004, \mnras, 351, 169
\bibitem[Martel et al.(2007)]{martel07} Martel, A.~R., Menanteau, F., Tozzi, P., Ford, H.~C., \& Infante, L.\ 2007, \apjs, 168, 19
\bibitem[Martini et al.(2007)]{mart07} Martini, P., Mulchaey, J.~S., \& Kelson, D.~D.\ 2007, \apj, 664, 761
\bibitem[McLean et al.(1998)]{mclean98} McLean, I.~S., Becklin, E.~E., Bendiksen, O., et al.\ 1998, \procspie, 3354, 566
\bibitem[McNamara \& Nulsen(2007)]{mc07} McNamara, B.~R., \& Nulsen, P.~E.~J.\ 2007, \araa, 45, 117 
\bibitem[Menanteau et al.(2001)]{men01} Menanteau, F., Jimenez, R., \& Matteucci, F.\ 2001, \apjl, 562, L23
\bibitem[Merloni \& Heinz(2008)]{merl08} Merloni, A., \& Heinz, S.\ 2008, \mnras, 388, 1011
\bibitem[Mihos \& Hernquist(1996)]{MiHern96} Mihos, J.~C., \& Hernquist, L.\ 1996, \apj, 464, 641 
\bibitem[Murdoch et al.(1973)]{murdoch73} Murdoch, H.~S., Crawford, D.~F., \& Jauncey, D.~L.\ 1973, \apj, 183, 1 
\bibitem[Nagao et al.(2006)]{nagao06} Nagao, T., Maiolino, R., \& Marconi, A.\ 2006, \aap, 459, 85
\bibitem[Nandra et al.(2007)]{nandra07} Nandra, K., Georgakakis, A., Willmer, C.~N.~A., et al.\ 2007, \apjl, 660, L11
\bibitem[Oemler et al.(2009)]{oemler09} Oemler, A., Jr., Dressler, A., Kelson, D., et al.\ 2009, \apj, 693, 152 
\bibitem[Oke et al.(1995)]{oke95} Oke, J.~B., Cohen, J.~G., Carr, M., et al.\ 1995, \pasp, 107, 375
\bibitem[Oke et al.(1998)]{oke98} Oke, J.~B., Postman, M., \& Lubin, L.~M.\ 1998, \aj, 116, 549 
\bibitem[Pierce et al.(2007)]{pierce07} Pierce, C.~M., Lotz, J.~M., Laird, E.~S., et al.\ 2007, \apjl, 660, L19
\bibitem[Poggianti \& Barbaro(1997)]{pogg97} Poggianti, B.~M., \& Barbaro, G.\ 1997, \aap, 325, 1025 
\bibitem[Poggianti et al.(1999)]{pogg99} Poggianti, B.~M., Smail, I., Dressler, A., et al.\ 1999, \apj, 518, 576 
\bibitem[Postman et al.(1998)]{post98} Postman, M., Lubin, L.~M., \& Oke, J.~B.\ 1998, \aj, 116, 560 
\bibitem[Postman et al.(2001)]{post01} Postman, M., Lubin, L.~M., \& Oke, J.~B.\ 2001, \aj, 122, 1125
\bibitem[Rieke et al.(2004)]{rieke04} Rieke, G.~H., Young, E.~T., Engelbracht, C.~W., et al.\ 2004, \apjs, 154, 25
\bibitem[Rosario et al.(2011)]{ros11} Rosario, D.~J., Mozena, M., Wuyts, S., et al.\ 2011, arXiv:1110.3816 
\bibitem[Rosati et al.(2002)]{rosati02} Rosati, P., Tozzi, P., Giacconi, R., et al.\ 2002, \apj, 566, 667
\bibitem[Rose(1985)]{rose85} Rose, J.~A.\ 1985, \aj, 90, 1927 
\bibitem[Rutledge et al.(2000)]{rut00} Rutledge, R.~E., Brunner, R.~J., Prince, T.~A., \& Lonsdale, C.\ 2000, \apjs, 131, 335
\bibitem[S{\'a}nchez et al.(2005)]{sanchez05} S{\'a}nchez, S.~F., Becker, T., Garcia-Lorenzo, B., et al.\ 2005, \aap, 429, L21 
\bibitem[S{\'a}nchez et al.(2004)]{sanchez04} S{\'a}nchez, S.~F., Jahnke, K., Wisotzki, L., et al.\ 2004, \apj, 614, 586
\bibitem[Schawinski et al.(2007)]{scha07} Schawinski, K., Thomas, D., Sarzi, M., et al.\ 2007, \mnras, 382, 1415
\bibitem[Schawinski et al.(2009)]{scha09} Schawinski, K., Virani, S., Simmons, B., et al.\ 2009, \apjl, 692, L19
\bibitem[Schiavon et al.(2006)]{sch06} Schiavon, R.~P., Faber, S.~M., Konidaris, N., et al.\ 2006, \apjl, 651, L93 
\bibitem[Shields et al.(2003)]{shields03} Shields, G.~A., Gebhardt, K., Salviander, S., et al.\ 2003, \apj, 583, 124 
\bibitem[Silverman et al.(2009a)]{silver09} Silverman, J.~D., Kova{\v c}, K., Knobel, C., et al.\ 2009a, \apj, 695, 171
\bibitem[Silverman et al.(2009b)]{silver09b} Silverman, J.~D., Lamareille, F., Maier, C., et al.\ 2009b, \apj, 696, 396 
\bibitem[Silverman et al.(2008)]{silver08} Silverman, J.~D., Mainieri, V., Lehmer, B.~D., et al.\ 2008, \apj, 675, 1025
\bibitem[Simcoe et al.(2000)]{simcoe00} Simcoe, R.~A., Metzger, M.~R., Small, T.~A., \& Araya, G.\ 2000, Bulletin of the American Astronomical Society, 32, 758
\bibitem[Somerville et al.(2008)]{somer08} Somerville, R.~S., Hopkins, P.~F., Cox, T.~J., Robertson, B.~E., \& Hernquist, L.\ 2008, \mnras, 391, 481
\bibitem[Springel et al.(2005)]{spring05} Springel, V., Di Matteo, T., \& Hernquist, L.\ 2005, \mnras, 361, 776 
\bibitem[Stott et al.(2009)]{stott09} Stott, J.~P., Pimbblet, K.~A., Edge, A.~C., Smith, G.~P., \& Wardlow, J.~L.\ 2009, \mnras, 394, 2098
\bibitem[Strateva et al.(2001)]{strat01} Strateva, I., Ivezic, Z., Knapp, G.~R., et al.\ 2001, \aj, 122, 1861
\bibitem[Sutherland \& Saunders(1992)]{SS92} Sutherland, W., \& Saunders, W.\ 1992, \mnras, 259, 413
\bibitem[Taylor et al.(2005)]{taylor05} Taylor, E.~L., Mann, R.~G., Efstathiou, A.~N., et al.\ 2005, \mnras, 361, 1352 
\bibitem[Tozzi et al.(2001)]{tozzi01} Tozzi, P., Rosati, P., Nonino, M., et al.\ 2001, \apj, 562, 42
\bibitem[Tremaine et al.(2002)]{trem02} Tremaine, S., Gebhardt, K., Bender, R., et al.\ 2002, \apj, 574, 740
\bibitem[Trouille et al.(2011)]{tro11} Trouille, L., Barger, A.~J., \& Tremonti, C.\ 2011, \apj, 742, 46
\bibitem[Weiner et al.(2005)]{weiner05} Weiner, B.~J., Phillips, A.~C.. Faber, S.~M., et al.\ 2005, \apj, 620, 595
\bibitem[Wild et al.(2010)]{wild10} Wild, V., Heckman, T., \& Charlot, S.\ 2010, \mnras, 405, 933
\bibitem[Xue et al.(2010)]{xue10} Xue, Y.~Q., Brandt, W.~N., Luo, B., et al.\ 2010, \apj, 720, 368
\bibitem[Yan et al.(2006)]{yan06} Yan, R., Newman, J.~A., Faber, S.~M., et al.\ 2006, \apj, 648, 281

\end{thebibliography}
\end{document}